 \documentclass[preprint2]{aastex}
\usepackage{longtable,lscape,rotating,pict2e}

\graphicspath{{figures-paper/}{figures-lab/}}

\newcommand{\cs}  {cm$^{-2}$}
\newcommand{\wn}  {cm$^{-1}$}

\newcommand{\hii}{\ion{H}{2}}

\newcommand{\m}{CH$_3$OH}
\newcommand{\ammonia}{NH$_3$}
\newcommand{\methane}{CH$_4$}
\newcommand{\ammonium}{NH$_4^+$}
\newcommand{\ocn}{OCN$^-$}
\newcommand{\water}{H$_2$O}
\newcommand{\dime}{CH$_3$OCH$_3$}

\newcommand{\mc}{CH$_3$CN}

\newcommand{\wav}{cm$^{-1}$}
\newcommand{\methanol}{CH$_3$OH }
\newcommand{\modeCO}{$\nu\rm_{CO}$ }
\newcommand{\modeNH}{$\nu\rm_{NH3}$ }
\newcommand{\modeCH}{$\nu\rm_{CH3}$ }

\newcommand{\spitzer}{{\it Spitzer}}

\shorttitle{\ammonia\ and \m\ in ices around low-mass YSOs}
\shortauthors{Bottinelli et al.}

\begin{document}

\title{
The c2d \spitzer\ spectroscopic survey of ices 
around low-mass young stellar objects. IV. NH$_3$ and \m\
}

\author{Sandrine~Bottinelli\altaffilmark{1,2,3},
{A.~C.~Adwin~Boogert\altaffilmark{4}},
{Jordy~Bouwman\altaffilmark{5}},
{Martha~Beckwith\altaffilmark{5,6}},
{Ewine~F.~van Dishoeck\altaffilmark{1,7}},
{Karin~I.~\"Oberg\altaffilmark{5,8}},
{Klaus~M.~Pontoppidan\altaffilmark{9}},
{Harold~Linnartz\altaffilmark{5}},
{Geoffrey~A.~Blake\altaffilmark{9}},
{Neal~J.~Evans II\altaffilmark{10}},
{Fred~Lahuis\altaffilmark{11,1}}
}
\altaffiltext{1}{Leiden Observatory, Leiden University, 
P.O. Box 9513, NL 2300 RA Leiden, The Netherlands.
}
\altaffiltext{2}{Present address: Centre d'Etude Spatiale des Rayonnements (CESR),
CNRS-UMR 5187,
9 avenue du Colonel Roche, BP 4346, 31028 Toulouse Cedex 4, France.\\
{\tt sandrine.bottinelli@cesr.fr}
}
\altaffiltext{3}{CESR, Universit\'e de Toulouse [UPS], France}
\altaffiltext{4}{IPAC, NASA Herschel Science Center, Mail Code 100-22, 
California Institute of Technology, Pasadena, CA 91125, USA.}
\altaffiltext{5}{Raymond and Beverly Sackler Laboratory for Astrophysics, Leiden Observatory, 
Leiden University, P.O. Box 9513, NL 2300 RA Leiden, the Netherlands.}
\altaffiltext{6}{Present address: Department of Chemistry and Chemical Biology,
Cornell University, Ithaca, New York 14853}
\altaffiltext{7}{Max-Planck Institute f\"ur Extraterrestrische Physik,
Giessenbachstr. 1, D-85748 Garching, Germany}
\altaffiltext{8}{Present address: Harvard-Smithsonian Center for Astrophysics, 60 Garden Street, Cambridge, MA 02138, USA}
\altaffiltext{9}{California Institute of Technology, Division of Geological and Planetary Sciences, 
Pasadena, CA 91125, USA.}
\altaffiltext{10}{Department of Astronomy, University of Texas at Austin, 
1 University Station C1400, Austin, TX 78712-0259, USA.}
\altaffiltext{11}{SRON Netherlands Institute for Space Research, 
PO Box 800, NL 9700 AV Groningen, The Netherlands.}

\begin{abstract}
\ammonia\ and \m\ are key molecules in astrochemical networks leading
to the formation of more complex N- and O-bearing molecules, such as
\mc\ and \dime.  Despite a number of recent studies, little is
known about their abundances in the solid state. This is particularly the case for
low-mass protostars, for which only the launch of
the \spitzer\ Space Telescope has permitted high sensitivity
observations of the ices around these objects.  In this work, we
investigate the $\sim 8-10$~\micron\ region in the \spitzer\ IRS (InfraRed
Spectrograph) spectra of 41 low-mass young stellar objects (YSOs).  
These data are
part of a survey of interstellar ices in a sample of low-mass YSOs
studied in earlier papers in this series.  We used both an empirical
and a local continuum method to correct for the contribution from the
10~\micron\ silicate absorption in the recorded spectra.  In addition,
we conducted a systematic laboratory study of \ammonia- and
\m-containing ices to help interpret the astronomical spectra. 
We clearly detect a feature at 
$\sim$9~\micron\ in 24 low-mass YSOs. 
Within the uncertainty in continuum determination, we identify
this feature with the \ammonia\ $\nu_2$ umbrella mode, and derive
abundances with respect to water
between $\sim$2 and 15\%.  Simultaneously, we also revisited the case
of \m\ ice by studying the $\nu_4$ C--O stretch mode of this molecule
at $\sim$9.7~\micron\ in 16 objects, yielding abundances consistent
with those derived by \citet{boogert-etal08} (hereafter paper I) based on a simultaneous 9.75
and 3.53~$\mu$m data analysis.  
Our study indicates that
\ammonia\ is present primarily in \water-rich ices, but that in some
cases, such ices are insufficient
to explain the observed narrow FWHM. 
The laboratory data point to \m\ being
in an almost pure methanol ice, or mixed mainly with CO or
CO$_2$, consistent with its formation through hydrogenation
on grains.
Finally, we use our derived \ammonia\ abundances in combination with
previously published abundances of other solid N-bearing species to find that
up to 10--20~\% of nitrogen is locked up in known ices.
\end{abstract}


\section{Introduction}

Ammonia and methanol are among the most ubiquitous and abundant 
(after H$_2$ and CO)
molecules in space. Gaseous \ammonia\ and \m\ are found in a variety
of environments such as infrared dark clouds, dense gas surrounding ultra-compact \hii\
regions, massive hot cores, hot corinos, and comets. Solid \m\ has
been observed in the ices surrounding massive YSOs
(e.g. \citealt{schutte-etal91,dartois-etal99,gibb-etal04}) and more
recently toward low-mass protostars \citep{pontoppidan-etal03}. The
presence of solid \ammonia\ has been claimed toward massive YSOs only
\citep{lacy-etal98,dartois-etal02,gibb-etal04,guertler02}, with the exception
of a possible detection in the low-mass object IRAS 03445+3242 
\citep{guertler02}. However, these detections
are still controversial and ambiguous \citep{taban-etal03}.

Both molecules are key participants in gas-grain chemical networks
resulting in the formation of more complex N- and O-bearing molecules,
such as \mc\ and \dime\ (e.g. \citealt{rodgers+charnley01}).
Moreover, UV processing of \ammonia- and \m-containing ices has been
proposed as a way to produce amino-acids and other complex organic
molecules (e.g.,
\citealt{munozcaro+schutte03,bernstein-etal02,oberg-etal09}). In
addition, the amount of \ammonia\ in ices has a direct impact on the
content of ions such as NH$_4^+$ and OCN$^-$, which form reactive
intermediates in solid-state chemical networks. A better knowledge of
the \ammonia\ and \m\ content in interstellar ices will thus help to
constrain chemical models and to gain a better understanding of the
formation of more complex, prebiotic, molecules.

During the pre-stellar phase, \ammonia\ is known to freeze out on grains 
(if the core remains starless long enough -- \citealt{lee-etal04}). Moreover,
\m\ is known to have gas-phase abundances with
respect to H$_2$ in hot cores/corinos that are much larger than in
cold dense clouds: $\sim (1-10)\times 10^{-6}$ vs $\leq 10^{-7}$, with
the former values most likely representing evaporated ices in warm
regions
\citep[e.g.][]{genzel-etal82,blake-etal87,federman-etal90}. Together, these findings
suggest that ices are an important reservoir of \ammonia\ and \m\ and
that prominent features should be seen in the absorption spectra
toward high- and low-mass protostars. Unfortunately, as summarized in
Table~\ref{tab:features}, \ammonia\ and \m\ bands, with the exception
of the 3.53~\micron\ \m\ feature, are often blended with deep water
and/or silicate absorptions, complicating unambiguous identifications
and column density measurements. This is particularly true for
\ammonia\ whose abundance determination based on the presence of an
ammonium hydrate feature at 3.47~\micron\ remains controversial
(e.g. \citealt{dartois+dhendecourt01}).
Nonetheless, it is important to use all available constraints to
accurately determine the abundances of these two molecules. Despite
the overlap with the 10~\micron\ silicate (Si--O stretch)
feature, the \ammonia\ $\nu_2$ umbrella mode at $\sim$9~\micron\
($\sim$1110~\wn) offers a strong intrinsic
absorption cross section and appears as the most promising feature to
determine the abundance of this species in the solid phase. Moreover,
the \m\ $\nu_4$ C--O stretch at $\sim$9.7~\micron\ ($\sim$1030~\wn)
provides a good check on the validity of the different methods we
will use to subtract the 10-\micron\ silicate absorption,
since the abundance of this molecule
has been accurately determined previously
from both the 3.53 and 9.75~\micron\ features (see Paper I).

\begin{deluxetable}{lccp{6cm}}
\tablewidth{0pt}
\tablecaption{Selected near- and mid-infrared features of \ammonia\ and \m. \label{tab:features}}
\tablehead{Mode	& $\lambda$ (\micron)	& $\bar{\nu}$ (cm$^{-1}$) & Problem}
\startdata
\sidehead{\ammonia\ features:} 
$\nu_3$ N--H stretch	& \phn 2.96	& 3375	& Blended with \water\ (O--H stretch, 3.05~\micron/3275~\wn) \\
$\nu_4$ H--N--H bend	& \phn 6.16	& 1624	& Blended with \water\ (H--O--H bend, 5.99~\micron/1670~\wn), HCOOH\\
{\bf \boldmath{$\nu_2$} umbrella}		& {\bf \phn 9.00}	& {\bf 1110}	& {\bf Blended with silicate} \\
\hline
\sidehead{\m\ features:} 
$\nu_2$ C--H stretch	& \phn 3.53	& 2827	& --\\
$\nu_6$ \& $\nu_3$ --CH$_3$ deformation	& \phn 6.85	& 1460	& Blended (e.g. with NH$_4^+$)\\
$\nu_7$ --CH$_3$ rock	& \phn 8.87	& 1128	& Weak; blended with silicate\\
{\bf \boldmath{$\nu_4$} C--O stretch}	& {\bf \phn9.75}	& {\bf 1026}	& {\bf Blended with silicate} \\
Torsion		& 14.39		& \phn695	& Blended with \water\ libration mode\\
\enddata
\tablecomments{The bold-faced lines indicate the features studied here.}
\tablecomments{The nomenclature for the NH$_3$ and CH$_3$OH vibrational modes are adopted from
\citet{herzberg45}. }
\end{deluxetable}

More detailed spectroscopic information is particularly interesting
for low-mass protostars as the ice composition reflects the conditions
during the formation of Sun-like stars.  Such detections have only become
possible with \spitzer, whose sensitivity is necessary to observe low
luminosity objects even in the nearest star-forming clouds.
The spectral resolution of
the \spitzer\ Infrared Spectrograph (IRS, \citealt{houck-etal04}) of $\Delta
\lambda/\lambda \sim 100$ in this wavelength range is comparable to
that of the Infrared Space Observatory (ISO) PHOT-S instrument but
lower than that of the ISO-SWS and other instruments used to identify solid \ammonia\
toward high luminosity sources.
The spectral appearance of ice absorption features,
such as band shape, band position and integrated band strength, is
rather sensitive to the molecular environment. 
Thus, the interpretation of the astronomical spectra should be supported by a
systematic laboratory study of interstellar ice analogues containing
\ammonia\ and \m.
Changes in the lattice
geometry and physical conditions of an ice are directly reflected by
variations in these spectral properties. In the laboratory, it is
possible to record dependencies over a wide range of astrophysically
relevant parameters, most obviously ice composition, mixing ratios,
and temperature. Such laboratory data exist for pure and some
H$_2$O-rich NH$_3$- and CH$_3$OH-containing ices
\citep[e.g.,][]{dhendecourt+allamandola86,hudgins-etal93,kerkhof-etal99,taban-etal03},
but a systematic study and comparison with observational spectra is
lacking.

In principle, the molecular environment also provides information on
the formation pathway of the molecule. For example, NH$_3$ ice is
expected to form simultaneously with H$_2$O and CH$_4$ ice in the
early, low-density molecular cloud phase from hydrogenation of N atoms
\citep[e.g.,][]{tielens+hagen82}.  In contrast, solid CH$_3$OH is
thought to result primarily from hydrogenation of solid CO, a process
which has been confirmed to be rapid at low temperatures in several
laboratory experiments
\citep[e.g.][]{watanabe+kouchi02,hidaka-etal04,fuchs-etal09}. A
separate, water-poor layer of CO ice is often found on top of the
water-rich ice layer in low-mass star-forming regions due to the
`catastrophic' freeze-out of gas-phase CO at high densities
\citep{pontoppidan-etal03,pontoppidan06}. Hydrogenation of this CO
layer should lead to a nearly pure CH$_3$OH ice layer
\citep[e.g.,][]{cuppen-etal09}, which will have a different
spectroscopic signature from that of CH$_3$OH embedded in a water-rich
matrix.  The latter signature would be expected if CH$_3$OH ice were
formed by hydrogenation of CO in a water-rich environment or by
photoprocessing of H$_2$O:CO ice mixtures, another proposed route
\citep[e.g.,][]{moore+hudson98}.

Here, we present 
\spitzer\ spectra between 5 and 35~\micron\ of ices surrounding
41 low-mass protostars, focusing on the $\sim 8-10$~\micron\ region that
contains the $\nu_2$ umbrella and $\nu_4$ C--O stretch modes of
\ammonia\ and \m, respectively. This work is the fourth paper in a
series of ice studies
\citep{boogert-etal08,pontoppidan-etal08-co2,oberg-etal08-ch4} carried
out in the context of the \spitzer\ Legacy Program ``From Molecular
Cores to Planet-Forming Disks'' (``c2d''; \citealt{evans-etal03}). 
In Section~\ref{sec:astro},
we carry out the analysis of the \spitzer\ data in $8-10~\micron$ range. In
Section~\ref{sec:lab}, we present the laboratory data specifically
obtained to help interpret the data that are discussed in 
Section~\ref{sec:comparison}. Finally, we conclude  in Section~\ref{sec:ccl} 
with a short discussion of the joint
astronomy-laboratory work
(including the overall continuum determination).

\section{Astronomical observations and analysis}
\label{sec:astro}

The source sample consists of 41 low-mass YSOs that were selected
based on the presence of ice absorption features. The entire sample
spans a wide range of spectral indices $\alpha$ = $-0.25$ to $+2.70$,
with $\alpha$ defined as $d\log (\lambda F_\lambda) / d\log
(\lambda)$, where $d$ indicates the derivative, and
$F_\lambda$ represents all the photometric fluxes available between
$\lambda$ = 2.17~\micron\ (2MASS K$_{\rm s}$-band) and 
$\lambda$ = 24~$\mu$m (\spitzer/MIPS band). 
In the infrared broad-band
classification scheme, 35 out of 41 objects fall in the embedded Class
0/I category ($\alpha > 0.3$). The remaining 6 objects are
flat-spectrum type objects ($-0.3 < \alpha < 0.3$;
\citealt{greene-etal94}).\\ \spitzer/IRS spectra (5-35\,\micron) were
obtained as part of the c2d Legacy program (PIDs 172 and 179), as well
as a dedicated open time program (PID 20604), and several previously 
published GTO spectra \citep{watson-etal04}.  We refer the
reader to Table~1 and Section~3 of \citet{boogert-etal08} for the
source coordinates and a description of the data reduction process
(including overall continuum determination).\\

As mentioned previously, spectral signatures in the $\sim
8-10$~\micron\ region are dominated by the Si--O stretching mode of
silicates.  The overall shape as well as the sub-structure of the silicate feature
depend on grain size, mineralogy, level of crystallinity. These effects are degenerate and 
so these different factors cannot be easily separated.  
For example, large grains and the presence of SiC both
produce a shoulder at 11.2~\micron\
\citep[e.g.,][]{min-etal07}. Therefore, trying to fit the 10~\micron\
silicate feature by determining the composition and size of the grains
is a complex process.  For this reason, we use two alternative
methods to model the silicate profile and extract the \ammonia\ (and \m) feature(s) from the
underlying silicate absorption.

\subsection{Local continuum}

\begin{figure*}[ht]
\centering
\includegraphics[width=0.95\textwidth,bb=12 175 599 595]{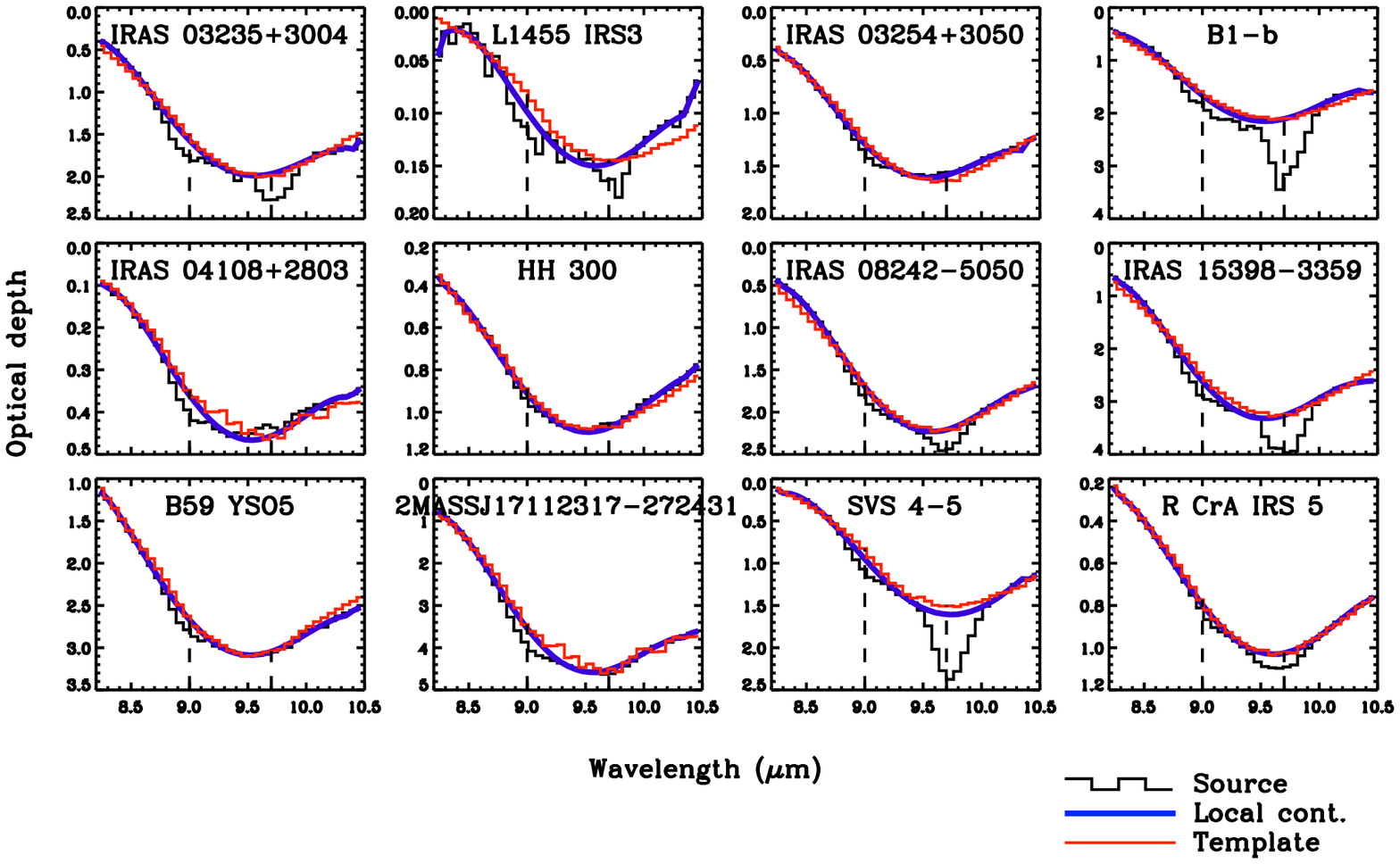}
\includegraphics[width=0.95\textwidth,bb=12 175 599 575]{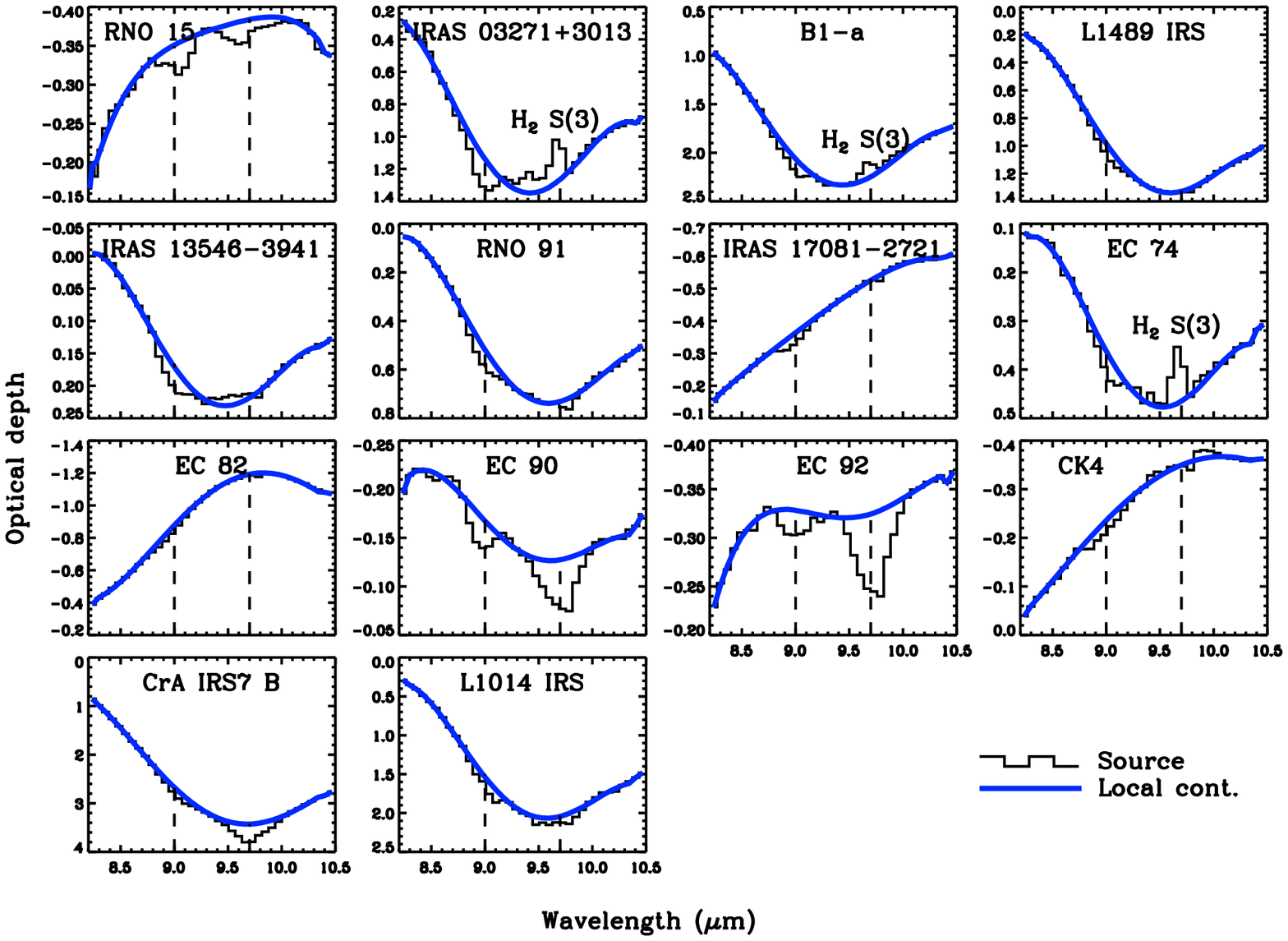}
\caption{\footnotesize ({\it Top}) Local continuum (thick blue/black lines) and template (red/grey lines) fits to all sources
for which a template could be found. (See Section \ref{sec:template} for details) ---
({\it Bottom}) Local continuum fits to emission sources or sources for
which no reasonable template could be found.
}
\label{fig:continuum}
\end{figure*}

The first method uses a local continuum to fit the shape of the
silicate absorption. For this, we fit a fourth order polynomial over
the wavelength ranges 8.25--8.75, 9.23--9.37, and 9.98--10.4~\micron,
avoiding the positions where \ammonia\ and \m\ absorb around 9 and
9.7~\micron. These fits are shown with thick black lines in
Fig.~\ref{fig:continuum}.  After subtraction of the local continuum
from the observations, we fit a Gaussian to the remaining \ammonia\
and/or \m\ feature, when present, as shown in
Fig.~\ref{fig:continuum-subtracted}. The results of the Gaussian fits
are listed in Table~\ref{tab:gaussian param} of
Appendix~\ref{ap:gaussian param}.

\setlength{\unitlength}{1cm}
\thicklines

\begin{figure*}[ht]
\centering
\includegraphics[width=0.95\textwidth,bb=12 205 599 595]{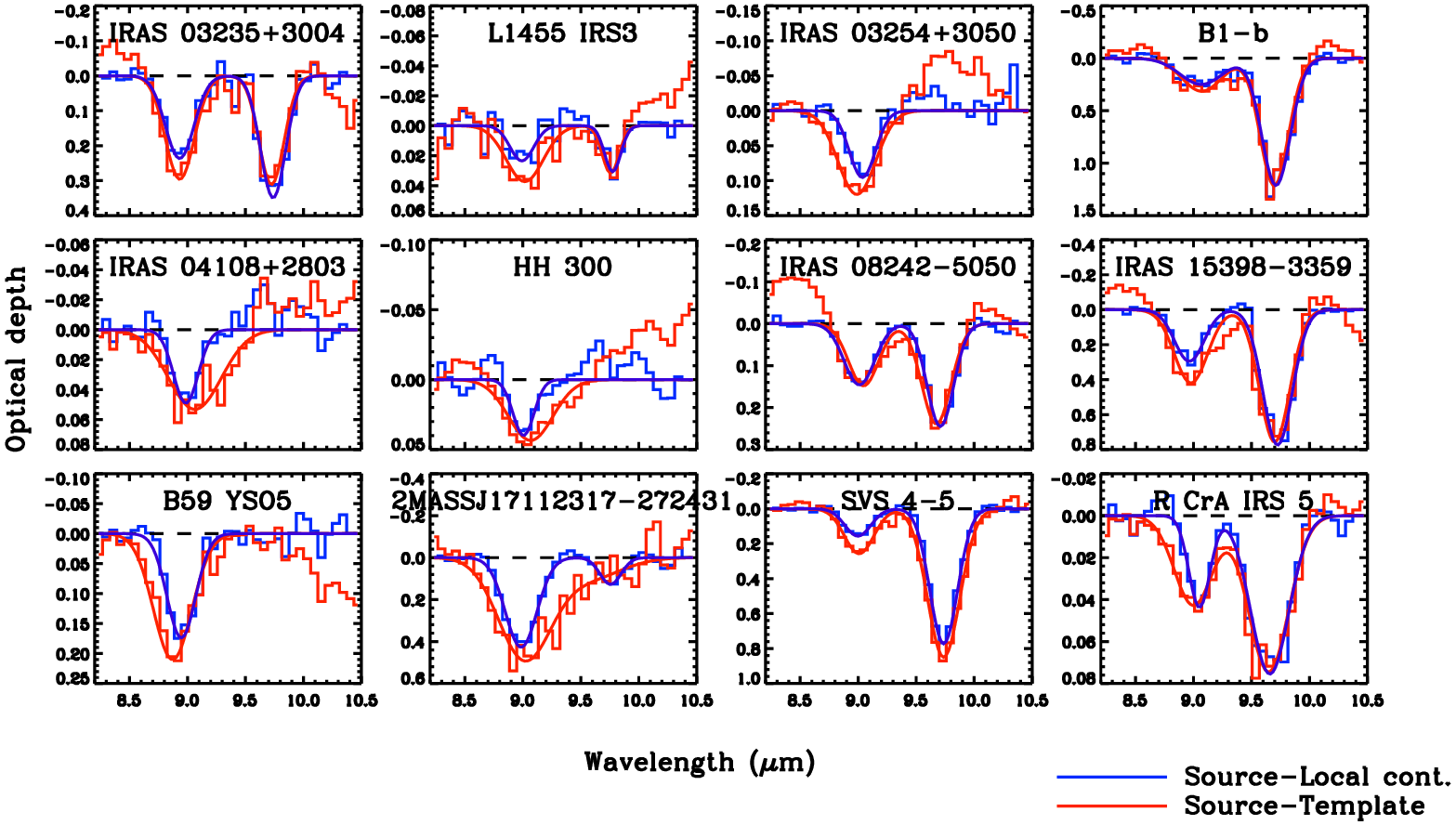}
\includegraphics[width=0.95\textwidth,bb=12 180 599 585]{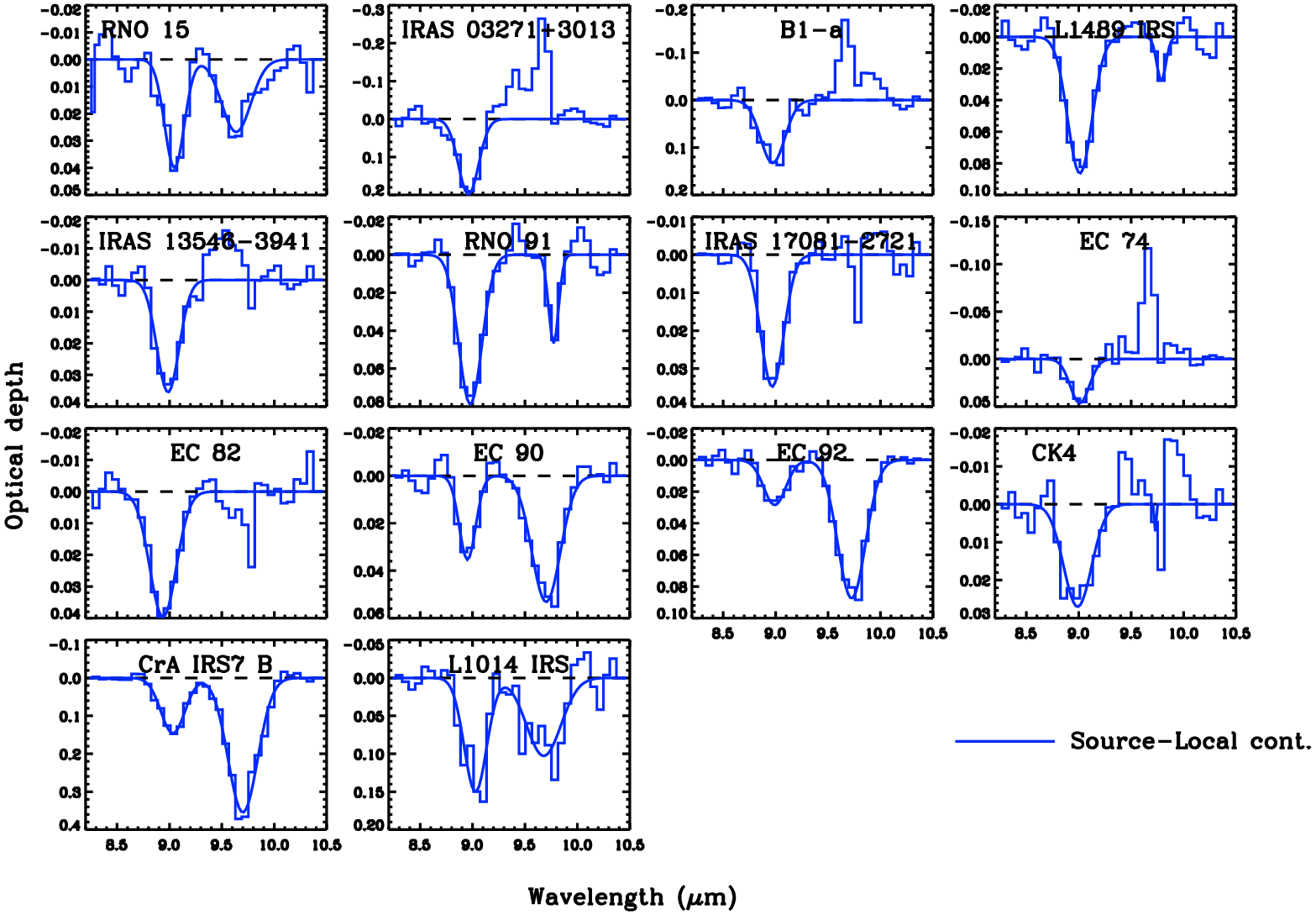}

\caption{\footnotesize ({\it Top}) Residual after removal of local continuum and template fits for all sources
for which a template could be found.  (See Section \ref{sec:template} for details) ---
({\it Bottom}) Residual after removal of local continuum fits for emission sources or sources for
which no reasonable template could be found.
\label{fig:continuum-subtracted}
}
\end{figure*}

\subsection{Template}
\label{sec:template} 

\begin{figure*}[ht]
\includegraphics[height=4.4cm,bb= 25 209 544 611,clip=true]{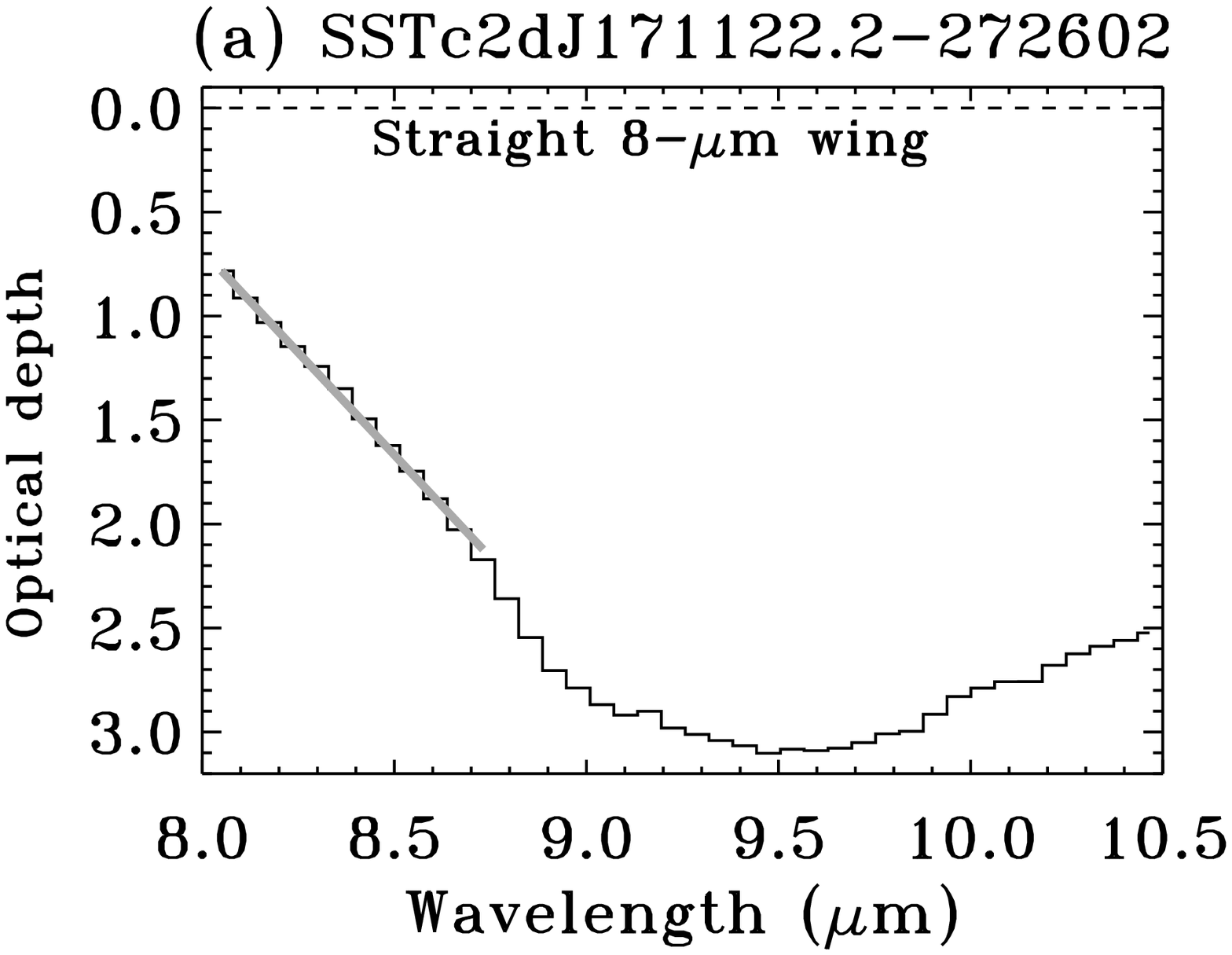}%
\includegraphics[height=4.4cm,bb= 50 209 544 611,clip=true]{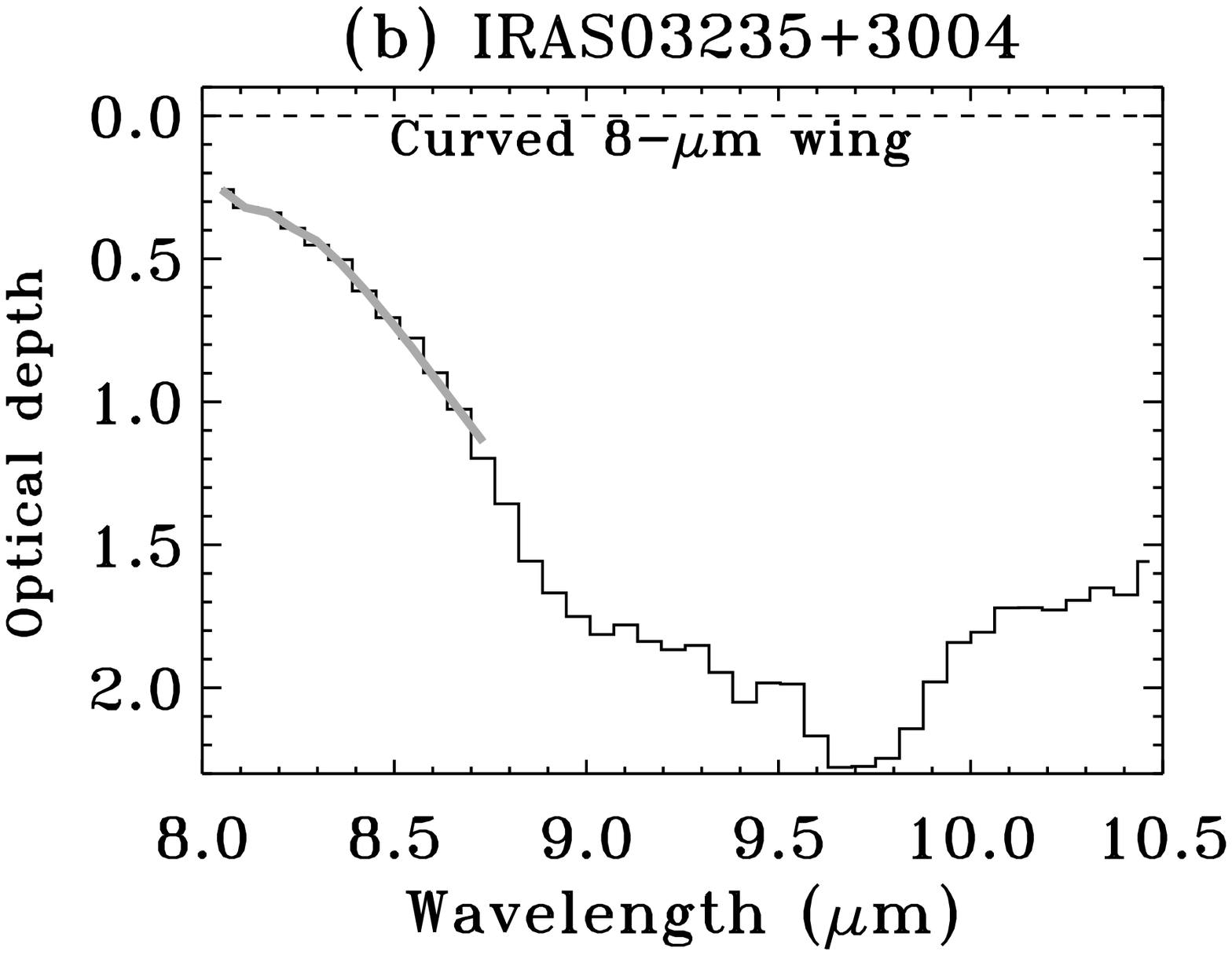}%
\includegraphics[height=4.4cm,bb= 30 209 544 611,clip=true]{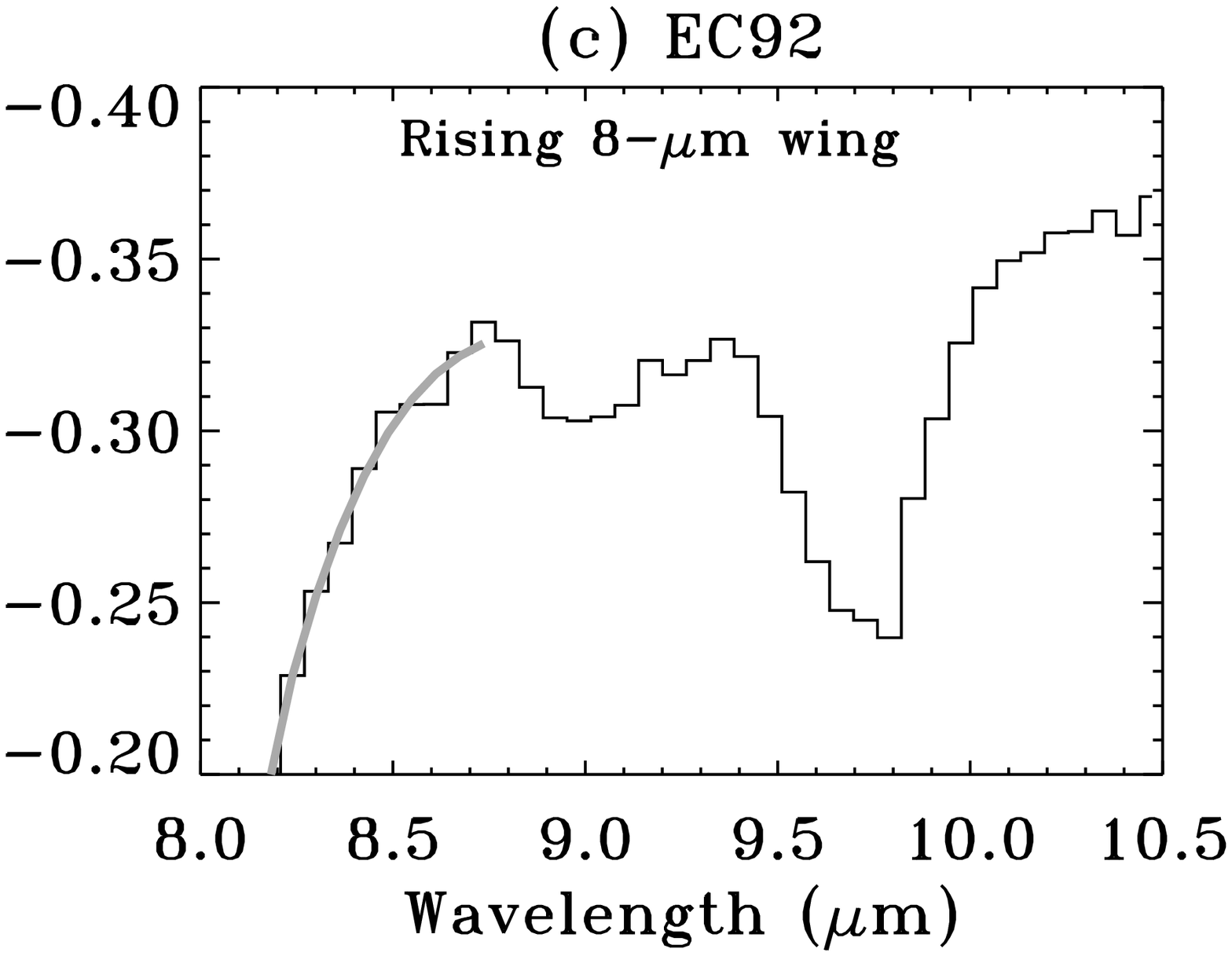}
\setlength{\unitlength}{1cm}
\put(-15.5,3.25){ \vector(1,-1){1} }
\qbezier(-10,3.65)(-9.5,3.65)(-9,2.75)\put(-9.2,2.85){ \vector(1,-2){.1} }
\qbezier(-4.6,1.65)(-4.35,2.55)(-3.85,2.9)\put(-3.9,3.){ \vector(10,10){.01} }
\caption{Examples illustrating the three shapes of the 8~\micron\ wing
shown by the thick grey line: 
(a) straight, (b) curved, and (c) rising.}
\label{fig:wing type}
\end{figure*}

The second method assumes that the 8--10~\micron\ continuum 
can be represented by a template silicate absorption feature, selected
among the observed sources. A comparison of the results obtained using a template 
to those obtained using a simple local continuum provides an estimate of the influence 
of the continuum choice on the shape and depth of the \ammonia\ and \methanol\ features.
The templates were chosen using an empirical method. Upon examination of the 10~\micron\ feature 
of the entire sample, the sources could be separated into three general categories, 
depending on the shape of the wing of the silicate absorption between $\sim$8 and 8.7~\micron\
(which we will refer to as the 8~\micron\ wing): 
 (i) sources with a straight 8~\micron\ wing (Fig.~\ref{fig:wing type}-a),
(ii) sources with a curved 8~\micron\ wing (Fig.~\ref{fig:wing type}-b), and
(iii) sources with a rising 8~\micron\ wing (``emission'' sources, Fig.~\ref{fig:wing type}-c).

Note that, since radiative transfer in the 8--10~\micron\ region can
be complicated by the presence of silicate emission, we only consider
sources that are the least affected by emission, that is those falling
in one of the first two categories.  Nevertheless, non-rising silicate
profiles might still suffer from the presence of emission. 
To try and estimate the impact of this potential effect, we
used two silicate emission sources 
from \citet{kessler-silacci-etal06}, and subtracted these emission profiles
from our absorption profiles, assuming that the emission represented
10 to 50\% of the observed absorption.
After removal of a local continuum, we determined the 
integrated optical depths of the 
\ammonia\ and \m\ 
features in the spectra corrected for emission, and
compared these to the integrated optical depths of the uncorrected spectra.
We find that the difference can be up to a factor of two and therefore
identify this possible presence of underlying emission as the largest 
source of uncertainty in our abundance determinations.

\begin{figure*}[ht]
\centering
\begin{minipage}[c]{0.66\textwidth}
\includegraphics[bb=22 82 599 689,clip=true,width=0.5\textwidth]{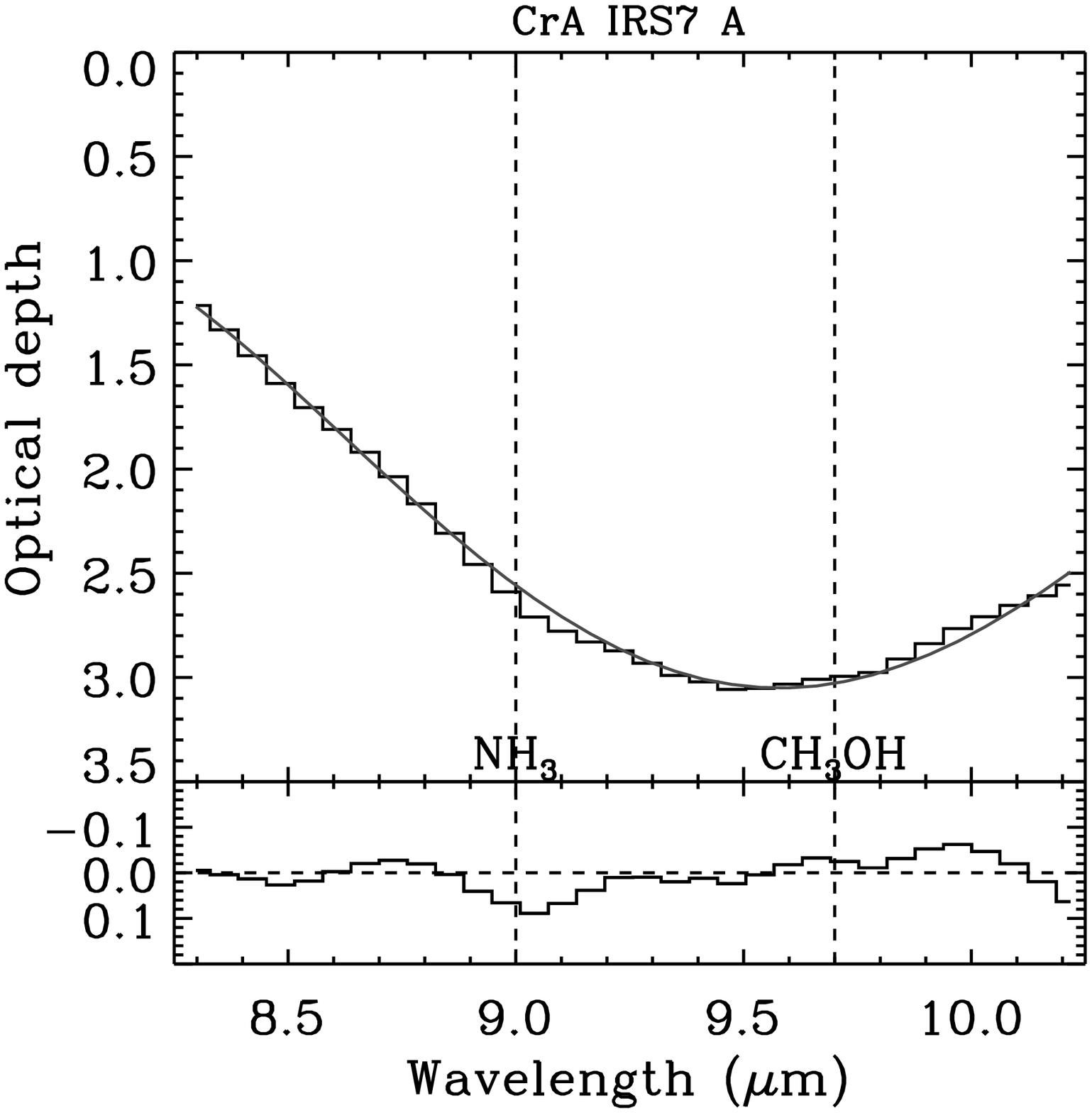}%
\includegraphics[bb=22 82 599 689,clip=true,width=0.5\textwidth]{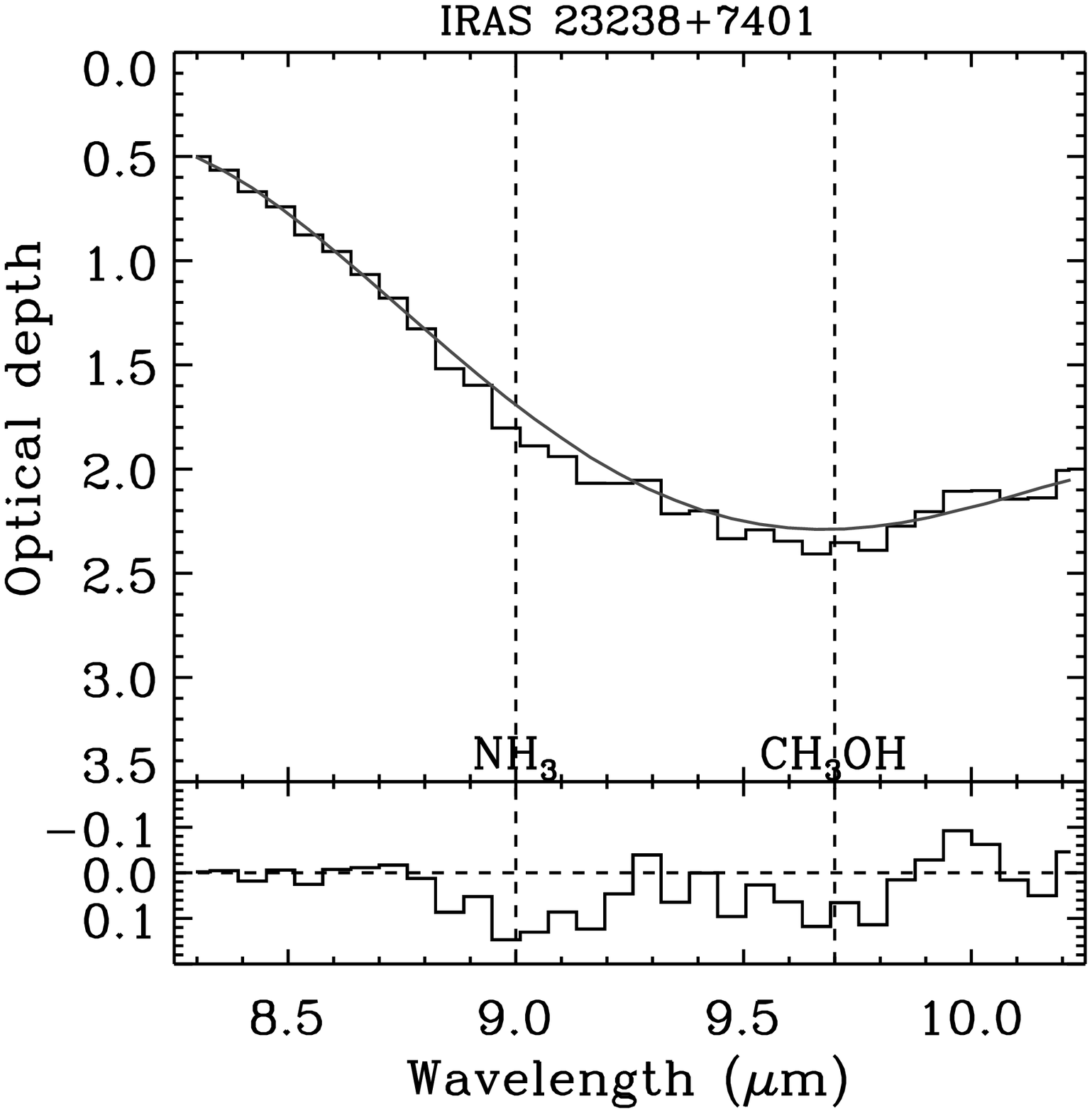}
\includegraphics[bb=22 82 599 689,clip=true,width=0.5\textwidth]{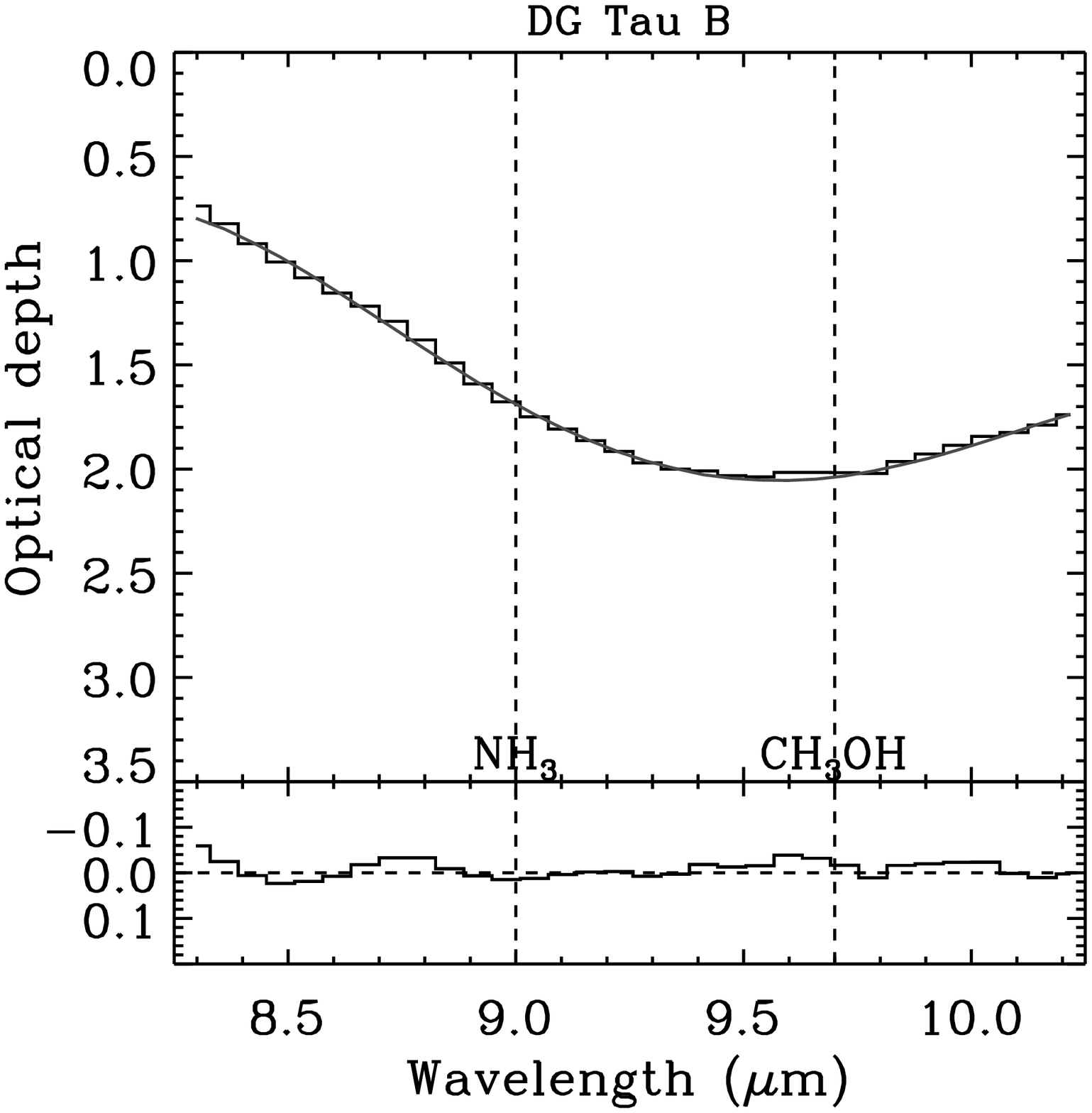}%
\includegraphics[bb=22 82 599 689,clip=true,width=0.5\textwidth]{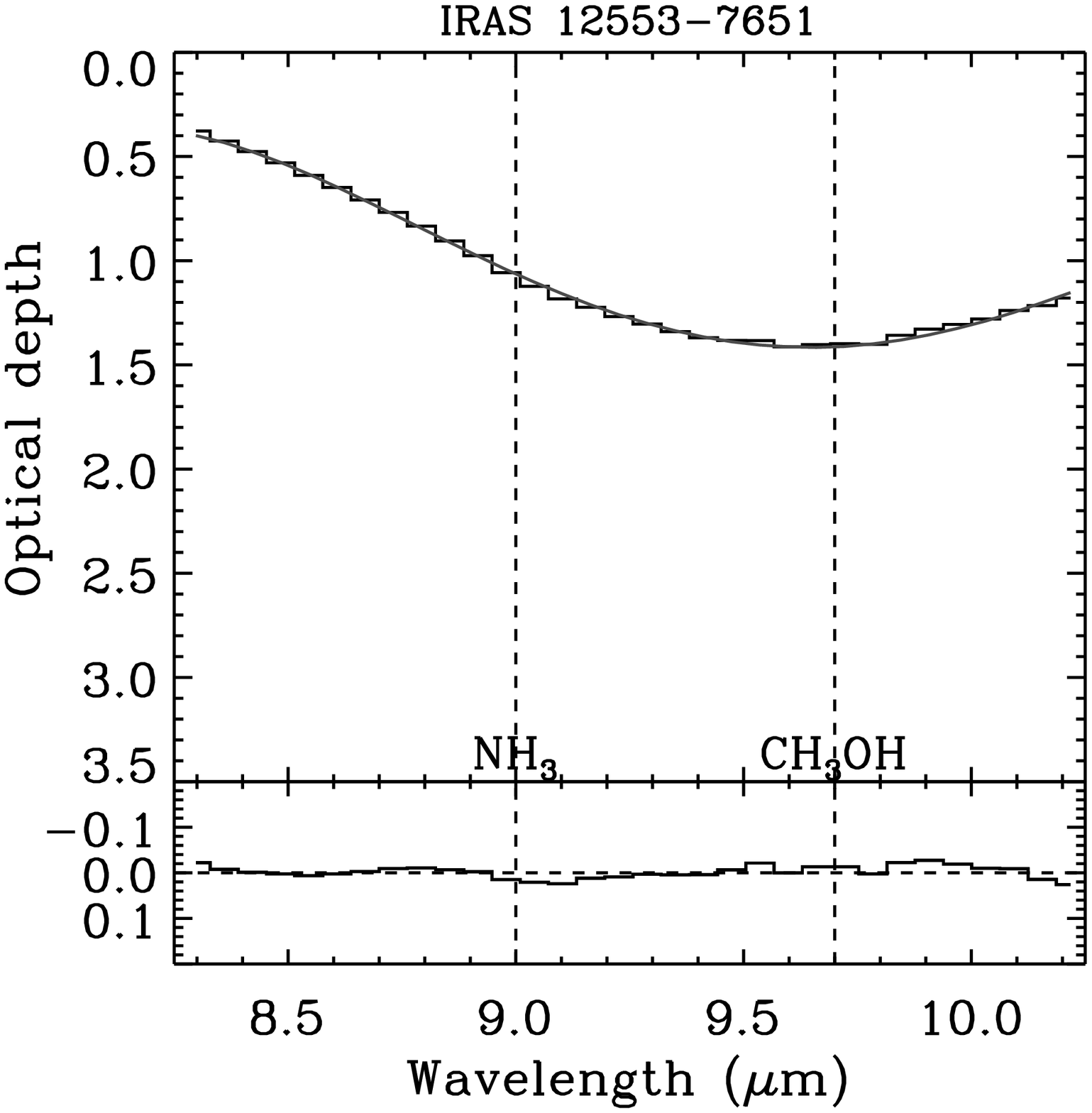}
\end{minipage}\begin{minipage}[c]{0.33\textwidth}
\includegraphics[bb=22 82 599 689,clip=true,width=\textwidth]{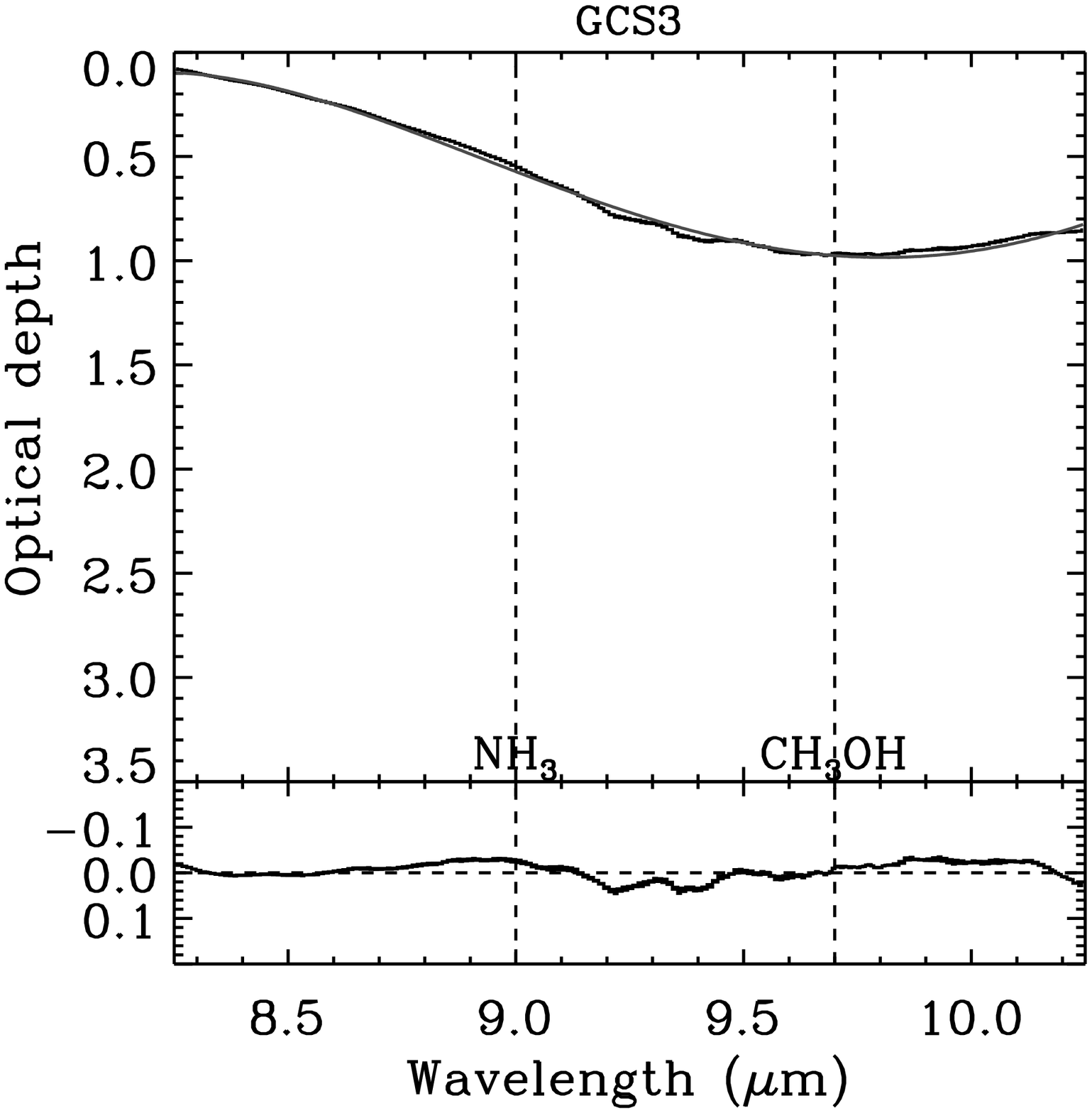}
\end{minipage}
\caption{Silicate features of the sources used as templates for
  a straight 8~\micron\ wing (left), curved 8-\micron\ wing (middle),
  and GCS3 (right). The bottom panels of each plot are the residuals
  after removal of the local continuum shown in grey in the top
  panels. The optical depth scale is kept fixed for comparison. These
  sources are selected to have no or at most weak NH$_3$ and CH$_3$OH
  absorptions.}
\label{fig:templates}
\end{figure*}

For each of the straight and curved 8 \micron\ wings, two sources (in
order to test for template-dependent effects) were selected as
possible templates for the silicate feature.  The selection criteria
were: (i) a silicate feature as deep as possible to minimize the
effects of silicate emission and (ii) little \ammonia\ and \m\ signal, as
estimated after subtraction of a local continuum.  Additionally, we
added to this list the GCS3 spectrum observed by \citet{kemper-etal04}
toward the Galactic Center.  The spectra of these templates in the
8--10~\micron\ region are displayed in Fig.~\ref{fig:templates}.

For all the other sources in our sample, the best template was determined 
by scaling the possible templates to the observed optical depth at different wavelengths 
(8.75, 9.30, 9.37, 9.70, 9.98~\micron) and finding the combination (template+scaling point) 
that gave the least residuals over the same wavelength ranges used to estimate the local continuum 
(8.25--8.75, 9.23--9.37, 9.98--10.4~\micron).
The result of this process is displayed for each source in the top
part of Fig.~\ref{fig:continuum}, where the best template is shown by
a grey line.  The bottom panels of Fig.~\ref{fig:continuum} show
sources for which no reasonable template could be found, as well as
emission sources, in which case only the local continuum is
overlaid. As in the case of the local continuum method, the spectra
obtained after subtraction of the templates are shown in
Fig.~\ref{fig:continuum-subtracted}. Taken together, NH$_3$ features
are detected in 24 out of 41 sources.

The top panel of Figure~\ref{fig:continuum-subtracted} shows that the
CH$_3$OH feature is not affected by the continuum choice, whereas the
width of the NH$_3$ band is somewhat sensitive to this choice,
especially
if there is no \m\ absorption, in which case the local continuum yields 
a wider \ammonia\ profile. 
For both continua, there is a clear feature around 9 \micron, 
which we attribute to \ammonia, with the characteristics and limitations
given and discussed in the following sections.

\subsection{NH$_3$ ice column densities and abundances \label{sec:nh3 results}}

Gaussian fits were performed to the \ammonia\ and/or \m\ features when present, 
and derived parameters for NH$_3$ are listed in Table~\ref{tab:gaussian param} 
(Appendix~\ref{ap:gaussian param}). Table~\ref{tab:colden} gives 
the column densities derived for \ammonia\ for each of the two methods 
employed to determine the continuum, using a band strength of 
1.3$\times$10$^{-17}$~cm~molecule$^{-1}$ 
for the NH$_3$ $\nu_2$ umbrella mode appropriate for a water-rich
ice \citep{dhendecourt+allamandola86,kerkhof-etal99}.
The two methods generally agree to within a factor of 2 or better.  
A similar factor of $\leq$2 overall uncertainty is estimated 
for those sources for which only the local continuum has been used.

The position of the NH$_3$ $\nu_2$ umbrella mode is very close to that
of the $\nu_7$ CH$_3$-rock mode of CH$_3$OH.  As illustrated by our
laboratory data (see Section~\ref{sec:lab}), sources with an
absorption depth at $\sim$9.7~\micron\ (CO-stretch mode of \m) at least
twice as large as the absorption depth at $\sim$9~\micron\ (blend of
CH$_3$-rock mode of \m\ and \ammonia\ umbrella mode) have a
significant contribution to the 9-\micron\ integrated optical depth
from the CH$_3$-rock mode of \m. 
In these cases (sources followed by an asterisk in
Table~\ref{tab:colden} and in Table~\ref{tab:gaussian param} of 
Appendix~\ref{ap:gaussian param}),
we performed the following correction: 
we scaled a \water:\m=9:1 laboratory spectrum to the observed
optical depth of the CO-stretch mode of \m, determined
the integrated optical depth of the CH$_3$-rock mode of \m\
in that scaled spectrum, and subtracted it
from the total observed optical depth at 9~\micron.
This correction is justified by the fact that the \water:\m:\ammonia=10:4:1 spectrum,
 a typical interstellar abundance mixture, is well reproduced 
around 8--10~\micron\ 
by a combination of \water:\m=9:1 and \water:\ammonia=9:1 (see Sec.~\ref{sec:lab}).

The inferred NH$_3$ ice abundances range from $\lesssim 1$\% to 15\%
with respect to H$_2$O ice,
excluding the abnormally high value of EC 82.
When considering all values (except that of EC 82)
determined with the local continuum method,
this relative abundance is centered on 5.3\% with a standard
deviation of 2.0\%. 
If we use values determined with the template method whenever
available, we find a mean of 7.0$\pm$3.2\%.
Either way, within the errors, 
this is similar to what was obtained by
\citet{oberg-etal08-ch4} for CH$_4$ (4.7$\pm$1.6\%), another ice component that should
form via hydrogenation.  For 6 out of the 8 sources where both \ammonia\
and CH$_4$ are detected, the \ammonia-to-\methane\ abundance ratio is slightly
larger than 1 ($\sim$1.2). Based on elemental abundance ratios, one
would expect \ammonia/\methane\ smaller than 1, but since two thirds
of the carbon is in refractory grains and some fraction of the gaseous
CO locked up in CO at the ice formation threshold,
\ammonia-to-\methane\ ratios larger than 1 are consistent with both
\ammonia\ and \methane\ being formed by hydrogenation of N and C,
respectively. 
Further comparison between these and other ice species will be 
addressed in an upcoming paper (\"Oberg et al. in preparation).\\

Regarding \m, we only report values for the Gaussian parameters and derived column
densities in the appendix (see Table~\ref{tab:gaussian param ch3oh}),
to show that the numbers we obtain in this independent study are consistent with those
reported in Paper I. Our recommended abundances are those from paper
I, based on the combined 9.75 and 3.53 $\mu$m analysis.
The inferred CH$_3$OH abundances range from $<1\%$ to $>25$\% with
respect to H$_2$O ice, indicating significant CH$_3$OH/NH$_3$
abundance variations from source-to-source.  Such relative abundance
variations can already be clearly seen from the changing relative
depths of the 9.0 and 9.7 $\mu$m features (see also Paper I).  Thus,
NH$_3$ and CH$_3$OH ice are likely formed through different formation
pathways and/or in different ice environments.

\clearpage

\begin{deluxetable}{l rr c rr c lc}
\tablecolumns{9}
\tabletypesize{\scriptsize}
\tablecaption{\ammonia\ column densities$^a$ and abundances with respect to 
H$_2$O ice$^b$
\label{tab:colden} }
\tablewidth{0pt}
\tablehead{
Source & \multicolumn{2}{c}{\ammonia, local} &
       & \multicolumn{2}{c}{\ammonia, template} && Template & Scaling point\\
\cline{2-3}\cline{5-6}
       & $\times10^{17}$ \cs & \multicolumn{1}{c}{\% \water$^b$} &
       & $\times10^{17}$ \cs & \multicolumn{1}{c}{\% \water$^b$} &&  & \micron 
}
\startdata
       IRAS 03235+3004 &  6.83 ( 0.98) &  4.71 ( 1.00) &&  8.94 ( 1.03) &  6.17 ( 1.20) &&      IRAS 12553  &   9.30 \\ 
            L1455 IRS3 &  0.57 ( 0.23) &  6.21 ( 3.51) &&  1.41 ( 0.27) & 15.37 ( 6.86) &&            GCS3  &   9.37 \\ 
       IRAS 03254+3050 &  2.44 ( 0.39) &  6.66 ( 1.37) &&  4.58 ( 0.49) & 12.52 ( 2.10) &&      IRAS 12553  &  10.40 \\ 
                  B1-b$^*$ &  \multicolumn{1}{c}{$\sim$7.3} &  \multicolumn{1}{c}{$\sim$4.2} && \multicolumn{1}{c}{$\sim$9.8} &  \multicolumn{1}{c}{$\sim$5.6} &&      IRAS 12553  &   9.70 \\ 
       IRAS 04108+2803 &  1.23 ( 0.24) &  4.29 ( 1.03) &&  2.07 ( 0.39) &  7.21 ( 1.69) &&      IRAS 23238  &   9.70 \\ 
                HH 300 &  0.90 ( 0.22) &  3.46 ( 0.90) &&  2.23 ( 0.37) &  8.60 ( 1.65) &&        DG Tau B  &   9.70 \\ 
       IRAS 08242$-$5050 &  4.77 ( 0.46) &  6.13 ( 0.85) &&  4.41 ( 0.54) &  5.66 ( 0.89) &&      IRAS 12553  &   9.70 \\ 
       IRAS 15398$-$3359 &  8.73 ( 1.18) &  5.90 ( 1.77) && 13.80 ( 1.35) &  9.33 ( 2.65) &&      IRAS 12553  &   9.70 \\ 
              B59 YSO5 &  4.92 ( 0.72) &  3.53 ( 0.88) &&  6.37 ( 0.99) &  4.57 ( 1.17) &&      CrA IRS7 A  &   9.70 \\ 
 2MASSJ17112317$-$272431 & 13.10 ( 1.06) &  6.70 ( 0.54) && 20.60 ( 2.76) & 10.58 ( 1.42) &&      IRAS 23238  &   9.70 \\ 
               SVS 4-5$^*$ & \multicolumn{1}{c}{$\sim$2.4} &  \multicolumn{1}{c}{$\sim$4.3} &&  \multicolumn{1}{c}{$\sim$5.8} & \multicolumn{1}{c}{$\sim$10.3} &&            GCS3  &   8.75 \\ 
           R CrA IRS 5 &  0.91 ( 0.23) &  2.54 ( 0.67) &&  1.49 ( 0.31) &  4.15 ( 0.92) &&      IRAS 12553  &   9.70 \\ 
\hline
                RNO 15$^c$ &  0.80 ( 0.21) & 11.58 ( 3.18) && \multicolumn{1}{c}{--} & \multicolumn{1}{c}{--}  && \multicolumn{1}{c}{--} & --  \\ 
       IRAS 03271+3013 &  4.90 ( 0.88) &  6.37 ( 1.86) && \multicolumn{1}{c}{--} & \multicolumn{1}{c}{--}  && \multicolumn{1}{c}{--} & --  \\ 
                  B1-a &  3.46 ( 0.69) &  3.33 ( 0.98) && \multicolumn{1}{c}{--} & \multicolumn{1}{c}{--}  && \multicolumn{1}{c}{--} & --  \\ 
             L1489 IRS &  2.31 ( 0.30) &  5.42 ( 0.96) && \multicolumn{1}{c}{--} & \multicolumn{1}{c}{--}  && \multicolumn{1}{c}{--} & --  \\ 
       IRAS 13546$-$3941 &  0.94 ( 0.16) &  4.56 ( 0.87) && \multicolumn{1}{c}{--} & \multicolumn{1}{c}{--}  && \multicolumn{1}{c}{--} & --  \\ 
                RNO 91 &  2.03 ( 0.30) &  4.78 ( 0.81) && \multicolumn{1}{c}{--} & \multicolumn{1}{c}{--}  && \multicolumn{1}{c}{--} & --  \\ 
       IRAS 17081$-$2721 &  0.86 ( 0.16) &  6.54 ( 1.39) && \multicolumn{1}{c}{--} & \multicolumn{1}{c}{--}  && \multicolumn{1}{c}{--} & --  \\ 
                 EC 74$^c$ &  1.00 ( 0.29) &  9.35 ( 3.13) && \multicolumn{1}{c}{--} & \multicolumn{1}{c}{--}  && \multicolumn{1}{c}{--} & --  \\ 
                 EC 82 &  1.22 ( 0.14) & 31.31 ( 6.65) && \multicolumn{1}{c}{--} & \multicolumn{1}{c}{--}  && \multicolumn{1}{c}{--} & --  \\ 
                 EC 90 &  0.67 ( 0.20) &  3.94 ( 1.24) && \multicolumn{1}{c}{--} & \multicolumn{1}{c}{--}  && \multicolumn{1}{c}{--} & --  \\ 
                 EC 92$^*$ & \multicolumn{1}{c}{$\sim$0.5} & \multicolumn{1}{c}{$\sim$3.0} && \multicolumn{1}{c}{--} & \multicolumn{1}{c}{--}  && \multicolumn{1}{c}{--} & --  \\ 
            CrA IRS7 B$^*$ &   \multicolumn{1}{c}{$\sim$3.0} & \multicolumn{1}{c}{$\sim$2.8} && \multicolumn{1}{c}{--} & \multicolumn{1}{c}{--}  && \multicolumn{1}{c}{--} & --  \\ 
             L1014 IRS &  3.72 ( 0.91) &  5.20 ( 1.43) && \multicolumn{1}{c}{--} & \multicolumn{1}{c}{--}  && \multicolumn{1}{c}{--} & --  \\ 
                   CK4 &  0.84 ( 0.13) &  5.37 ( 0.86) && \multicolumn{1}{c}{--} & \multicolumn{1}{c}{--}  && \multicolumn{1}{c}{--} & --  \\ 
\hline
\multicolumn{9}{c}{3-$\sigma$ upper limits} \\
\hline
LDN 1448 IRS1          & \multicolumn{1}{c}{\phn0.20}  & \multicolumn{1}{c}{\phn4.15}  && \multicolumn{1}{c}{--} & \multicolumn{1}{c}{--}  && \multicolumn{1}{c}{--} & --  \\  
IRAS 03245+3002        & \multicolumn{1}{c}{   17.28}  & \multicolumn{1}{c}{\phn4.40}  && \multicolumn{1}{c}{--} & \multicolumn{1}{c}{--}  && \multicolumn{1}{c}{--} & --  \\  
L1455 SMM1             & \multicolumn{1}{c}{   15.10}  & \multicolumn{1}{c}{\phn8.29}  && \multicolumn{1}{c}{--} & \multicolumn{1}{c}{--}  && \multicolumn{1}{c}{--} & --  \\  
IRAS 03301+3111        & \multicolumn{1}{c}{\phn0.24}  & \multicolumn{1}{c}{\phn5.93}  && \multicolumn{1}{c}{--} & \multicolumn{1}{c}{--}  && \multicolumn{1}{c}{--} & --  \\  
B1-c                   & \multicolumn{1}{c}{   11.93}  & \multicolumn{1}{c}{\phn4.04}  && \multicolumn{1}{c}{--} & \multicolumn{1}{c}{--}  && \multicolumn{1}{c}{--} & --  \\  
IRAS 03439+3233        & \multicolumn{1}{c}{\phn0.31}  & \multicolumn{1}{c}{\phn3.10}  && \multicolumn{1}{c}{--} & \multicolumn{1}{c}{--}  && \multicolumn{1}{c}{--} & --  \\  
IRAS 03445+3242        & \multicolumn{1}{c}{\phn0.47}  & \multicolumn{1}{c}{\phn2.09}  && \multicolumn{1}{c}{--} & \multicolumn{1}{c}{--}  && \multicolumn{1}{c}{--} & --  \\  
{\bf DG Tau B}    & \multicolumn{1}{c}{\phn0.47}  & \multicolumn{1}{c}{\phn2.05}  && \multicolumn{1}{c}{--} & \multicolumn{1}{c}{--}  && \multicolumn{1}{c}{--} & --  \\  
{\bf IRAS 12553-7651}     & \multicolumn{1}{c}{\phn0.61}  & \multicolumn{1}{c}{\phn2.04}  && \multicolumn{1}{c}{--} & \multicolumn{1}{c}{--}  && \multicolumn{1}{c}{--} & --  \\  
Elias 29               & \multicolumn{1}{c}{\phn0.28}  & \multicolumn{1}{c}{\phn0.93}  && \multicolumn{1}{c}{--} & \multicolumn{1}{c}{--}  && \multicolumn{1}{c}{--} & --  \\  
CRBR 2422.8$-$342        & \multicolumn{1}{c}{\phn0.52}  & \multicolumn{1}{c}{\phn1.23}  && \multicolumn{1}{c}{--} & \multicolumn{1}{c}{--}  && \multicolumn{1}{c}{--} & --  \\  
HH 100 IRS             & \multicolumn{1}{c}{\phn0.46}  & \multicolumn{1}{c}{\phn1.89}  && \multicolumn{1}{c}{--} & \multicolumn{1}{c}{--}  && \multicolumn{1}{c}{--} & --  \\  
{\bf CrA IRS7 A}  & \multicolumn{1}{c}{\phn0.97}  & \multicolumn{1}{c}{\phn0.89}  && \multicolumn{1}{c}{--} & \multicolumn{1}{c}{--}  && \multicolumn{1}{c}{--} & --  \\  
CrA IRAS32             & \multicolumn{1}{c}{\phn5.44}  & \multicolumn{1}{c}{   10.35}  && \multicolumn{1}{c}{--} & \multicolumn{1}{c}{--}  && \multicolumn{1}{c}{--} & --  \\  
{\bf IRAS 23238+7401}   & \multicolumn{1}{c}{\phn1.60}  & \multicolumn{1}{c}{\phn1.24}  && \multicolumn{1}{c}{--} & \multicolumn{1}{c}{--}  && \multicolumn{1}{c}{--} & --  \\  
\enddata
\tablecomments{Sources in bold were used as templates. 
Uncertainties quoted in parenthesis are statistical errors from the Gaussian fits while absolute errors are up to a factor of 2.}
\tablenotetext{a}{Derived using a bandstrength of 1.3$\times 10^{-17}$~cm molecule$^{-1}$.}
\tablenotetext{b}{Using the H$_2$O ice column densities listed in Paper I.}
\tablenotetext{c}{Values are likely upper limits (see Section~\ref{sec: 3-6 micron} for details).}
\tablenotetext{*}{Sources with $\tau_{9.7\micron}>2\times\tau_{9.0\micron}$, for which an estimated contribution from the CH$_3$-rock mode
of \m\ was subtracted (see text for details).}
\end{deluxetable}

\clearpage

\section{Laboratory work and analysis}
\label{sec:lab}

The band profiles presented in Fig.~\ref{fig:continuum-subtracted}
contain information on the ice environment in which NH$_3$ and
CH$_3$OH are located, and thus their formation and processing
history. To extract this information, a systematic laboratory study of
the NH$_3$ and CH$_3$OH features in a variety of ices has been carried
out. Specifically, three features between 8 and 10 $\mu$m have
been analyzed:

\begin{enumerate}
\item the \ammonia\ $\nu_2$ umbrella mode, at $\sim$9.35~$\mu$m or 1070~cm$^{-1}$ 
in pure \ammonia\ ice, and with band strength $A\rm_{pure}$
=1.7$\times$10$^{-17}$~cm~molecule$^{-1}$ \citep{dhendecourt+allamandola86},
\item the \methanol\ $\nu_4$ CO-- stretching mode, at $\sim$9.74~$\mu$m or 1027~cm$^{-1}$ 
in pure \m\ ice, and with $A\rm_{pure}$ = 1.8$\times$10$^{-17}$~cm~molecule$^{-1}$
\citep{dhendecourt+allamandola86},
\item the \methanol\ $\nu_7$ CH$_3$ rocking mode, at $\sim$8.87~$\mu$m or 1128~cm$^{-1}$ 
in pure \m\ ice, and with $A\rm_{pure}$ = 1.8$\times$10$^{-18}$~cm~molecule$^{-1}$
\citep{hudgins-etal93}.
\end{enumerate}

It should be noted that, as mentioned in the above list, the quoted positions are for pure ices only  
and therefore slightly deviate from the astronomical values given in Table~\ref{tab:features}.

This laboratory study targeted pure, binary and tertiary interstellar
ice analogs consisting of different mixtures of \water, \ammonia, \m, CO and CO$_2$, the
major ice components.  All measurements were performed under high
vacuum conditions ($\sim 10^{-7}$~mbar) using an experimental approach
described in \citet{gerakines-etal95}, \citet{bouwman-etal07} and
\citet{oberg-etal07-h2o-co2}. The ice spectra were recorded in
transmission using a Fourier transform infrared spectrometer covering
25--2.5~\micron\ (400--4000~\wn) with 1~\wn\ resolution and by
sampling relatively thick ices, typically several thousands Langmuir\footnote{One 
Langmuir corresponds to a pressure of 10$^{-6}$ torr
for 1 second and measures the exposure of a surface to adsorption of gases.
One Langmuir is equivalent to about
10$^{15}$~molecules~cm$^{-2}$. }
(L) thick. These ices were grown at a speed of
$\sim$10$^{16}$~molecules~cm$^{-2}$~s$^{-1}$ on a
temperature-controlled CsI window.  

\begin{figure}[ht]
\centering
\includegraphics[angle=90,width=\columnwidth]{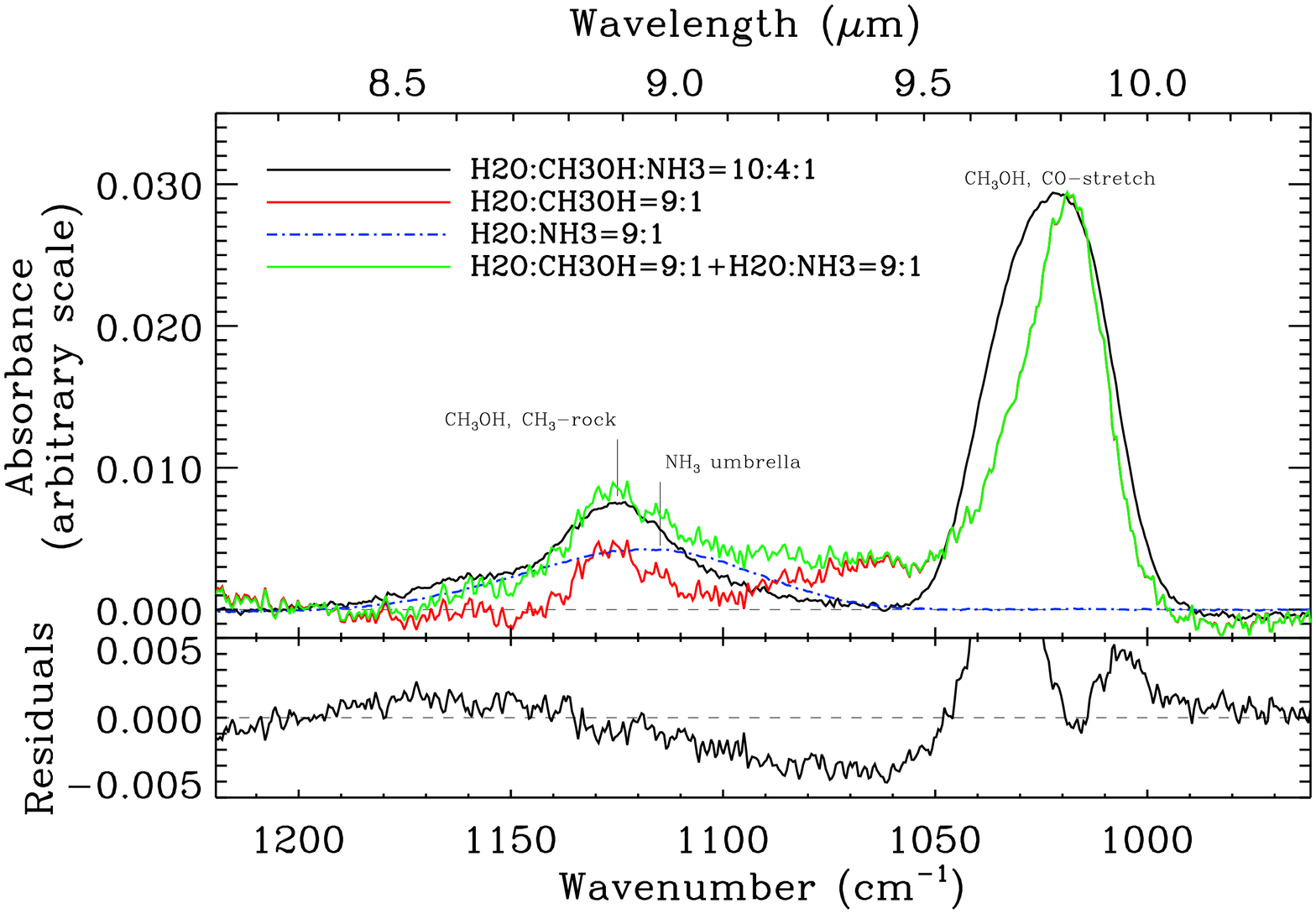}
\caption{ 
Example of a reduced laboratory spectrum (solid black line) for a
 \water:\m:\ammonia = 10:4:1 ice mixture at 15~K,
in the 8--10~\micron\ / 960--1220~\wn\ range.
This spectrum can be approximated as
the sum (solid green/dark grey line) of \water:\m=9:1 (solid red/light grey line)
and \water:\ammonia=9:1 (dash-dot blue/grey line). 
The bottom plot is the difference between the two, showing
that the feature at 9~\micron\ (blend of \ammonia\ and \m\ CH$_3$-rock modes)
is well reproduced by the sum of the two individual signatures.
This figure also illustrates the fact that the positions of the features in mixed ices
differ from that in pure ices 
(see list at the beginning of this section).
\label{tertiaryspectrum}
}
\end{figure}

A typical reduced spectrum for an ice
mixture containing \water:\m:\ammonia = 10:4:1 at 15~K is shown in
Fig.~\ref{tertiaryspectrum}. Since band profiles and strengths change with ice composition
and also with temperature, the three fundamentals mentioned above were
investigated as a function of temperature ranging from 15 to 140~K
with regular temperature steps for a number of binary and tertiary
mixtures (listed in Appendix~\ref{ap:lab}). An IDL routine was used to
determine the location of the band maximum, FWHM and integrated
absorbance of the individual absorption bands.  For the asymmetric
\ammonia\ $\nu_2$ umbrella mode the band position has been determined
by the maximum absorbance and for the symmetric profiles the spectral
parameters have been determined from Gaussian fits of baseline
subtracted spectra. The resulting absolute frequency uncertainty is of the order of
1~\wn.
The measurements are presented in Table~\ref{table_lab_overview} of
Appendix B, and are included in the Leiden laboratory
database\footnote{\tt http://www.strw.leidenuniv.nl/$\sim$lab/}.

\begin{figure*}[ht!]
\includegraphics[angle=90,width=0.48\textwidth]{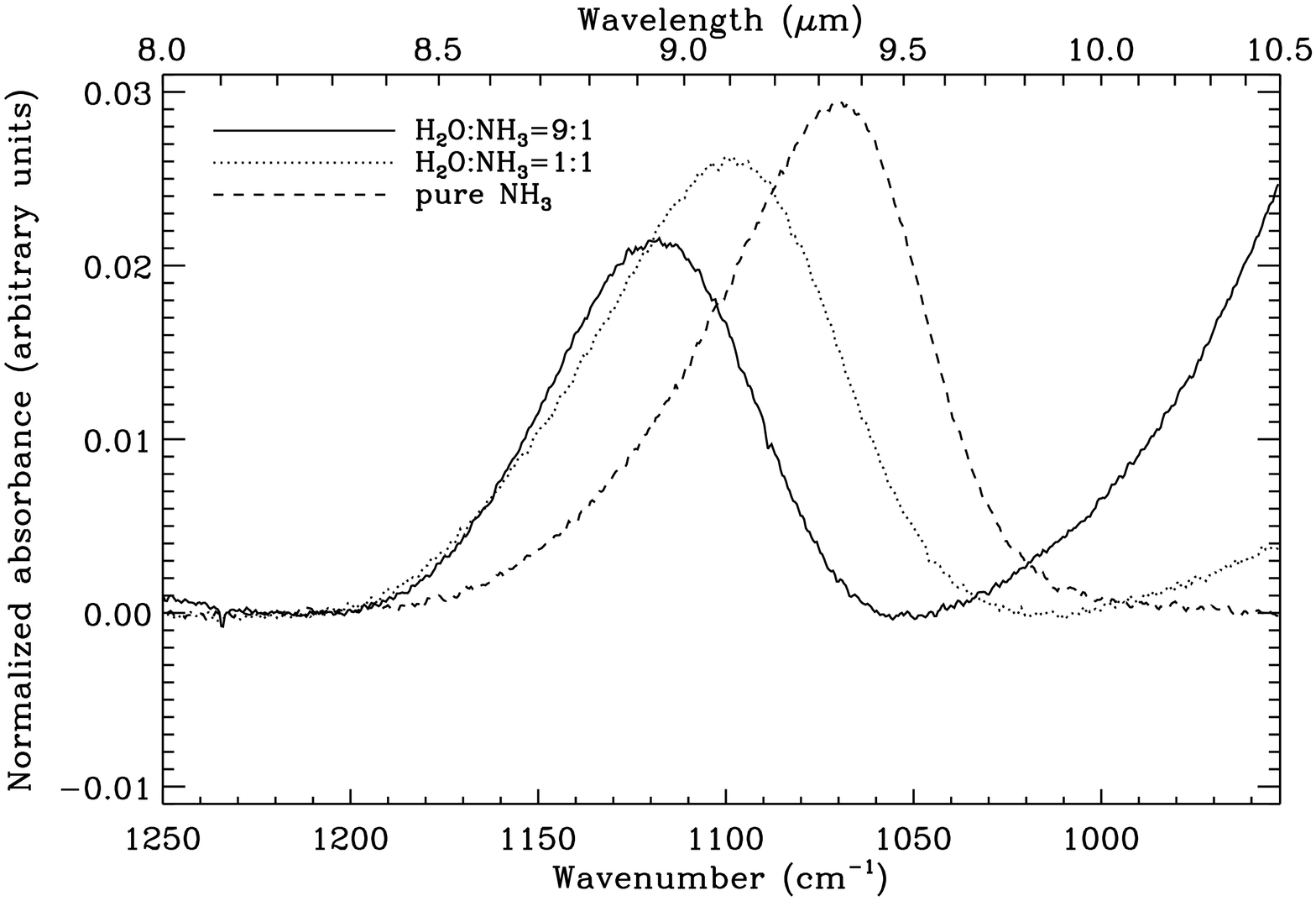}\hfill%
\includegraphics[angle=90,width=0.48\textwidth]{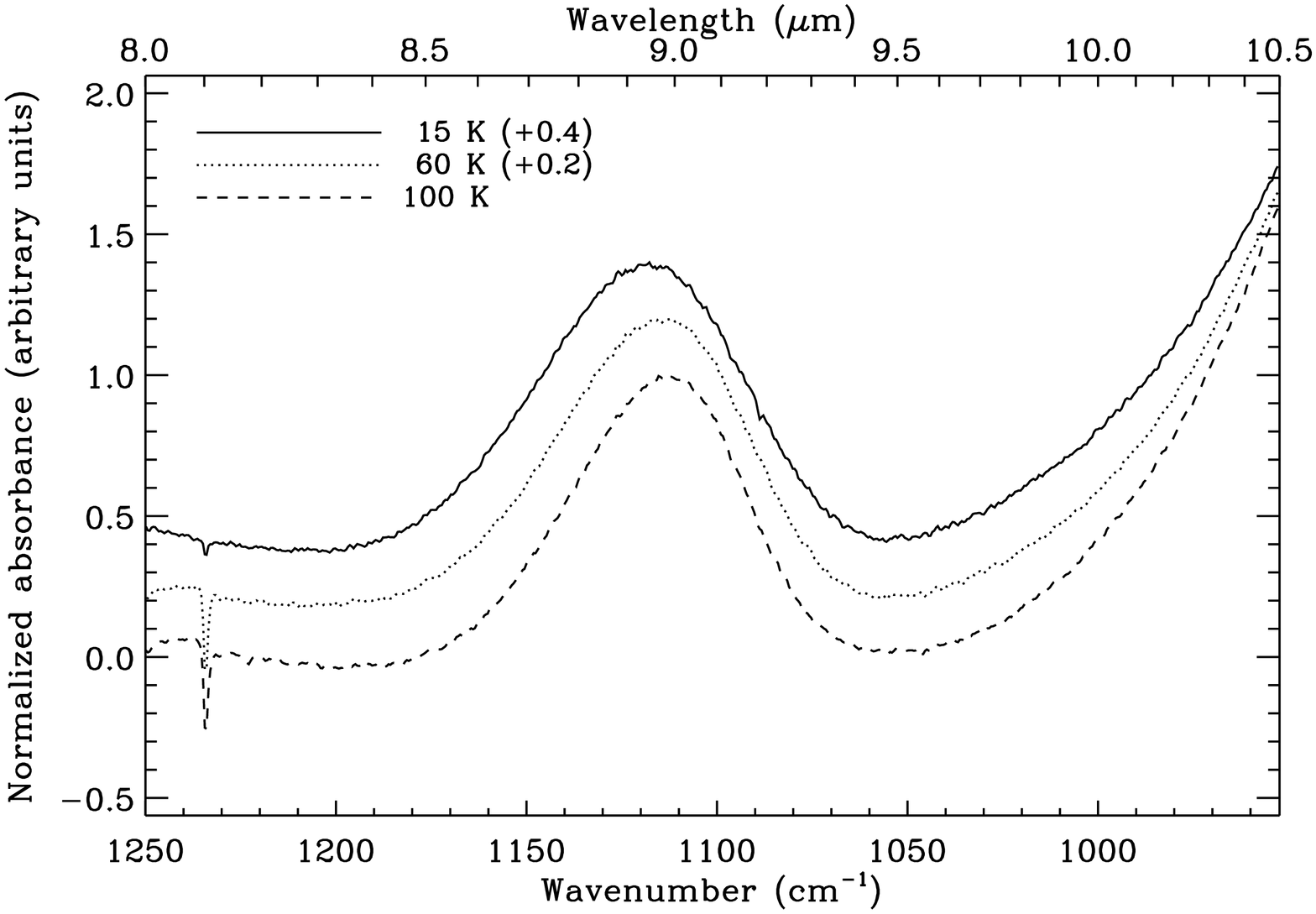}
\caption{({\it Left}) FTIR ice spectra of the \modeNH\ mode for pure NH$_3$, a 
H$_2$O:NH$_3$=1:1 and a H$_2$O:NH$_3$=9:1 mixture at a temperature of 15~K. 
At the low frequency side of the spectrum the \water\ libration mode
(centered around 770~\wn, or 13~\micron)
starts showing up for the \water-containing mixtures. 
--- ({\it Right}) Temperature effect on a H$_2$O:NH$_3$=9:1 mixture:
decreasing FWHM with increasing temperature.
\label{ammonia1}
}
\end{figure*}

\ammonia\ and \m\ both have the ability to form hydrogen bonds in
water-rich matrices, so it is not surprising that the band profile
changes compared with pure ices because of the various
molecular interactions \citep[e.g.,][]{dhendecourt+allamandola86} 

\begin{figure*}[ht!]
\centering
\includegraphics[width=\textwidth]{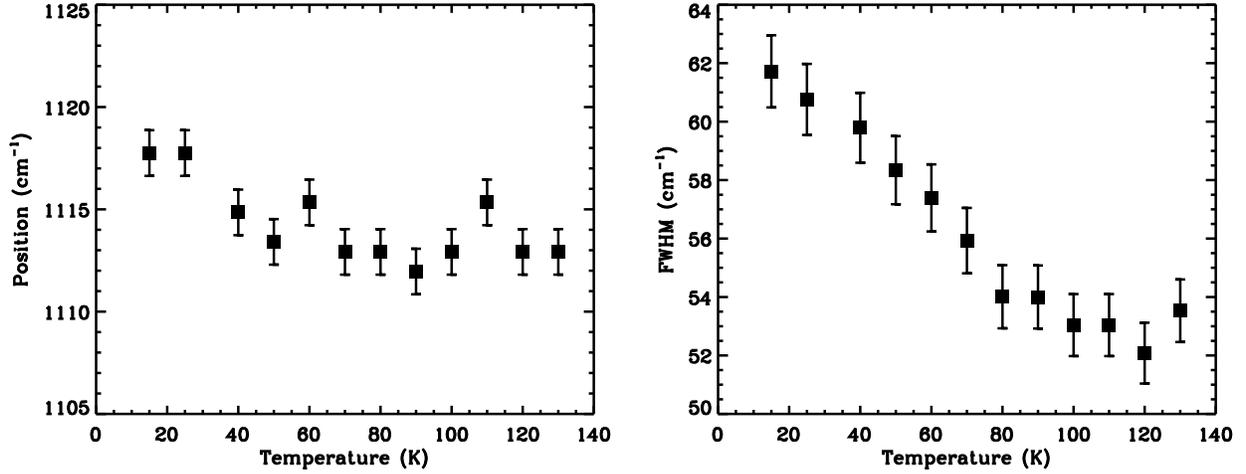}
\caption{A plot indicating the changes in peak position (left) and FWHM (right) of the 
NH$_3$ $\nu_{2}$ umbrella mode as a function of temperature in a 
9:1 \water:\ammonia\ ice. 
\label{AmmoniavsTemp}}
\end{figure*}

In addition to profiles, band strengths can change with environment
and with temperature, as discussed for the cases of CO and CO$_2$ in
water-rich ices in
\citet{kerkhof-etal99,oberg-etal07-h2o-co2,bouwman-etal07}.
Figure~\ref{ammonia1} shows how the \ammonia\ $\nu_2$ umbrella mode
absorption maximum shifts from 1070~\wn\ (9.35 $\mu$m) 
for pure \ammonia\ ice to 1118~\wn\ (8.94 $\mu$m) for an
astronomically more realistic \water:\ammonia=9:1 (hereafter 9:1)
mixture, for which the FWHM and integrated band strength also change
significantly. For example, the band strength is lowered in the 9:1
mixture to 70\% of its initial value in pure \ammonia\ ice. This is in
good agreement with previous experiments performed by
\citet{kerkhof-etal99}. The spectral appearance also depends on
temperature; for the 9:1 mixture a temperature increase from 15 to
120~K results in a redshift of the peak position from 1118 to
1112~\wn\ (8.94 to 8.99~\micron) and the FWHM decreases from 62 to
52~\wn\ (0.50 to 0.42~$\mu$m) (see Fig.~\ref{AmmoniavsTemp}). The
NH$_3$ bandstrength, on the other hand, does not show any temperature
dependence.

If NH$_3$ is in a water-poor environment with CO and/or CO$_2$, the
$\nu_2$ peak position shifts to the red compared with pure NH$_3$, to
as much as 1062 cm$^{-1}$ (9.41 $\mu$m). The FWHM is not much affected
whereas the band strength is lowered by 20\%. Because of the intrinsically large
difference in band maximum position
between NH$_3$ in a water-poor and water-rich
environment, the astronomical observations can distinguish between these two
scenarios.

\begin{figure*}[ht!]
\includegraphics[angle=90,width=\textwidth]{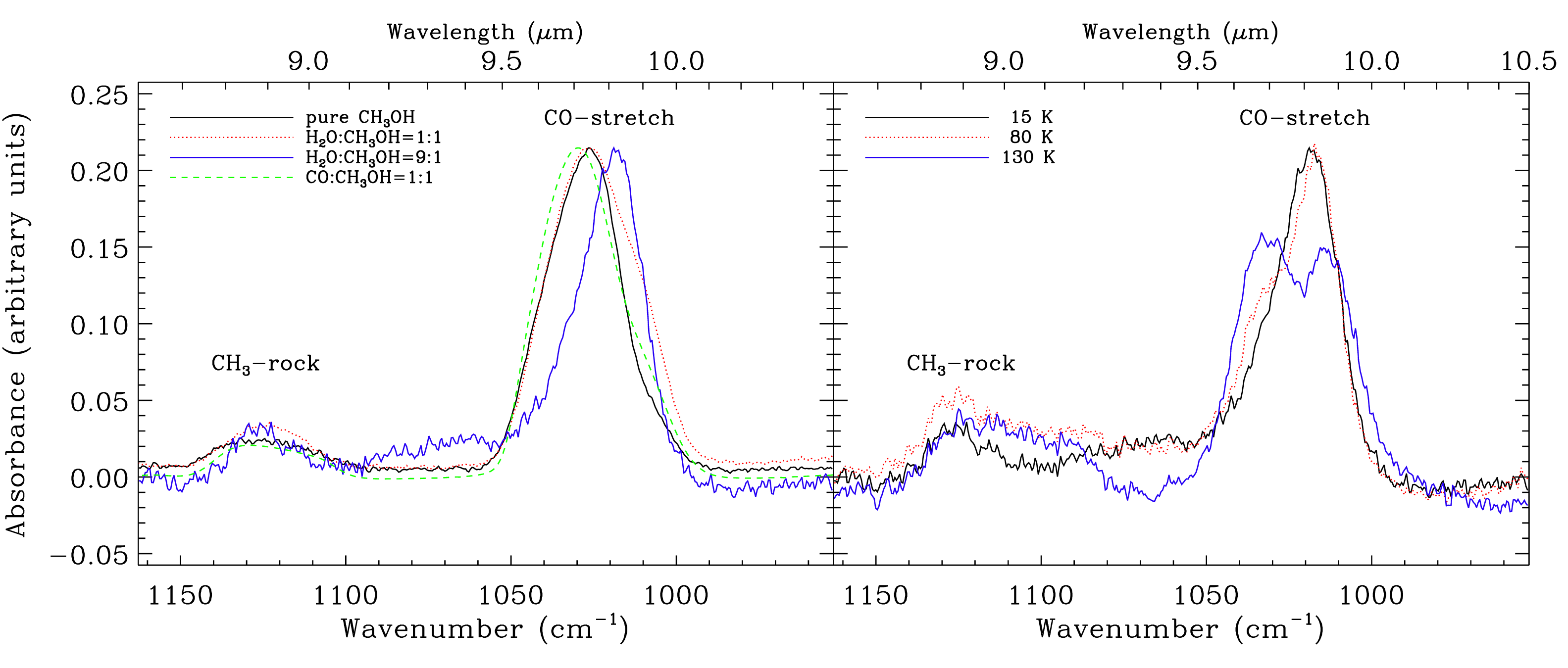}
\caption{({\it Left}) Spectra of the \methanol\ \modeCO\ modes and \modeCH\ modes 
for pure CH$_3$OH, a H$_2$O:CH$_3$OH=1:1, a H$_2$O:CH$_3$OH=9:1 and a CO:CH$_3$OH=1:1   
ice mixture at a temperature of 15~K. --- ({\it Right}) Temperature effect on the CO-stretch
mode of a H$_2$O:CH$_3$OH=9:1 mixture.
\label{methanol1}
}
\end{figure*}

Methanol-containing ices have been studied in a similar way (see
Fig.~\ref{methanol1}). The weakly absorbing $\nu_7$ CH$_3$ rocking
mode at $\sim$1125~\wn\ (8.89~$\mu$m) is rather insensitive to \water\
mixing, but the $\nu_4$ CO stretch vibration shifts to the red from
1028 to 1020~\wn\ (9.73 to 9.80~$\mu$m) when changing from a pure \m\
ice to a \water:\m=9:1 mixture. 
In the latter spectrum the \methanol\
$\nu_4$ CO stretch mode needs to be fitted with a double
Gaussians. 
A substructure appears for a temperature of 80~K (right panel of Fig.~\ref{methanol1})
while for even higher temperatures, a clearly double peaked structure
becomes visible (as previously seen in e.g. Fig.~2 of
\citealt{schutte-etal91}).  
This splitting hints at different physical sites and has been
previously ascribed to type II clathrate formation in the ice
\citep{blake-etal91-clathrate}.

\begin{figure}[ht]
\centering
\includegraphics[angle=90,width=\columnwidth]{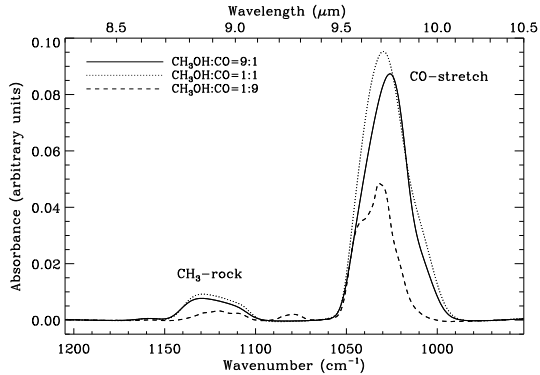}
\caption{Spectra of \m:CO mixtures in the range of the methanol CO
stretch mode and the methanol CH$_3$ rock mode. A small blue shift
together with a clear substructure are seen upon mixing in more CO. 
\label{binaryCOCH3OH} }
\end{figure}

When CH$_3$OH is mixed with CO, the band maximum shifts from 1028 to
1034~\wav\ (9.73 to 9.67~$\mu$m) when going from a 9:1 to a 1:9 \m:CO
mixture.  When 50\% or more CO is mixed in, the \m\ $\nu_4$ CO stretch
mode starts to show a shoulder and cannot be fitted correctly by a
single Gaussian component (see Fig.~\ref{binaryCOCH3OH}).  Such a
two-component profile would not be recognized, however, at the
spectral resolution and signal/noise of our {\it Spitzer} data, so for
the comparison between laboratory and observational data a single
Gaussian is used. Overall, the shifts of the \m\ $\nu_4$ mode between
water-rich and CO-rich mixtures are much smaller than in the case of
the NH$_3$ $\nu_2$ mode.

The effect of \m\ on the 4.7$\mu$m $\nu_1$ stretch mode of CO has also
been investigated. The band maximum shifts from 2139~\wav\ (4.68~$\mu$m)
for the nearly pure 9:1 CO:\m\ mixture to 2136 and 2135~\wav\ for the
1:1 and 1:9 mixtures, respectively. The CO band located at 2136~\wav\
is often referred to as CO residing in a polar, mainly \water\ ice,
environment. Clearly, the polar \m\ molecules can also contribute to
CO absorption at 2136~\wav\, when intimately mixed in an astronomical
ice.

Binary mixtures of \ammonia\ and \m\ have been studied as well. The
\m\ modes behave very much as they do in a pure methanol ice, but the
\ammonia\ $\nu_2$ umbrella mode is clearly suppressed. Its integrated
absorbance is readily reduced to 70\% of the integrated absorbance of
pure \ammonia\ in a \m:\ammonia=1:1 mixture and becomes even lower for
a 4:1 binary composition. The \ammonia\ band also broadens compared to pure
\ammonia\ or \water:\ammonia\ mixtures and strongly overlaps with the
CO stretching mode of \m, to the level that it becomes difficult to
measure.

\begin{figure}[ht]
\centering
\includegraphics[angle=90,width=\columnwidth]{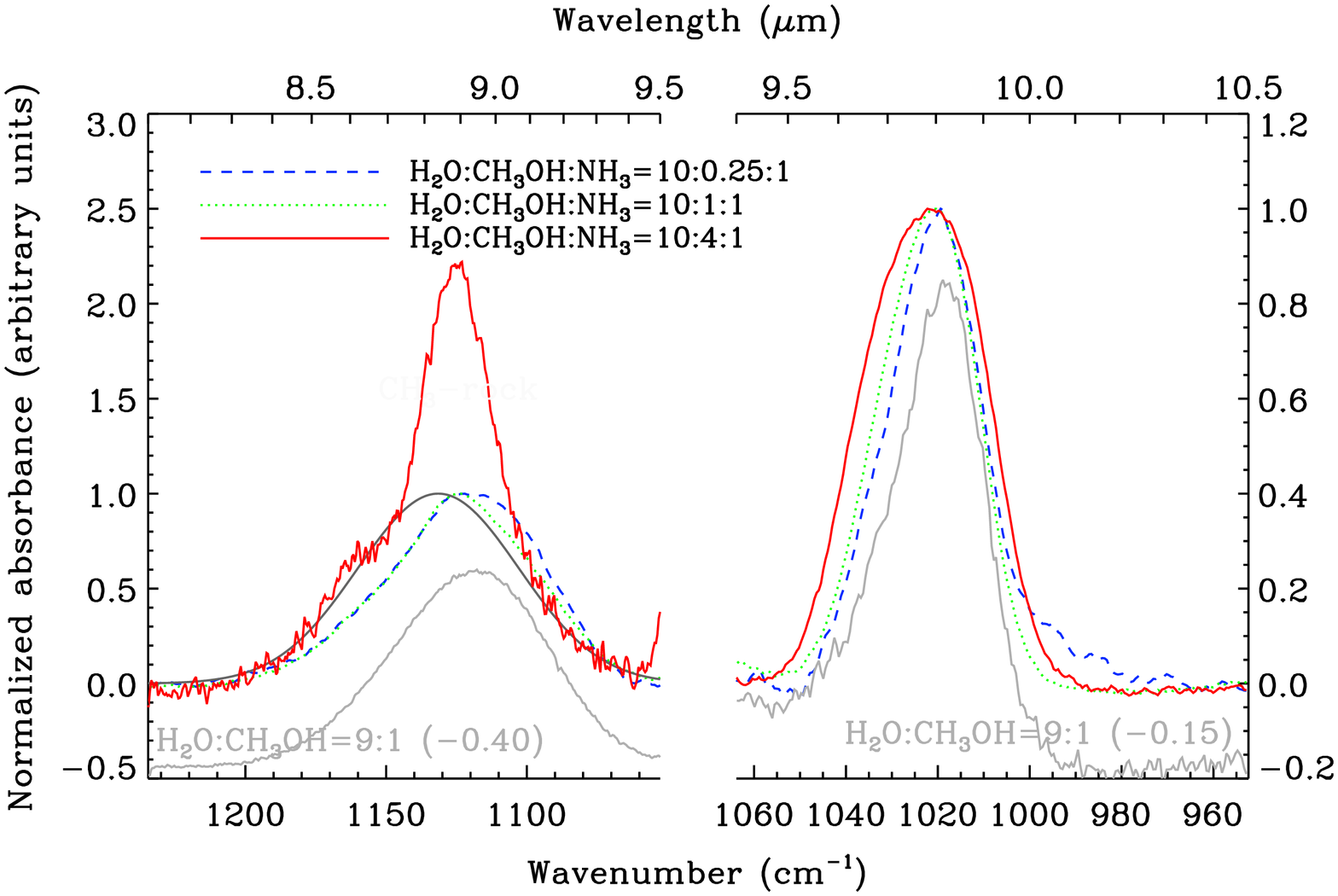}
\caption{Normalized spectra of the \methanol\ $\nu_4$ C--O mode (right panel), 
and NH$_3$ $\nu_2$ umbrella mode (left panel) 
for a H$_2$O:CH$_3$OH:NH$_3$=10:0.25:1, a H$_2$O:CH$_3$OH:NH$_3$=10:1:1 
and a H$_2$O:CH$_3$OH:NH$_3$=10:4:1 mixture at a temperature of 15~K. 
These mixture ratios span the range of observed interstellar column density ratios.
Spectra were normalized to better show the changes in band maximum position and FWHM
of each feature.
Spectra of a H$_2$O:CH$_3$OH=9:1 and a H$_2$O:NH$_3$=1:1 mixture
were offset and overlaid in light grey in the right and left panel, respectively.
In the case of H$_2$O:CH$_3$OH:NH$_3$=10:4:1, 
the NH$_3$ $\nu_2$ umbrella mode is heavily blended with the
\methanol\ $\nu_7$ CH$_3$ rocking mode, so that
the dark grey line actually shows the Gaussian fit to the underlying \ammonia\
feature, whereas the full 9-\micron\ feature is shown in black.
\label{binarytertiary} }
\end{figure}

A qualitative comparison with the astronomical data (see Section \ref{sec:comparison})
indicates that neither pure \ammonia, \m, nor mixed
\m:\ammonia\ or \water-diluted binary ices can simultaneously explain
the different \ammonia\ profiles in the recorded \spitzer\ spectra.
Thus, a series of tertiary mixtures with \water:\m:\ammonia\ in ratios
10:4:1, 10:1:1 and 10:0.25:1 have been measured, because \m\ is the next
major ice component. These ratios roughly span the range of observed
interstellar column density ratios. In Fig.~\ref{binarytertiary}, the
spectra of \water:\m:\ammonia\ tertiary mixtures are plotted and
compared to binary \water:\m\ and \water:\ammonia\ data. The \ammonia\
$\nu_2$ umbrella mode shifts slightly to the blue in the presence of
both \water\ and \m, with an absorption maximum at 1125~\wn\
(8.90~$\mu$m) for the 10:4:1 \water:\m:\ammonia\ mixture (compared to
1118~\wn\ (8.94~$\mu$m) in the \water:\ammonia=9:1 mixture). The peak
intensity of the \ammonia\ $\nu_2$ umbrella mode band in this tertiary
mixture is small compared with that of the \m\ CH$_3$ rock mode, but
its integrated intensity is a factor of two larger because of the larger
NH$_3$ width.

\begin{figure*}[ht!]
\begin{minipage}{0.48\textwidth}
\includegraphics[width=\columnwidth,bb=46 399 608 710,clip=true]{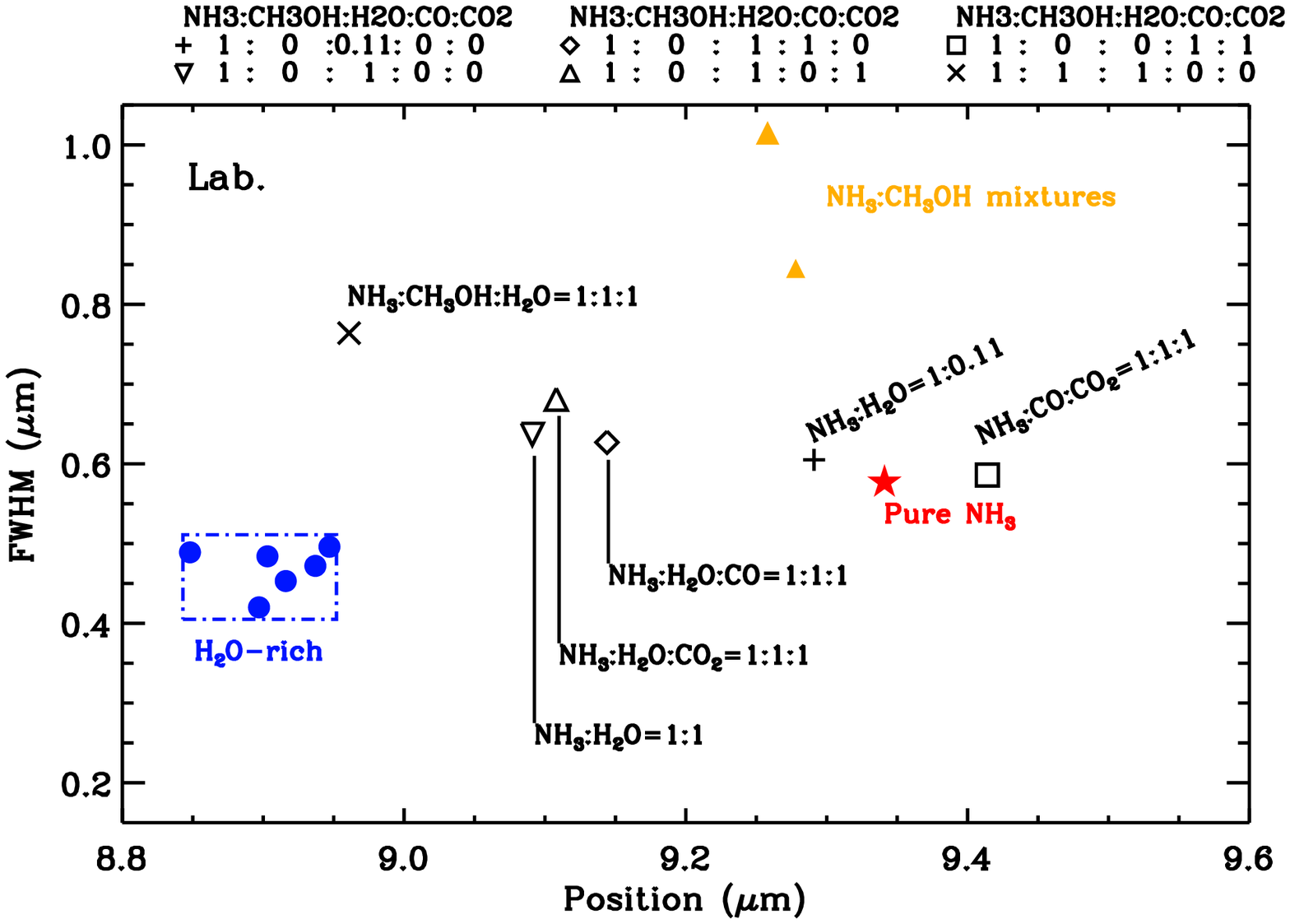}\\
\includegraphics[width=\columnwidth,bb=46 360 608 709,clip=true]{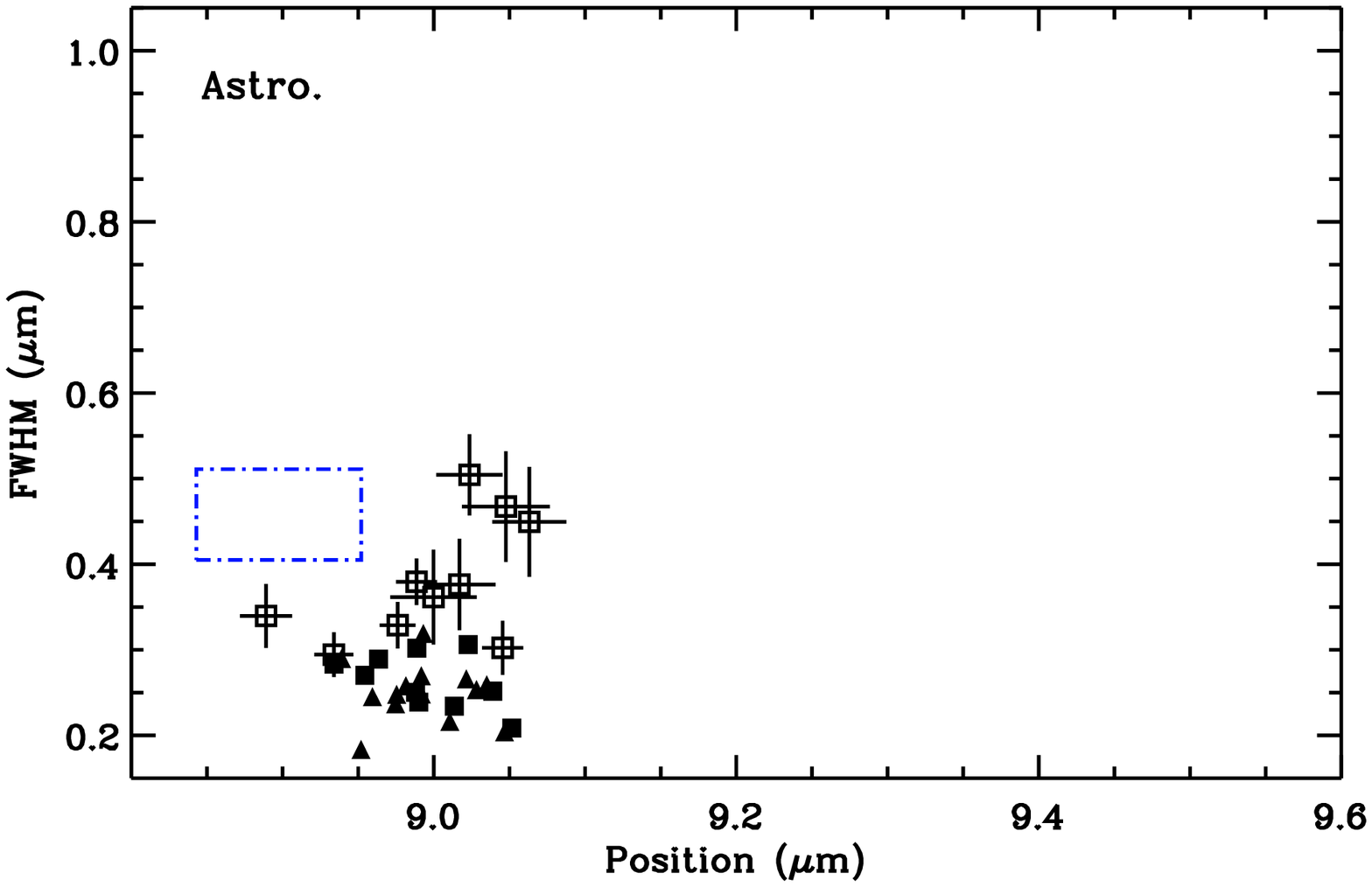}
\caption{\footnotesize 
FWHM and band maximum position of the \ammonia\ feature
measured in
the laboratory mixtures at 15 K (``Lab.'', top panel)
and in the \spitzer\ spectra (``Astro.'', bottom panel).  
In the top panel, the symbols are labelled with their
corresponding mixtures ; for the \ammonia:\m\ mixtures (orange triangles),
an increasing symbol size is indicative of increasing
\m\ content. 
In the bottom panel, open and filled squares indicate values
obtained with the template and local continuum method, respectively.
In both panels, the dash-dot polygons delimitate the parameter space of FWHM and positions 
corresponding to \water-rich mixtures.
\label{fig:fwhm vs pos nh3} }
\end{minipage}\hfill%
\begin{minipage}{0.48\textwidth}
\epsscale{1}
\includegraphics[width=\columnwidth,bb=46 399 608 710,clip=true]{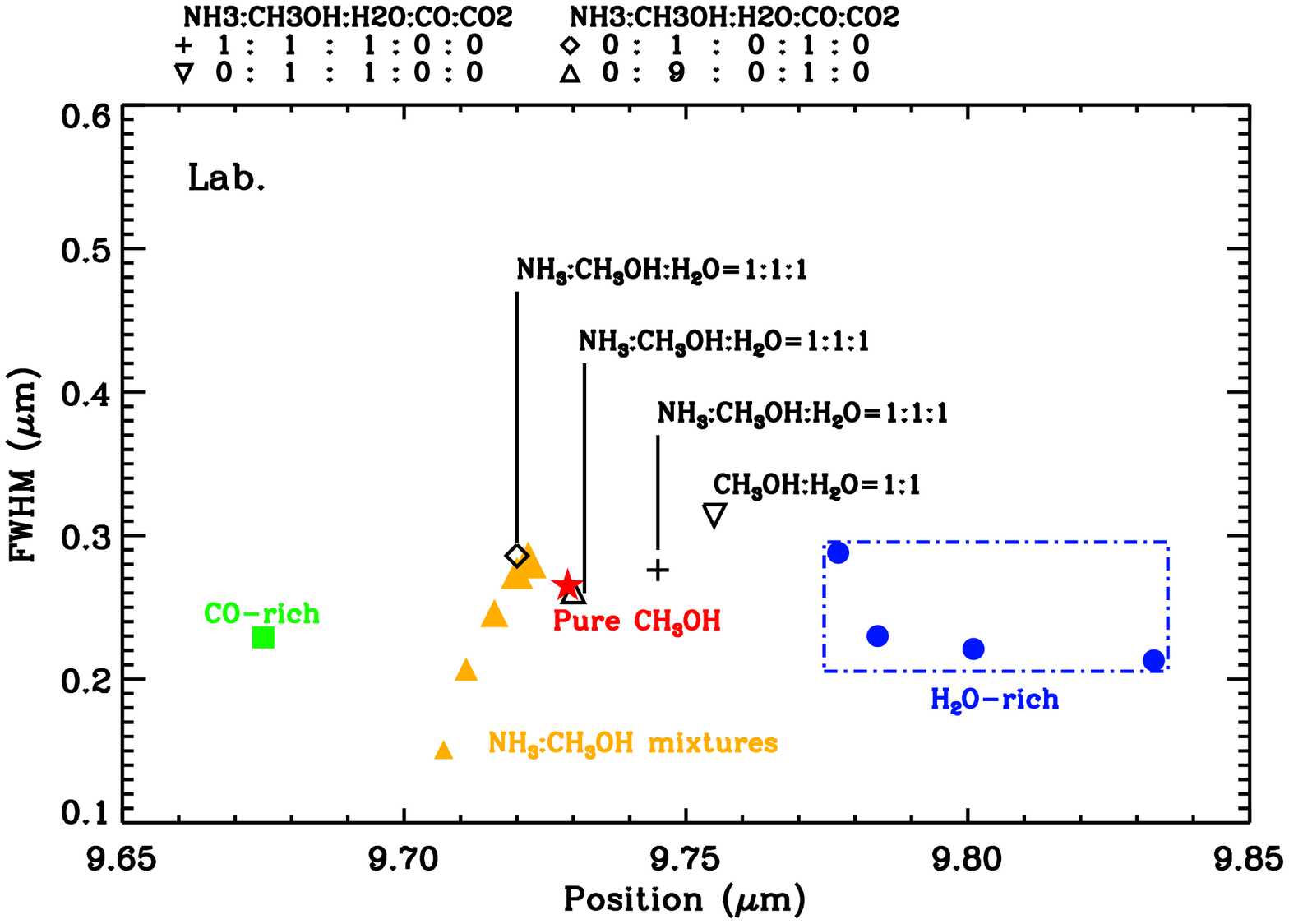}\\
\includegraphics[width=\columnwidth,bb=46 360 608 709,clip=true]{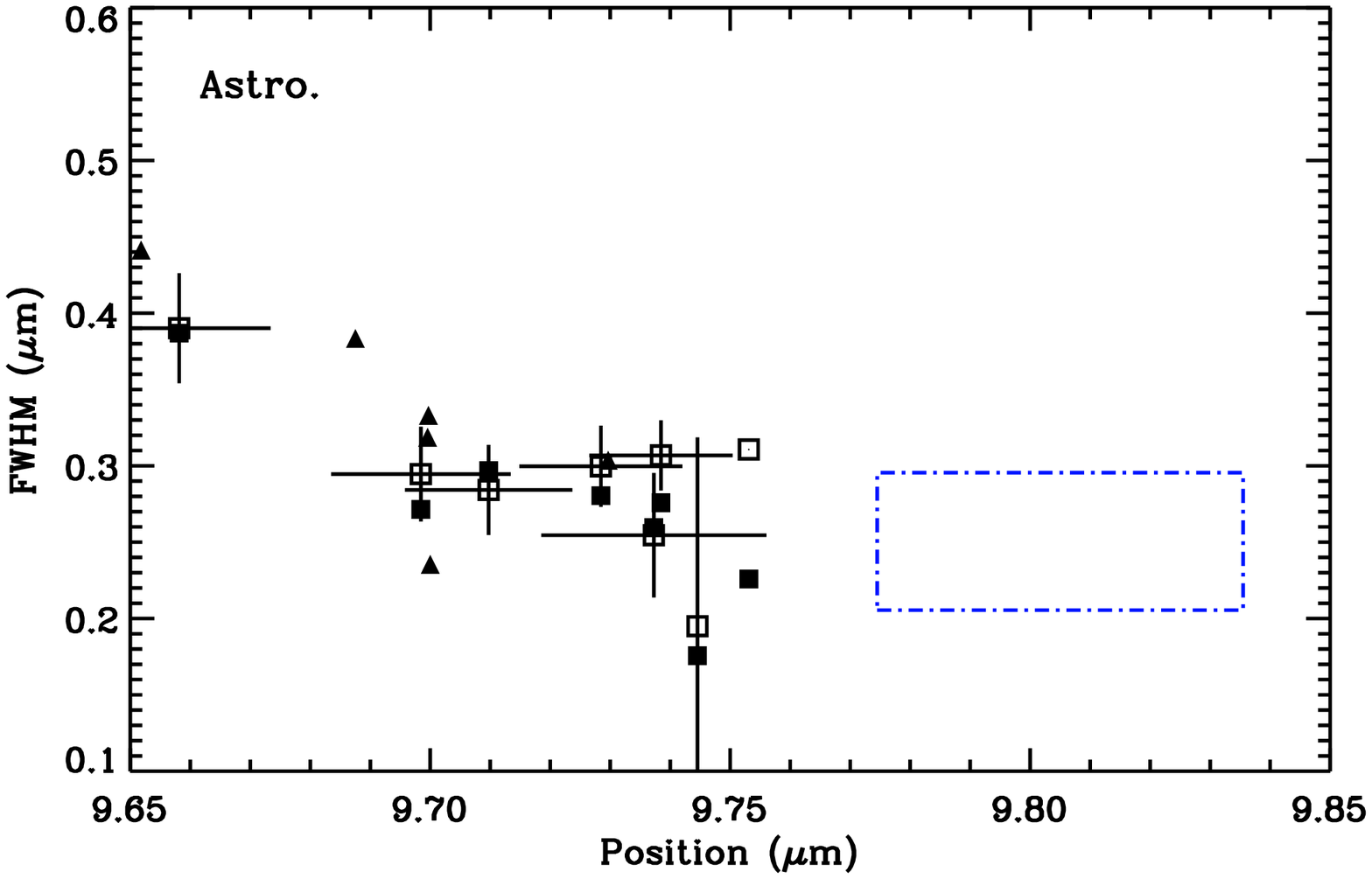}
\caption{\footnotesize 
Same as Fig.~\ref{fig:fwhm vs pos nh3} but for \m.
%
\protect\rule[-9.5\baselineskip]{0pt}{1.75\baselineskip} 
\label{fig:fwhm vs pos ch3oh} }
\end{minipage}
\end{figure*}

The $\nu_4$ C--O stretching vibration profile of CH$_3$OH in the
tertiary mixture does not differ much from the binary values for the
highest water content. The position of the absorption maximum is also
only marginally affected by the temperature. The FWHM decreases from
30~\wn\ (0.29~$\mu$m) for the 10:4:1 mixture to 22~\wn\ (0.21~$\mu$m)
for the 10:0.25:1 mixture.

Besides \water, other species may also be regarded as potential
candidates for changing the spectral appearance of the \ammonia\
and/or \m\ features. Chemically linked is HCOOH
\citep{bisschop-etal07-hcooh} which unfortunately cannot be deposited
in the present setup because of its reactive behavior when mixed with
NH$_3$. Tertiary mixtures with CO and CO$_2$, two other important
constituents in interstellar ices, have been measured (see
Appendix~\ref{ap:lab}) but here the differences are small compared
with the observed binary water-rich or CO-rich mixtures,
and do not offer an alternative explanation.


\section{Comparison between astronomical and laboratory data}
\label{sec:comparison}

\subsection{8-10~\micron\ range}

The FWHM and band positions of the \ammonia\ and \m\ features measured
in the laboratory and astronomical spectra are shown in
Figs.~\ref{fig:fwhm vs pos nh3} (for \ammonia) and \ref{fig:fwhm vs pos ch3oh}
(for \m).  For the YSOs, the values obtained after removal of the
silicate absorption (see Section~\ref{sec:astro}) using the local
continuum method are indicated by filled squares,
whereas those obtained from the template method are plotted with open
squares. Note that the presence of significant amounts of CH$_3$OH may
artificially lower the inferred NH$_3$ $\nu_2$ width in CH$_3$OH rich
sources (indicated with * in Table~\ref{tab:colden}) because of the contribution of
the narrower $\nu_7$ CH$_3$-rock mode.

\begin{figure*}[ht!]
\centering
\begin{minipage}{0.5\textwidth}
\includegraphics[width=\textwidth]{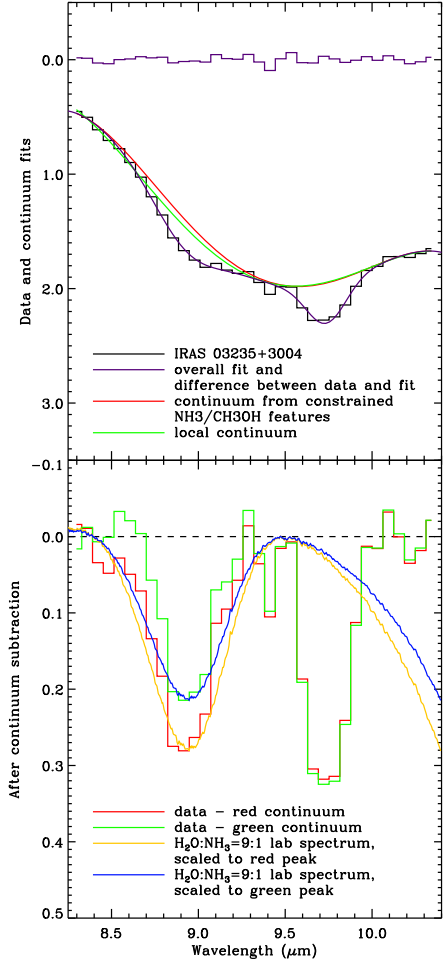}
\end{minipage}\hfill%
\begin{minipage}{0.45\textwidth}
\caption{{\small Example of the effect of continuum determination.
{\it Upper panel:} Spectrum of IRAS~03235+3004 in the 8.25--10.4~\micron\ region
overlaid with the overall fit (smooth purple line --  sum of a 4$^{\rm th}$ order polynomial and two gaussians with positions and widths 
constrained by the values measured in the laboratory spectra of \water:\ammonia\ and \m\ mixtures),
the 4$^{\rm th}$ order polynomial obtained by the overall fit (red) and the local continuum previously determined (green).
The purple histogram is the difference between the data and the overall fit. ---
{\it Bottom panel:} Residuals after subtraction of the two continua, in respective colors. 
The yellow and blue lines are \water:\ammonia=9:1 laboratory spectra scaled to the red and green residuals, respectively.
Note the good agreement between the feature
extracted with the red continuum and the lab
data (orange) showing that astronomical and laboratory data of \ammonia\ ice
mixtures can be consistent if a slightly different continuum
determination is adopted.
} \label{fig: new cont}}
\end{minipage}
\end{figure*}

Regardless of the method used to subtract the continuum, or the type
of source (\m-rich/poor), we find that the observational band positions and FWHM of the
$\nu_2$ \ammonia\ umbrella mode absorptions vary, within the errors,
between 8.9 and 9.1~\micron\ and between $\sim$0.2 and 0.5 \micron, respectively. 
These position and width are not well simultaneously reproduced by 
any of the investigated mixtures. Regarding the positions, those measured in
water-rich ice mixtures are the closest, whereas the positions in pure NH$_3$ or CO/CO$_2$ rich
ices are too far away to be representative of the astronomical positions.
The derived \spitzer\ FWHM values range between 0.23 and 0.32~\micron\
(except for B1-b : 0.39~\micron), when using the local continuum
method, not depending on whether the target is \m-rich or -poor.  For
the template method, \m-rich sources 
generally tend to have a narrower inferred FWHM,
0.3--0.5~\micron, contrary to what would be expected if the NH$_3$
mode is contaminated by the CH$_3$-rock feature. In any case, most of
these widths are still narrower than the laboratory FWHM values. 
To investigate further the effect of the continuum on the positions and widths of
the bands, we performed the following alternative analysis to check whether a continuum
could be found that would yield \ammonia\ and \m\ features with parameters
within the laboratory measurements. 
To do  that, we fitted the data between 8.25 and 10.4~\micron\ with a function that is the 
sum of a 4$^{\rm th}$ order polynomial and two gaussians ; 
positions and widths of the gaussians were constrained 
with limits taken from the laboratory data of binary water mixtures
(8.9--8.95~\micron\ for the \ammonia\ position, 0.42--0.52~\micron\ for its width; 
9.67--9.77~\micron\ for the \m\ position, 0.2--0.3~\micron\ for its width). 
As illustrated in Fig.~\ref{fig: new cont},
we found that the continuum derived in this way is different from those determined via the other
two methods. This result supports the fact that the difference between astronomical and
laboratory data could be attributed to the uncertainty in the continuum determination.

Taking the above considerations into account,
Figs.~\ref{fig:fwhm vs pos nh3} and \ref{fig:fwhm vs pos ch3oh} 
suggest that the template method for
subtraction of the 10~\micron\ silicate absorption is more consistent
with the laboratory measurements, but both methods probably miss some
weak NH$_3$ absorption features in the broad line wings where they blend with
the continuum at the $S/N$ of the data.  
If so, the too small line widths inferred from the data
(most probably due to the uncertainty in the continuum determination)
would mean that we have underestimated \ammonia\ 
abundances by a up to a factor of 2.

The observational band position and FWHM of the \m\ features derived with either the
local continuum or template method are clustered around
9.7--9.75~\micron, with the exception of R CrA IRS 5 at 9.66~\micron.
Similarly the FWHM of the \m\ features are all very similar between
$\sim$0.22 and 0.32~\micron, except for R CrA IRS 5 with 0.39~\micron.
These values agree (with a few exceptions) with the values obtained
from the laboratory spectra.
Note that the observed positions of the
\m\ feature are all on the low side of the laboratory range. Since the
position of this feature shifts to higher wavelengths with increasing
water content, the observed low values could therefore indicate that
\m\ and \water\ are not well mixed and that there exists a separate
\m-rich component, as suggested in previous work
\citep[e.g.][]{pontoppidan-etal03,skinner-etal92}. Alternatively, the
low values could be due to the presence of CO as indicated by the \m\
feature shift to 9.70~\micron\ in \m:CO=1:1.  Both interpretations
would be consistent with the bulk of the CH$_3$OH formation coming
from hydrogenation of a CO-rich layer, rather than photochemistry in a
water-rich matrix. However, the shift from the water-rich mixtures is
small, and some water-rich fraction cannot be excluded with the
current spectral resolution.

\subsection{The 3 and 6~\micron\ ranges}
\label{sec: 3-6 micron}

\begin{figure*}
\includegraphics[angle=0,width=0.95\textwidth,bb=54 397 649 532,clip=true]{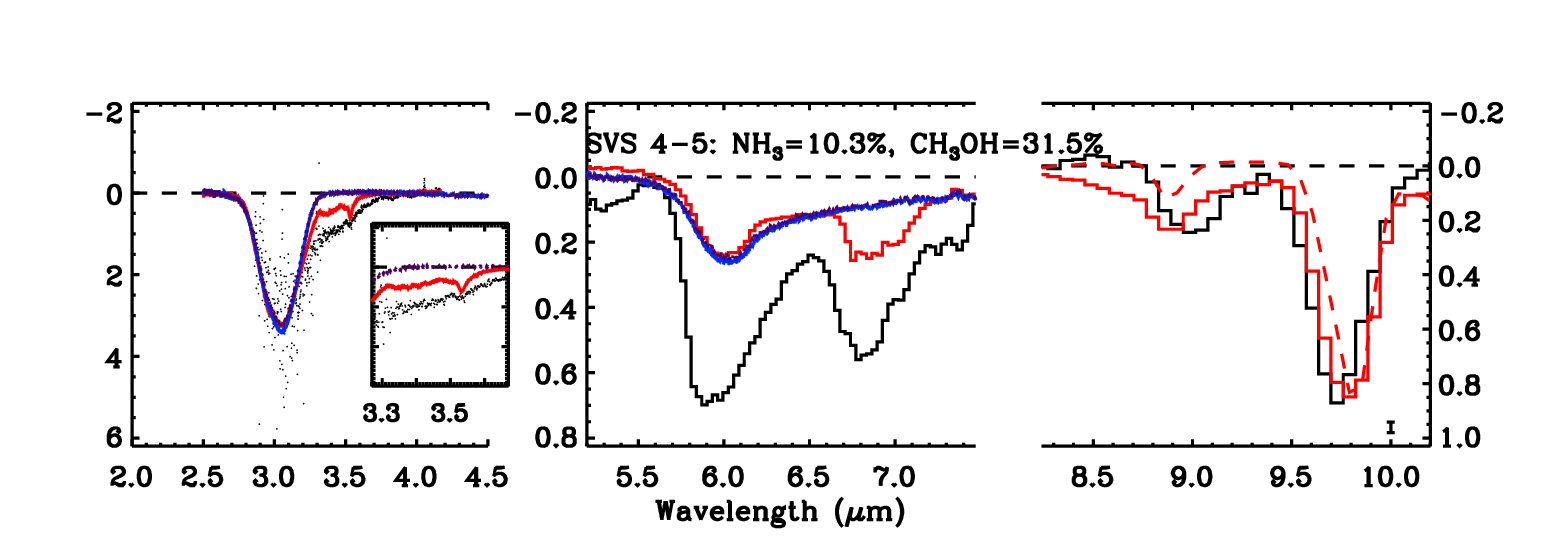}\\
\includegraphics[angle=0,width=0.95\textwidth,bb=54 360 649 528]{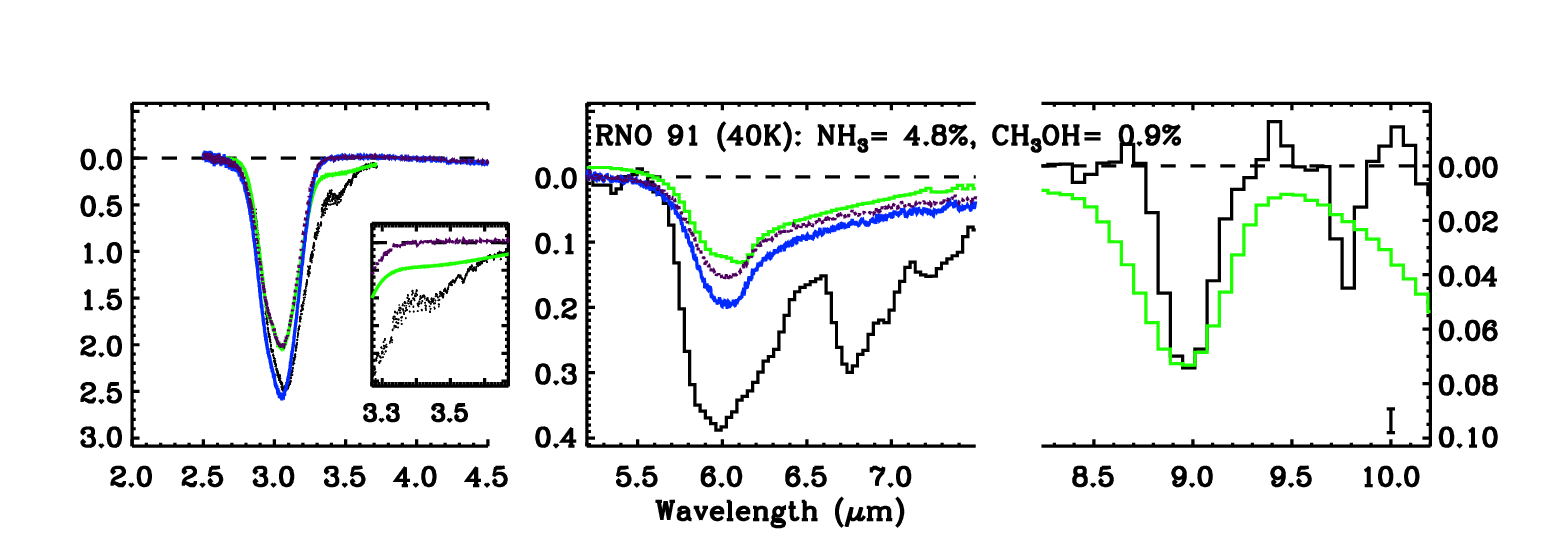}
\caption{ Comparison of astronomical data (VLT or Keck measurements at short wavelengths, IRS \spitzer\ observations elsewhere)
and laboratory spectra in selected wavelength ranges:
$2.0-4.5$~\micron\ (left panels),
$5.2-7.5$~\micron\ (middle panels) and $8.2-10.2$~\micron\ (right panels, silicate absorption 
subtracted via the template method). Error bars are indicated in the bottom-right corner.
Overlaid in red and green are laboratory spectra corresponding to
\water:\m:\ammonia=10:4:1 and \water:\ammonia=9:1, respectively,
scaled to the 9-\micron\ \ammonia\ umbrella mode, and smoothed to the \spitzer\ resolution
in the $\sim 5-10~\micron$ range.
The dark blue line represents the pure water laboratory spectrum scaled to the water column density 
taken in paper I.
The dotted purple line corresponds to a pure water spectrum scaled to the 3-\micron\ water feature of the mixed
ice spectrum, showing the contribution of \ammonia\ features around 3.47 and 6.1~\micron. 
Finally, the red dashed line in the right panel of SVS 4-5 represents a \water:\m=9:1 laboratory spectrum
scaled to the 9.7-\micron\ \m\ CO-stretch mode: this gives an indication of the contribution of the 9-\micron\
\m\ CH$_3$-rock mode to the total 9-\micron\ feature.
The laboratory spectra are recorded at 15~K unless indicated differently.
\label{fig:obs lab example} }
\end{figure*}

\citet{dartois+dhendecourt01} discussed the possibility of a
3.47~\micron\ absorption band which could be related to the formation
of an ammonia hydrate in the ice mantles: they found that if this band
were mostly due to this hydrate, then ammonia abundances would be at
most 5\% with respect to water ice.  Considering the fact that our
derived abundances are larger than 10\% in some sources, it is
necessary to investigate the effect of such a high abundance on the
ammonia features in other spectral ranges.  For this, depending on the
\ammonia-to-\m\ abundance ratio observed in the \spitzer\ spectra, we
scale one of the following laboratory spectra to the 9~\micron\
\ammonia\ feature: \water:\ammonia=9:1, \water:\ammonia=4:1,
\water:\m:\ammonia=10:1:1,
\water:\m:\ammonia=10:4:1. 
Figure~\ref{fig:obs lab example} illustrates the comparison 
between the \spitzer\ and scaled laboratory spectra for the
relevant wavelength ranges for a couple
of sources, while Fig.~\ref{fig:obs lab}-a and -b 
(see Appendix~\ref{ap:lab}) show the
comparison for all sources where \ammonia\ was tentatively detected. \\
For further
comparison, we also overplotted in Fig.~\ref{fig:obs lab example} and \ref{fig:obs lab}, the
following spectra: (i)
the pure \water\ ice spectrum derived from the \water\ column
density quoted in \citet{boogert-etal08} (deep blue); and (ii)
for sources with 3~\micron\ data, the pure \water\ spectrum scaled to
the optical depth of the 3-\micron\ feature of the mixed ice
laboratory spectrum (purple-dotted).
The difference between this scaled pure water spectrum and the
mixed ice spectrum gives an indication of the contribution of ammonia features around 3.47 and 6.1~\micron.

We then determined the contributions from the \ammonia\ features to
the integrated optical depths of the 3 and 6~\micron\ bands and to the
optical depth of component C2, a feature at $6.0-6.4$~\micron\ arising
from a blend of several species, including \ammonia, \water, CO$_2$,
HCOO$^-$ (see Paper I for more details).  These contributions are
reported in Table~\ref{tab:contrib}.

\begin{deluxetable}{l rr rrrr}
\tablewidth{0pt}
\tabletypesize{\scriptsize}
\tablecaption{\ammonia\ contribution to the 3 and 6~\micron\ bands
\label{tab:contrib} }
\tablehead{
Source & $\displaystyle \frac{\int \tau_{\rm H_2O, 3.0}}{\int \tau_{\rm 3.0}}$ 
       & $\displaystyle \frac{\int \tau_{\rm mix, 3.0} }{\int \tau_{\rm 3.0}}$ 
       & $\displaystyle \frac{\int_{1562}^{1785} \tau_{\rm H_2O}}{\int_{1562}^{1785} \tau }$ 
       & $\displaystyle \frac{\int_{1562}^{1785} \tau_{\rm mix}  }{\int_{1562}^{1785} \tau}$ 
       & $\displaystyle \frac{\int \tau_{\rm NH_3, 6.16}}{\int_{1562}^{1785} \tau_{\rm H_2O}}$ 
       & $\displaystyle \frac{\tau_{\rm NH_3, 6.16}}{\tau_{\rm C2}}$  \\ }
\startdata
\multicolumn{7}{c}{Sources with template} \\
\hline
       IRAS 03235+3004 &       -- &       -- &     0.50 &     0.24 &     0.02 &     0.61 \\
       IRAS 03254+3050 &     0.73 &     1.30 &     0.56 &     0.92 &     0.12 &     1.72 \\
       IRAS 04108+2803 &     0.70 &     0.67 &     0.58 &     0.53 &     0.06 &     0.49 \\
                HH 300 &     0.70 &     0.57 &     0.50 &     0.39 &     0.05 &     0.45 \\
       IRAS 08242-5050 &     0.76 &     0.72 &     0.50 &     0.45 &     0.06 &     0.46 \\
       IRAS 08242-5050 &     0.76 &     0.56 &     0.50 &     0.35 &     0.05 &     0.36 \\
 2MASSJ17112317-272431 &      --  &      --  &     0.69 &     0.53 &     0.05 &     4.23 \\
               SVS 4-5 &     0.91 &     0.94 &     0.42 &     0.29 &     0.00 &     0.08 \\
           R CrA IRS 5 &     0.85 &     0.42 &     0.63 &     0.29 &     0.03 &     0.21 \\
\hline
\multicolumn{7}{c}{Sources with no associated template} \\
\hline
                RNO 15 &     0.80 &     1.97 &     0.53 &     1.23 &     0.16 &     0.45 \\
       IRAS 03271+3013 &      --  &      --  &     0.36 &     0.44 &     0.05 &     0.60 \\
                  B1-a &      --  &      --  &     0.67 &     0.43 &     0.03 &     0.57 \\
             L1489 IRS &     0.78 &     0.88 &     0.60 &     0.56 &     0.04 &     0.83 \\
                RNO 91 &     0.94 &     0.94 &     0.53 &     0.45 &     0.04 &     0.53 \\
       IRAS 17081-2721 &     0.65 &     0.95 &     0.62 &     0.75 &     0.05 &     1.64 \\
                 EC 74 &     0.95 &     2.34 &     0.57 &     1.18 &     0.09 &     0.76 \\
                 EC 92 &     0.90 &     0.35 &     0.38 &     0.10 &     0.00 &     0.01 \\
            CrA IRS7 B &      --  &      --  &     0.81 &     0.19 &     0.00 &     0.08 \\
             L1014 IRS &      --  &      --  &     0.62 &     0.55 &     0.06 &     0.34 \\
\enddata
\tablecomments{A dash indicates that the ratio was not calculated due to the high noise in the 3-\micron\ spectrum.\\
Parameters are: \\
$ \int \tau_{\rm H_2O, 3.0}$ = integrated optical depth of pure water at 3~\micron, 
determined from the column density of paper I and a band strength of 2.0$\times 10^{-16}$~\wn. \\
$ \int \tau_{\rm 3.0}$, $ \int \tau_{\rm mix, 3.0}$ = integrated optical depth over the entire 3~\micron\ region for, 
respectively, the considered source and the corresponding laboratory mixture 
(selected from the \ammonia\ feature at 9~\micron). \\
$ \int_{1562}^{1785} \tau_{\rm H_2O}$, $ \int_{1562}^{1785} \tau$, $ \int_{1562}^{1785} \tau_{\rm mix}$ =  
integrated optical depth of, respectively, pure water, source spectrum, and laboratory mixture, 
between 1562 and 1785~\wn\ (5.6 to 6.4~\micron). \\
$ \int \tau_{\rm NH_3, 6.16}$, $ \tau_{\rm NH_3, 6.16}$ = 
integrated and peak optical depth of the 6.16~\micron\ feature of ammonia 
obtained after subtraction of a pure water spectrum scaled to the optical depth at 3~\micron\ of the laboratory mixture.\\
$ \tau_{\rm C2}$ = peak optical depth of the C2 component from paper I.\\ }
\end{deluxetable}

Figure~\ref{fig:obs lab example},~\ref{fig:obs lab} and Table~\ref{tab:contrib} show that (i) the
scaled laboratory spectra {\it generally} do not overestimate the observed
absorption features, and (ii) for most sources, the presence of
\ammonia\ at the level we determine from the 9~\micron\ feature does
not explain by itself the depth of the C2 component and of the red
wing of the 3~\micron\ band.  Hence, our inferred NH$_3$ abundances up
to 15\% from the 9.7 $\mu$m data are not in conflict with the lack of
other NH$_3$ features.  The only exceptions are two sources (RNO 15
and EC 74), for which the scaled mixed ice spectrum exceeds the data
in the 3-\micron\ range. In the case of RNO 15, the \ammonia\
abundance could have been overestimated due to the contribution of the
\m\ CH$_3$-rock feature at $\sim$9~\micron. For EC 74, this
overestimate and the presence of emission weakens the identification
of the \ammonia\ signature. In both cases, the quoted \ammonia\
abundances should be considered as upper limits.

Overall, our reported \ammonia\ abundances are up to a factor of three
larger than the upper limits derived by \citet{dartois+dhendecourt01}.
Firstly, let's recall that the conclusions in their study and in ours
are drawn from the analysis of different samples.
Secondly, Dartois \& d'Hendecourt made an assumption that does 
not apply to our sample: indeed, they
considered a grain size distribution including also scattering from
larger grains, producing an enhanced 3-\micron\ wing,
whereas the results presented here can be taken
as representative of \ammonia\ absorption from small grains. 
It is beyond the scope of this paper to
investigate the effects of grain size distribution and scattering in as
much detail as did \citet{dartois+dhendecourt01}.

\subsection{Nitrogen ice inventory}

The confirmation of the presence of relatively large amounts of solid
\ammonia, up to 15\%, in interstellar ices solves a long-standing
problem.  Indeed, the detection of solid \ammonia\ has remained
elusive and/or controversial, despite a number of clues suggesting its
presence:
\begin{itemize}
\item High cosmic abundance of atomic nitrogen : $N_{\rm N}/N_{\rm H}
= 7.76\times10^{-5}$ 
\citep{savage+sembach96},
only a factor of a few below those of oxygen and carbon. Here
$N_H$ indicates the total number of hydrogen nuclei,
$N_H$=$N$(H)+2$N$(H$_2$).

\item High abundances of gaseous \ammonia\ of
$N_{\rm NH_3} /N_{\rm H_2} \sim 10^{-6}-10^{-5}$ in the Orion-KL
nebula \citep{barrett-etal77,genzel-etal82} and in other hot cores
such as G9.62+0.19, G29.96$-$0.02, G31.41+0.31
\citep{cesaroni-etal94-nh3-uchii}, and G10.47+0.03
\citep{cesaroni-etal94-nh3-uchii,osorio-etal09}.

\item Identification of substantial amounts of OCN$^-$
  \citep[e.g.][]{vanbroekhuizen-etal04,vanbroekhuizen-etal05} and
  NH$_4^+$ in ices \citep[e.g.][]{schutte+khanna03,boogert-etal08}:
  considering that these ions form via reactions involving \ammonia,
  the non-detection of solid \ammonia\ would be puzzling.
\end{itemize}

Our results can be used to draw up a possible nitrogen budget.
Assuming $N_{\rm H_2O}/N_{\rm H} \sim 5\times10^{-5}$
\citep{pontoppidan-etal04,boogert-etal04}, and average abundances
w.r.t. \water\ of 5.5\% for \ammonia\ (see  \S \ref{sec:nh3 results}), 7\%
for \ammonium\ (from Table~3 of Paper I), and 0.6\% for OCN$^-$
\citep{vanbroekhuizen-etal05}, then the \ammonia, \ammonium\ and \ocn\
abundances with respect to total H are 2.8, 3.5, and 0.3 $\times10^{-6}$
respectively.
This corresponds to, respectively, 
3.4, 4.4 and 0.4\%
of the atomic nitrogen cosmic abundance so that, in total, about 10\%
of the cosmically available nitrogen would be locked up in ices,
leaving solid and gaseous N$_2$, N and HCN as other substantial
nitrogen carriers. The main uncertainty in this determination is the
adopted H$_2$O ice abundance with respect to total H; in several sources
this may well be a factor of 2 larger, leading to about 20\% of the
nitrogen accounted for in ices.

\section{Conclusion}
\label{sec:ccl}

We have analyzed in detail the 8-10~\micron\ range of the spectra of
41 low-mass YSOs obtained with \spitzer\ and presented in
\citet{boogert-etal08}.  The sources are categorized into three types:
straight, curved and rising 8~\micron\ silicate wings, and for each
category template sources with little or no absorption from ices
around 9-10~\micron\ have been determined.  This has led to two ways
of subtracting the contribution from the 10~\micron\ silicate
absorption: first, by determining a local continuum, and second, by
scaling the templates to the optical depth at 9.7~\micron.  The two
methods give consistent band positions of the NH$_3$ features, but the
resulting widths can be up to a factor of two larger using the
template continuum method.  
Taking into account the uncertainty in continuum determination,
NH$_3$ ice is most likely detected in 24 of
the 41 sources with abundances of $\sim$2 to 15~\% w.r.t. \water, with
an average abundance of 5.5$\pm$2.0~\%. These abundances have estimated
uncertainties up to a factor of two and are not inconsistent with other
features in the 3 and 6~\micron\ ranges.  CH$_3$OH is often detected
as well, but the NH$_3$/CH$_3$OH abundance ratio changes strongly from
source to source. Our inferred CH$_3$OH column densities are
consistent with the values derived in paper I.

Targeted laboratory experiments have been carried out to characterize
the NH$_3$ and CH$_3$OH profiles (position, FWHM, integrated
absorbance).  Comparison with the observational data shows reasonable
agreement (within $\sim$1\%) for the position of the NH$_3$ feature in H$_2$O-rich ices,
but the observed widths are systematically smaller than the laboratory ones for
nearly all sources. The silicate template continuum method gives
widths that come closest to the laboratory values.  This difference in
width (i.e. widths derived from astronomical spectra smaller than those
in the laboratory spectra) suggests that the NH$_3$ abundances determined here
may be on the low side.

The CH$_3$OH profile is most consistent with a significant fraction of
the CH$_3$OH in a relatively pure or CO-rich phase, consistent with
its formation by the hydrogenation of CO ice.  
In contrast, the most likely formation
route of NH$_3$ ice remains hydrogenation of atomic N together with
water ice formation in a relatively low density molecular
phase. Finally, the nitrogen budget indicates that up to 10 to 20~\% of
nitrogen is locked up in known ices.


\begin{acknowledgments}
We thank Karoliina Isokoski (Leiden) for recording additional
laboratory spectra during the completion of this study.
We are also thankful to Helen Fraser and the c2d team for stimulating discussions and
useful comments on the manuscript. Support for this work, part of the
Spitzer Legacy Science Program, was provided by NASA through contracts
1224608, 1230779, 1230782, 1256316, and 1279952 issued by the Jet
Propulsion Laboratory, California Institute of Technology, under NASA
contract 1407. Astrochemistry in Leiden is supported by a Spinoza
grant of the Netherlands Organization for Scientific Research (NWO), 
and by a NOVA grant.
The laboratory work is financially supported by `Stichting voor
Fundamenteel Onderzoek der Materie' (FOM), and `the Netherlands
Research School for Astronomy' (NOVA). Funding for KI\"{O} was
provided by a grant from the European Early Stage Training
Network (MEST-CT-2004-504604). Support for KMP was provided by NASA
through Hubble Fellowship grant 1201.01 awarded by the Space Telescope
Science Institute, which is operated by the Association of
Universities for Research in Astronomy, Inc., for NASA, under contract
NAS 5-26555. 
\end{acknowledgments}

\clearpage

\appendix

\setcounter{table}{0}
\renewcommand{\thetable}{\thesection.\arabic{table}}

\setcounter{figure}{0}
\renewcommand{\thefigure}{\thesection.\arabic{figure}}

\section{Parameters of Gaussian fits}
\label{ap:gaussian param}

\begin{deluxetable}{l r@{$\pm$}l r@{$\pm$}l r@{$\pm$}l c r@{$\pm$}l r@{$\pm$}l r@{$\pm$}l}
\tablecaption{Parameters of Gaussian fits to the NH$_3$ feature.
\label{tab:gaussian param} }
\tabletypesize{\scriptsize}
\tablewidth{0pt}
\tablehead{
Source & \multicolumn{6}{c}{\ammonia, local} &
       & \multicolumn{6}{c}{\ammonia, template} \\ 
\cline{2-7}\cline{9-14}
       & \multicolumn{2}{c}{$\lambda$ (\micron)}
       & \multicolumn{2}{c}{FWHM (\micron)}
       & \multicolumn{2}{c}{$\tau_{\rm peak}$} &
       & \multicolumn{2}{c}{$\lambda$ (\micron)}
       & \multicolumn{2}{c}{FWHM (\micron)}
       & \multicolumn{2}{c}{$\tau_{\rm peak}$} }
\startdata
       IRAS 03235+3004 &  8.93 &  0.02 &  0.28 &  0.03 &  0.23 &  0.02 &  &  8.93 &  0.01 &  0.29 &  0.03 &  0.30 &  0.02 \\ 
            L1455 IRS3 &  8.99 &  0.03 &  0.24 &  0.07 &  0.02 &  0.01 &  &  9.02 &  0.02 &  0.38 &  0.05 &  0.04 &  0.01 \\ 
       IRAS 03254+3050 &  9.04 &  0.01 &  0.25 &  0.03 &  0.10 &  0.01 &  &  8.99 &  0.01 &  0.38 &  0.03 &  0.12 &  0.01 \\ 
B1-b\tablenotemark{*} &  9.05 &  0.03 &  0.39 &  0.06 &  0.25 &  0.02 &  &  9.07 &  0.03 &  0.40 &  0.06 &  0.31 &  0.03 \\ 
       IRAS 04108+2803 &  8.99 &  0.02 &  0.25 &  0.04 &  0.05 &  0.01 &  &  9.05 &  0.03 &  0.47 &  0.06 &  0.04 &  0.01 \\ 
                HH 300 &  9.01 &  0.02 &  0.23 &  0.05 &  0.04 &  0.01 &  &  9.06 &  0.02 &  0.45 &  0.06 &  0.05 &  0.01 \\ 
       IRAS 08242-5050 &  9.02 &  0.01 &  0.31 &  0.03 &  0.15 &  0.01 &  &  9.05 &  0.01 &  0.30 &  0.03 &  0.15 &  0.01 \\ 
       IRAS 15398-3359 &  8.96 &  0.01 &  0.29 &  0.03 &  0.30 &  0.02 &  &  8.98 &  0.01 &  0.33 &  0.03 &  0.41 &  0.02 \\ 
              B59 YSO5 &  8.95 &  0.01 &  0.27 &  0.03 &  0.18 &  0.02 &  &  8.89 &  0.02 &  0.34 &  0.04 &  0.18 &  0.02 \\ 
 2MASSJ17112317-272431 &  8.99 &  0.01 &  0.30 &  0.02 &  0.43 &  0.02 &  &  9.02 &  0.02 &  0.50 &  0.05 &  0.41 &  0.04 \\ 
SVS 4-5\tablenotemark{*} &  9.00 &  0.01 &  0.26 &  0.03 &  0.16 &  0.02 &  &  9.01 &  0.01 &  0.30 &  0.03 &  0.26 &  0.02 \\ 
           R CrA IRS 5 &  9.05 &  0.02 &  0.21 &  0.04 &  0.04 &  0.01 &  &  9.00 &  0.03 &  0.36 &  0.06 &  0.04 &  0.01 \\ 
\hline
                RNO 15 &  9.05 &  0.02 &  0.20 &  0.04 &  0.04 &  0.01 &  & \multicolumn{2}{c}{--}  & \multicolumn{2}{c}{--}  & \multicolumn{2}{c}{--}  \\ 
       IRAS 03271+3013 &  8.96 &  0.02 &  0.25 &  0.04 &  0.20 &  0.02 &  & \multicolumn{2}{c}{--}  & \multicolumn{2}{c}{--}  & \multicolumn{2}{c}{--}  \\ 
                  B1-a &  8.98 &  0.02 &  0.25 &  0.04 &  0.14 &  0.02 &  & \multicolumn{2}{c}{--}  & \multicolumn{2}{c}{--}  & \multicolumn{2}{c}{--}  \\ 
             L1489 IRS &  9.02 &  0.01 &  0.27 &  0.03 &  0.09 &  0.01 &  & \multicolumn{2}{c}{--}  & \multicolumn{2}{c}{--}  & \multicolumn{2}{c}{--}  \\ 
       IRAS 13546-3941 &  8.99 &  0.02 &  0.27 &  0.03 &  0.03 &  0.00 &  & \multicolumn{2}{c}{--}  & \multicolumn{2}{c}{--}  & \multicolumn{2}{c}{--}  \\ 
                RNO 91 &  8.98 &  0.01 &  0.26 &  0.03 &  0.08 &  0.01 &  & \multicolumn{2}{c}{--}  & \multicolumn{2}{c}{--}  & \multicolumn{2}{c}{--}  \\ 
       IRAS 17081-2721 &  8.97 &  0.02 &  0.24 &  0.04 &  0.04 &  0.00 &  & \multicolumn{2}{c}{--}  & \multicolumn{2}{c}{--}  & \multicolumn{2}{c}{--}  \\ 
                 EC 74 &  9.01 &  0.02 &  0.22 &  0.05 &  0.05 &  0.01 &  & \multicolumn{2}{c}{--}  & \multicolumn{2}{c}{--}  & \multicolumn{2}{c}{--}  \\ 
                 EC 82 &  8.94 &  0.01 &  0.29 &  0.03 &  0.04 &  0.00 &  & \multicolumn{2}{c}{--}  & \multicolumn{2}{c}{--}  & \multicolumn{2}{c}{--}  \\ 
                 EC 90 &  8.95 &  0.02 &  0.18 &  0.05 &  0.04 &  0.01 &  & \multicolumn{2}{c}{--}  & \multicolumn{2}{c}{--}  & \multicolumn{2}{c}{--}  \\ 
EC 92\tablenotemark{*} &  8.99 &  0.02 &  0.25 &  0.05 &  0.03 &  0.00 &  & \multicolumn{2}{c}{--}  & \multicolumn{2}{c}{--}  & \multicolumn{2}{c}{--}  \\ 
                   CK4 &  8.99 &  0.02 &  0.32 &  0.04 &  0.03 &  0.00 &  & \multicolumn{2}{c}{--}  & \multicolumn{2}{c}{--}  & \multicolumn{2}{c}{--}  \\ 
CrA IRS7 B\tablenotemark{*} &  9.04 &  0.01 &  0.26 &  0.03 &  0.15 &  0.01 &  & \multicolumn{2}{c}{--}  & \multicolumn{2}{c}{--}  & \multicolumn{2}{c}{--}  \\ 
             L1014 IRS &  9.03 &  0.02 &  0.25 &  0.05 &  0.15 &  0.02 &  & \multicolumn{2}{c}{--}  & \multicolumn{2}{c}{--}  & \multicolumn{2}{c}{--}  \\ 
\enddata
\tablecomments{Uncertainties are statistical errors from the Gaussian fits.}
\tablenotetext{*}{Sources with $\tau_{9.7\micron}>2\times\tau_{9.0\micron}$, 
for which the contribution from the \m\ CH$_3$-rock mode is significant. 
Since the latter and the \ammonia\ umbrella mode were difficult to disentangle, 
a single fit was performed (the reported parameters)
and the integrated optical depth of the ammonia feature was then
obtained from the total integrated optical depth at 9~\micron
by subtracting the estimated contribution of the \m\ CH$_3$-rock mode
(see \S\ref{sec:template}).
}
\end{deluxetable}

\clearpage

\begin{landscape}
\begin{deluxetable}{l r@{$\pm$}l r@{$\pm$}l r@{$\pm$}l r@{$\pm$}l c r@{$\pm$}l r@{$\pm$}l r@{$\pm$}l r@{$\pm$}l c r@{$\pm$}l}
\tablecaption{Parameters of Gaussian fits to the CH$_3$OH C-O stretch mode
(after subtraction of the continuum with the local and/or template method), 
and \m\ column densities (or 3-$\sigma$ upper limits).
\label{tab:gaussian param ch3oh}
}
\tabletypesize{\scriptsize}
\tablewidth{0pt}
\tablecolumns{21}
\tablehead{
Source & \multicolumn{8}{c}{Local continuum} &
       & \multicolumn{8}{c}{Template continuum} &
       & \multicolumn{2}{c}{Paper I} \\ 
\cline{2-9}\cline{11-18}
       & \multicolumn{2}{c}{$\lambda$}
       & \multicolumn{2}{c}{FWHM}
       & \multicolumn{2}{c}{$\tau_{\rm peak}$} 
       & \multicolumn{2}{c}{$X$} &
       & \multicolumn{2}{c}{$\lambda$}
       & \multicolumn{2}{c}{FWHM}
       & \multicolumn{2}{c}{$\tau_{\rm peak}$} 
       & \multicolumn{2}{c}{$X$} &
       & \multicolumn{2}{c}{$X$} \\
       & \multicolumn{2}{c}{(\micron)}
       & \multicolumn{2}{c}{(\micron)}
       & \multicolumn{2}{c}{} 
       & \multicolumn{2}{c}{(\% \water)} &
       & \multicolumn{2}{c}{(\micron)}
       & \multicolumn{2}{c}{(\micron)}
       & \multicolumn{2}{c}{} 
       & \multicolumn{2}{c}{(\% \water)} &
       & \multicolumn{2}{c}{(\% \water)} 
}
\startdata
       IRAS 03235+3004 &  9.74 &  0.02 &  0.26 &  0.03 &  0.35 &  0.04 &  4.40 &  1.04 &  &  9.74 &  0.02 &  0.25 &  0.04 &  0.31 &  0.04 &  3.84 &  0.99 &  & 4.20 &  1.20 \\ 
            L1455 IRS3 &  9.78 &  0.01 &  0.14 &  0.03 &  0.03 &  0.01 &  3.67 &  1.80 &  &  9.78 &  0.02 &  0.26 &  0.04 &  0.04 &  0.01 &  7.71 &  3.46 &  & \multicolumn{2}{c}{$<$12.5} \\
       IRAS 03254+3050 & \multicolumn{2}{c}{\nodata} & \multicolumn{2}{c}{\nodata} & \multicolumn{2}{c}{\nodata} & \multicolumn{2}{c}{$<$ 5.4} & & \multicolumn{2}{c}{\nodata} & \multicolumn{2}{c}{\nodata} & \multicolumn{2}{c}{\nodata} & \multicolumn{2}{c}{$<$ 5.4} & & \multicolumn{2}{c}{$<$ 4.6} \\
                  B1-b &  9.71 &  0.01 &  0.30 &  0.03 &  1.19 &  0.11 & 14.15 &  3.16 &  &  9.71 &  0.01 &  0.28 &  0.03 &  1.21 &  0.11 & 13.75 &  3.12 &  & 11.20 &  0.70 \\ 
       IRAS 04108+2803 & \multicolumn{2}{c}{\nodata} & \multicolumn{2}{c}{\nodata} & \multicolumn{2}{c}{\nodata} & \multicolumn{2}{c}{$<$ 2.7} & &  9.75 &  0.00 &  0.06 &  0.04 &  0.04 &  0.03 &  0.58 &  0.62 &  & \multicolumn{2}{c}{$<$ 3.5} \\
                HH 300 & \multicolumn{2}{c}{\nodata} & \multicolumn{2}{c}{\nodata} & \multicolumn{2}{c}{\nodata} & \multicolumn{2}{c}{$<$ 4.7} & &  9.74 &  0.00 &  0.19 &  0.12 &  0.01 &  0.01 &  0.78 &  0.52 &  & \multicolumn{2}{c}{$<$ 6.7} \\
       IRAS 08242-5050 &  9.70 &  0.01 &  0.27 &  0.03 &  0.25 &  0.02 &  6.12 &  1.01 &  &  9.70 &  0.01 &  0.29 &  0.03 &  0.24 &  0.02 &  6.39 &  1.09 &  &  5.50 &  0.30 \\ 
       IRAS 15398-3359 &  9.73 &  0.01 &  0.28 &  0.03 &  0.77 &  0.06 & 10.26 &  3.02 &  &  9.73 &  0.01 &  0.30 &  0.03 &  0.75 &  0.06 & 10.69 &  3.14 &  & 10.30 &  0.80 \\ 
              B59 YSO5 & \multicolumn{2}{c}{\nodata} & \multicolumn{2}{c}{\nodata} & \multicolumn{2}{c}{\nodata} & \multicolumn{2}{c}{$<$ 1.2} & & \multicolumn{2}{c}{\nodata} & \multicolumn{2}{c}{\nodata} & \multicolumn{2}{c}{\nodata} & \multicolumn{2}{c}{$<$ 1.2} & & \multicolumn{2}{c}{$<$ 1.3} \\
 2MASSJ17112317-272431 &  9.75 &  0.02 &  0.23 &  0.04 &  0.13 &  0.02 &  1.03 &  0.22 &  & \multicolumn{2}{c}{\nodata} & \multicolumn{2}{c}{\nodata} & \multicolumn{2}{c}{\nodata} & \multicolumn{2}{c}{$<$ 2.0} & & \multicolumn{2}{c}{$<$ 3.2} \\
               SVS 4-5 &  9.74 &  0.01 &  0.28 &  0.03 &  0.77 &  0.06 & 26.38 &  6.17 &  &  9.74 &  0.01 &  0.31 &  0.02 &  0.83 &  0.06 & 31.50 &  7.12 &  & 25.20 &  3.50 \\ 
           R CrA IRS 5 &  9.66 &  0.01 &  0.39 &  0.03 &  0.07 &  0.00 &  5.68 &  0.60 &  &  9.66 &  0.02 &  0.39 &  0.04 &  0.07 &  0.00 &  5.51 &  0.72 &  &  6.60 &  1.60 \\ 
\hline
                RNO 15 &  9.65 &  0.03 &  0.44 &  0.07 &  0.02 &  0.00 & 11.13 &  2.16 &  & \multicolumn{2}{c}{\nodata}  & \multicolumn{2}{c}{\nodata}  & \multicolumn{2}{c}{\nodata}  & \multicolumn{2}{c}{\nodata}  &  & \multicolumn{2}{c}{$<$ 5.0} \\
       IRAS 03271+3013 & \multicolumn{2}{c}{\nodata} & \multicolumn{2}{c}{\nodata} & \multicolumn{2}{c}{\nodata} & \multicolumn{2}{c}{$<$ 4.3} &  & \multicolumn{2}{c}{\nodata}  & \multicolumn{2}{c}{\nodata}  & \multicolumn{2}{c}{\nodata}  & \multicolumn{2}{c}{\nodata}  &  & \multicolumn{2}{c}{$<$ 5.6} \\
                  B1-a & \multicolumn{2}{c}{\nodata} & \multicolumn{2}{c}{\nodata} & \multicolumn{2}{c}{\nodata} & \multicolumn{2}{c}{$<$ 2.4} &  & \multicolumn{2}{c}{\nodata}  & \multicolumn{2}{c}{\nodata}  & \multicolumn{2}{c}{\nodata}  & \multicolumn{2}{c}{\nodata}  &  & \multicolumn{2}{c}{$<$ 1.9} \\
             L1489 IRS &  9.78 &  0.02 &  0.10 &  0.03 &  0.03 &  0.01 &  0.44 &  0.22 &  & \multicolumn{2}{c}{\nodata}  & \multicolumn{2}{c}{\nodata}  & \multicolumn{2}{c}{\nodata}  & \multicolumn{2}{c}{\nodata}  &  &  4.90 &  1.50 \\ 
       IRAS 13546-3941 & \multicolumn{2}{c}{\nodata} & \multicolumn{2}{c}{\nodata} & \multicolumn{2}{c}{\nodata} & \multicolumn{2}{c}{$<$ 2.0} &  & \multicolumn{2}{c}{\nodata}  & \multicolumn{2}{c}{\nodata}  & \multicolumn{2}{c}{\nodata}  & \multicolumn{2}{c}{\nodata}  &  & \multicolumn{2}{c}{$<$ 3.9} \\
                RNO 91 &  9.77 &  0.01 &  0.11 &  0.03 &  0.05 &  0.01 &  0.87 &  0.32 &  & \multicolumn{2}{c}{\nodata}  & \multicolumn{2}{c}{\nodata}  & \multicolumn{2}{c}{\nodata}  & \multicolumn{2}{c}{\nodata}  &  & \multicolumn{2}{c}{$<$ 5.6} \\
       IRAS 17081-2721 & \multicolumn{2}{c}{\nodata} & \multicolumn{2}{c}{\nodata} & \multicolumn{2}{c}{\nodata} & \multicolumn{2}{c}{$<$ 6.6} &  & \multicolumn{2}{c}{\nodata}  & \multicolumn{2}{c}{\nodata}  & \multicolumn{2}{c}{\nodata}  & \multicolumn{2}{c}{\nodata}  &  &  3.30 &  0.80 \\ 
                 EC 74 & \multicolumn{2}{c}{\nodata} & \multicolumn{2}{c}{\nodata} & \multicolumn{2}{c}{\nodata} & \multicolumn{2}{c}{$<$13.5} &  & \multicolumn{2}{c}{\nodata}  & \multicolumn{2}{c}{\nodata}  & \multicolumn{2}{c}{\nodata}  & \multicolumn{2}{c}{\nodata}  &  & \multicolumn{2}{c}{$<$ 9.3} \\
                 EC 82 & \multicolumn{2}{c}{\nodata} & \multicolumn{2}{c}{\nodata} & \multicolumn{2}{c}{\nodata} & \multicolumn{2}{c}{$<$24.6} &  & \multicolumn{2}{c}{\nodata}  & \multicolumn{2}{c}{\nodata}  & \multicolumn{2}{c}{\nodata}  & \multicolumn{2}{c}{\nodata}  &  & \multicolumn{2}{c}{$<$14.2} \\
                 EC 90 &  9.70 &  0.01 &  0.32 &  0.03 &  0.05 &  0.00 &  6.91 &  0.99 &  & \multicolumn{2}{c}{\nodata}  & \multicolumn{2}{c}{\nodata}  & \multicolumn{2}{c}{\nodata}  & \multicolumn{2}{c}{\nodata}  &  &  6.80 &  1.60 \\ 
                 EC 92 &  9.73 &  0.01 &  0.30 &  0.02 &  0.09 &  0.01 & 11.16 &  1.46 &  & \multicolumn{2}{c}{\nodata}  & \multicolumn{2}{c}{\nodata}  & \multicolumn{2}{c}{\nodata}  & \multicolumn{2}{c}{\nodata}  &  & 11.70 &  3.50 \\ 
                   CK4 & \multicolumn{2}{c}{\nodata} & \multicolumn{2}{c}{\nodata} & \multicolumn{2}{c}{\nodata} & \multicolumn{2}{c}{\nodata} &  &  \multicolumn{2}{c}{\nodata} &  \multicolumn{2}{c}{\nodata} &  \multicolumn{2}{c}{\nodata} &  \multicolumn{2}{c}{\nodata} &  & \multicolumn{2}{c}{\nodata} \\
            CrA IRS7 B &  9.70 &  0.01 &  0.33 &  0.02 &  0.36 &  0.02 &  7.74 &  1.56 &  & \multicolumn{2}{c}{\nodata}  & \multicolumn{2}{c}{\nodata}  & \multicolumn{2}{c}{\nodata}  & \multicolumn{2}{c}{\nodata}  &  &  6.80 &  0.30 \\ 
             L1014 IRS &  9.69 &  0.03 &  0.38 &  0.08 &  0.10 &  0.01 &  3.61 &  0.99 &  & \multicolumn{2}{c}{\nodata}  & \multicolumn{2}{c}{\nodata}  & \multicolumn{2}{c}{\nodata}  & \multicolumn{2}{c}{\nodata}  &  &  3.10 &  0.80 \\ 
\enddata
\tablecomments{This table shows that \m\ column densities obtained in this paper are consistent with those
in Paper I, which are our recommended values.}
\tablecomments{Uncertainties are statistical errors from the Gaussian fits.}
\end{deluxetable}

\end{landscape}

\clearpage

\section{Additional information on laboratory data}
\label{ap:lab}

\begin{deluxetable}{ccccc @{\extracolsep{5pt}} c cc cc c cl}
\tablecolumns{13}
\tabletypesize{\scriptsize}
\tablecaption{Ice composition, band maximum position (``peak position''), 
FWHM and band strength relative to the pure ice ($A_{\rm rel.}$), 
listed for a set of ice mixtures under investigation.
\label{table_lab_overview} } 
\tablewidth{0pt}
\tablehead{
\multicolumn{5}{c}{Ice mixture} & 
\colhead{Molecule} & 
\multicolumn{2}{c}{Peak position} & 
\multicolumn{2}{c}{FWHM} & 
\colhead{$A_{\rm rel.}$} & 
\multicolumn{2}{c}{Mode} \\	
\cline{1-5}\cline{7-8}\cline{9-10}\cline{12-13} 
\colhead{NH$_3$} &\colhead{CH$_3$OH} & \colhead{H$_2$O} &\colhead{CO} &\colhead{CO$_2$} & 
\colhead{}   &\colhead{cm$^{-1}$} & \colhead{$\mu$m} & \colhead{cm$^{-1}$} & \colhead{$\mu$m} & 
\colhead{} &  \colhead{} &  \colhead{}
} 
\startdata
1      &0      &0       &0  &0     &NH$_3$    &1070        &9.341       &66          & 0.577 &1      &$\nu_2$  & umbrella    \\
\hline                                                                                                                       
1      &0      &0.11    &0  &0     &NH$_3$    &1076        &9.291       &70          & 0.605 &1      &$\nu_2$  & umbrella    \\
1      &0      &1       &0  &0     &NH$_3$    &1100        &9.091       &77          & 0.637 &1      &$\nu_2$  & umbrella    \\
1      &0      &9       &0  &0     &NH$_3$    &1118        &8.947       &62          & 0.496 &0.7    &$\nu_2$  & umbrella    \\
\hline                                                                                                                       
1      &0      &10      &1  &0     &NH$_3$    &1124        &8.897       &53          & 0.420 &0.7    &$\nu_2$  & umbrella    \\
1      &0      &1       &1  &0     &NH$_3$    &1094        &9.144       &75          & 0.627 &1      &$\nu_2$  & umbrella    \\
\hline                                                                                                                       
1      &0      &10      &0  &2     &NH$_3$    &1122        &8.916       &57          & 0.453 &0.8    &$\nu_2$  & umbrella    \\
1      &0      &1       &0  &1     &NH$_3$    &1098        &9.108       &82          & 0.681 &0.9    &$\nu_2$  & umbrella    \\
\hline                                                                                                                       
1      &0      &0       &1  &1     &NH$_3$    &1062        &9.414       &66          & 0.586 &0.8    &$\nu_2$  & umbrella    \\
\hline                                                                                                                   
1      &4      &0       &0  &0     &NH$_3$    &1129$^a$    &8.856$^a$   &108$^a$     & 0.849 &0.4$^a$&$\nu_2$  & umbrella    \\
1      &4      &0       &0  &0     &CH$_3$OH  &1029        &9.722       &30          & 0.283 &--      &$\nu_{4}$& C-O stretch \\
1      &4      &0       &0  &0     &CH$_3$OH  &1128        &9.707       &35          & 0.275 &--      &$\nu_{7}$& CH$_3$ rock \\
1      &4      &0       &0  &0     &CH$_3$OH  &2823        &3.543       &28          & 0.035 &--     &$\nu_{2}$& C-H stretch \\
\hline                                                                                                                       
1      &2      &0       &0  &0     &NH$_3$    &1111$^a$    &8.994$^a$   &115$^a$     & 0.934 &0.6$^a$&$\nu_2$  & umbrella    \\
1      &2      &0       &0  &0     &CH$_3$OH  &1029        &9.720       &29          & 0.274 &--      &$\nu_{4}$& C-O stretch \\
1      &2      &0       &0  &0     &CH$_3$OH  &1132        &8.833       &35          & 0.273 &--      &$\nu_{7}$& CH$_3$ rock \\
1      &2      &0       &0  &0     &CH$_3$OH  &2820        &3.546       &26          & 0.033 &--      &$\nu_{2}$& C-H stretch \\
\hline                                                                                                                     
1      &1      &0       &0  &0     &NH$_3$    &1086        &9.209       &137         & 1.166 &0.8    &$\nu_2$  & umbrella    \\
1      &1      &0       &0  &0     &CH$_3$OH  &1029        &9.716       &26          & 0.246 &--      &$\nu_{4}$& C-O stretch \\
1      &1      &0       &0  &0     &CH$_3$OH  &1135        &8.813       &44          & 0.342 &--      &$\nu_{7}$& CH$_3$ rock \\
1      &1      &0       &0  &0     &CH$_3$OH  &2817        &3.550       &26          & 0.033 &--      &$\nu_{2}$& C-H stretch \\
\hline                                                                                                                    
1      &0.5    &0       &0  &0     &NH$_3$    &1080        &9.258       &118         & 1.015 &0.8    &$\nu_2$  & umbrella    \\
1      &0.5    &0       &0  &0     &CH$_3$OH  &1030        &9.711       &22          & 0.207 &--      &$\nu_{4}$& C-O stretch \\
1      &0.5    &0       &0  &0     &CH$_3$OH  &1128$^a$    &8.865$^a$   &35$^a$      & 0.275 &--      &$\nu_{7}$& CH$_3$ rock \\
1      &0.5    &0       &0  &0     &CH$_3$OH  &2813        &3.555       &27          & 0.034 &--      &$\nu_{2}$& C-H stretch \\
\hline                                                                                      
1      &0.25   &0       &0  &0     &NH$_3$    &1078        &9.278       &98          & 0.845 &0.9    &$\nu_2$  & umbrella    \\
1      &0.25   &0       &0  &0     &CH$_3$OH  &1030        &9.707       &16          & 0.151 &--	    &$\nu_{4}$& C-O stretch \\
1      &0.25   &0       &0  &0     &CH$_3$OH  &--$^a$ &--$^a$ &--$^a$ &--$^a$&--    &$\nu_{7}$& CH$_3$ rock  \\
1      &0.25   &0       &0  &0     &CH$_3$OH  &2808$^a$    &3.561$^a$   &17$^a$      & 0.022 &--$^a$  &$\nu_{2}$& C-H stretch \\
\hline                                                                                                                   
1      &1      &1       &0  &0     &NH$_3$    &1116$^a$    &8.961       &95          & 0.764 &0.7    &$\nu_2$ & umbrella    \\  
1      &1      &1       &0  &0     &CH$_3$OH  &1026        &9.745       &29          & 0.276 &--      &$\nu_{4}$& C-O stretch \\
1      &1      &1       &0  &0     &CH$_3$OH  &1125$^a$    &8.888$^a$   &32$^a$      & 0.253 &--      &$\nu_{7}$& CH$_3$ rock \\
1      &1      &1       &0  &0     &CH$_3$OH  &2824        &3.541       &26          & 0.033 &--      &$\nu_{2}$& C-H stretch \\
\hline                                                                                                                       
1      &0.25   &10      &0  &0     &NH$_3$    &1119        &8.937       &59          & 0.472 &1      &$\nu_2$  & umbrella    \\
1      &0.25   &10      &0  &0     &CH$_3$OH  &1017        &9.833       &22          & 0.213 &--      &$\nu_{4}$& C-O stretch \\
1      &0.25   &10      &0  &0     &CH$_3$OH  &--$^a$ &--$^a$ &--$^a$ &--$^a$ &--   &$\nu_{7}$& CH$_3$ rock \\
1      &0.25   &10      &0  &0     &CH$_3$OH  &2829$^a$    &3.534$^a$   &30$^a$      & 0.037 &--      &$\nu_{2}$& C-H stretch \\
\hline                                                                                                                       
1      &1      &10      &0  &0     &NH$_3$    &1123        &8.903       &61          & 0.484 &1      &$\nu_2$  & umbrella    \\
1      &1      &10      &0  &0     &CH$_3$OH  &1022        &9.784       &24          & 0.230 &--      &$\nu_{4}$& C-O stretch \\
1      &1      &10      &0  &0     &CH$_3$OH  &--$^a$ &--$^a$ &--$^a$ &--$^a$ &--  &$\nu_{7}$& CH$_3$ rock \\
1      &1      &10      &0  &0     &CH$_3$OH  &2830        &3.533       &15          & 0.019 &--      &$\nu_{2}$& C-H stretch \\
\hline                                                                                                                       
1      &4      &10      &0  &0     &NH$_3$    &1130        &8.848       &62          & 0.489   &--      &$\nu_2$  & umbrella    \\
1      &4      &10      &0  &0     &CH$_3$OH  &1023        &9.777       &30          & 0.288   &--      &$\nu_{4}$& C-O stretch \\
1      &4      &10      &0  &0     &CH$_3$OH  &1124        &8.896       &23          & 0.183   &--      &$\nu_{7}$& CH$_3$ rock \\
1      &4      &10      &0  &0     &CH$_3$OH  &2830        &3.534       &14          & 0.017   &--      &$\nu_{2}$& C-H stretch \\
\hline
0      &1      &0       &0  &0     &CH$_3$OH  &1028        &9.729       &28          & 0.265   &1      &$\nu_{4}$& C-O stretch \\
0      &1      &0       &0  &0     &CH$_3$OH  &1125        &8.888       &34          & 0.269   &1      &$\nu_{7}$& CH$_3$ rock \\	
0      &1      &0       &0  &0     &CH$_3$OH  &2828        &3.536       &33          & 0.041   &1      &$\nu_{2}$& C-H stretch \\
\hline                                                                                                                         
0      &1      &1       &0  &0     &CH$_3$OH  &1025        &9.755       &33          & 0.314   &--      &$\nu_{4}$& C-O stretch \\
0      &1      &1       &0  &0     &CH$_3$OH  &1124        &8.897       &40          & 0.317   &--      &$\nu_{7}$& CH$_3$ rock \\
0      &1      &1       &0  &0     &CH$_3$OH  &2828        &3.536       &23          & 0.029   &--      &$\nu_{2}$& C-H stretch \\
\hline                                                                                          
0      &1      &9       &0  &0     &CH$_3$OH  &1020        &9.801       &23          & 0.221   &--      &$\nu_{4}$& C-O stretch \\
0      &1      &9       &0  &0     &CH$_3$OH  &1126        &8.883       &13          & 0.103   &--      &$\nu_{7}$& CH$_3$ rock \\
0      &1      &9       &0  &0     &CH$_3$OH  &2828        &3.536       &23          & 0.029   &--      &$\nu_{2}$& C-H stretch \\
\hline                                                                                                                         
0      &1      &0       &9  &0     &CH$_3$OH  &1034        &9.675       &25          & 0.229   &--      &$\nu_{4}$& C-O stretch \\
0      &1      &0       &9  &0     &CH$_3$OH  &1119        &8.938       &30          & 0.242   &--      &$\nu_{7}$& CH$_3$ rock \\
0      &1      &0       &9  &0     &CH$_3$OH  &2831        &3.532       &--     & -- &--      &$\nu_{2}$& C-H stretch \\
0      &1      &0       &9  &0     &CO        &2138        &4.677       &7           & 0.014   &--     &$\nu_{1}$& C-O stretch \\
\hline                                                                                                                      
0      &1      &0       &1  &0     &CH$_3$OH  &1029        &9.720       &30          & 0.286   &--      &$\nu_{4}$& C-O stretch \\
0      &1      &0       &1  &0     &CH$_3$OH  &1124        &8.898       &32          & 0.258   &--      &$\nu_{7}$& CH$_3$ rock \\
0      &1      &0       &1  &0     &CH$_3$OH  &2830        &3.534       &--     & -- &--      &$\nu_{2}$& C-H stretch \\
0      &1      &0       &1  &0     &CO        &2136        &4.682       &9           & 0.020   &--      &$\nu_{1}$& C-O stretch \\
\hline                                                                                                                     
0      &9      &0       &1  &0     &CH$_3$OH  &1028        &9.730       &28          & 0.261   &--      &$\nu_{4}$& C-O stretch \\
0      &9      &0       &1  &0     &CH$_3$OH  &1125        &8.890       &32          & 0.255   &--      &$\nu_{7}$& CH$_3$ rock \\
0      &9      &0       &1  &0     &CH$_3$OH  &2824        &3.541       &--     & -- &--      &$\nu_{2}$& C-H stretch \\
0      &9      &0       &1  &0     &CO        &2135        &4.685       &9           & 0.021   &--      &$\nu_{1}$& C-O stretch \\
\enddata
\tablenotetext{a}{Band is weak and spectral overlap prohibits accurate fitting.}
\end{deluxetable}

\clearpage

\section{Comparison between astronomical and laboratory data for all sources}
\label{ap: compare astro lab}

\begin{figure*}[ht]
\centering
\includegraphics[width=0.5\textwidth]{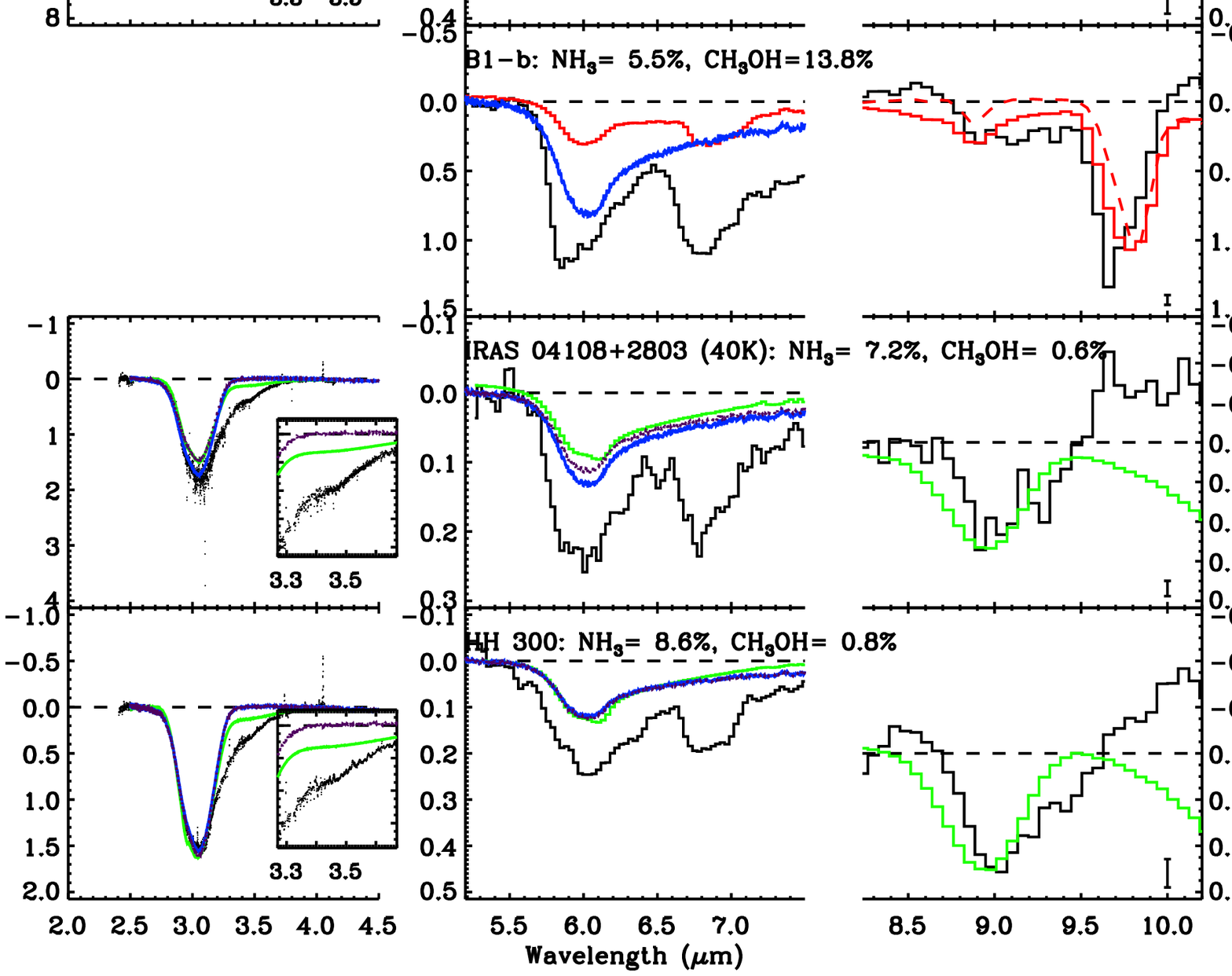}%
\includegraphics[width=0.5\textwidth]{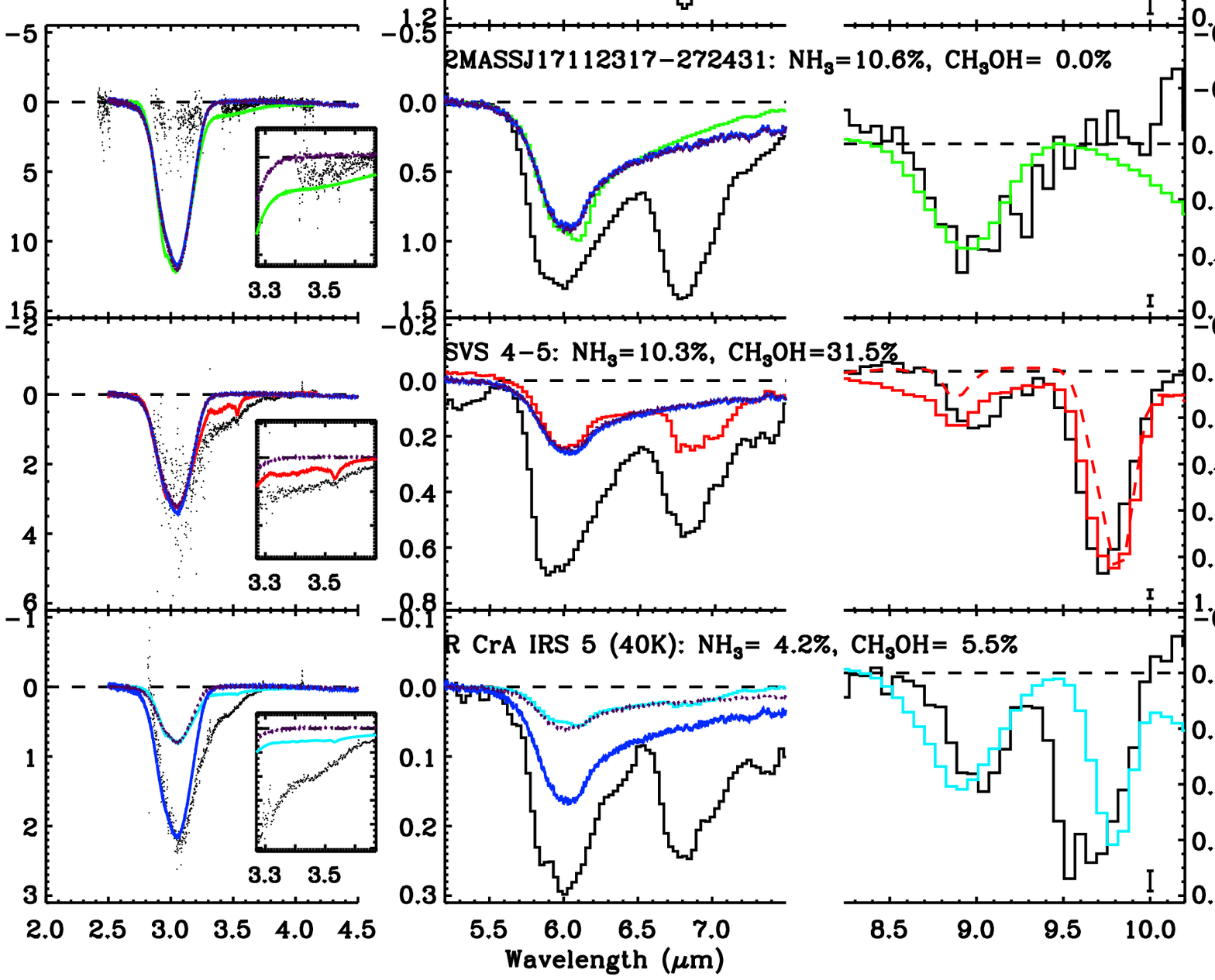}
\caption{\footnotesize (a) Comparison between astronomical and laboratory data for sources whose
silicate absorption feature was fitted with a template.
For a given source (displayed in either the left or right column of the figure), the middle and right panels show
5.2-7.5 and 8.2-10.2~\micron\ 
regions from IRS \spitzer\ spectra overlaid with laboratory spectra, 
scaled to the 9-\micron\ \ammonia\ umbrella mode, and smoothed to the \spitzer\ resolution.
Error bars for the \spitzer\ spectra are indicated in the bottom-right corner.
The dark blue line represents the pure water laboratory spectrum scaled to the water column density 
taken in paper I.
Other colors are representative of laboratory spectra obtained for the following mixtures: \water:\ammonia=9:1 (green), 
\water:\m:\ammonia=10:0.25:1 (orange), \water:\m:\ammonia=10:1:1 (cyan), and \water:\m:\ammonia=10:4:1 (red).
When available (see \citealt{boogert-etal08}), VLT or Keck data (2.0-4.5~\micron, left panel) are also plotted.
In this case, we overplotted (dotted purple line) a pure water spectrum scaled to the 3-\micron\ water feature of the mixed
ice spectrum. Whenever present, a red dashed line in the right panel of a given source represents a \water:\m=9:1 laboratory spectrum
scaled to the 9.7-\micron\ \m\ CO-stretch mode: this gives an indication of the contribution of the 9-\micron\
\m\ CH$_3$-rock mode to the total 9-\micron\ feature.
The laboratory spectra are recorded at 15~K unless indicated differently.
\label{fig:obs lab} }
\end{figure*}

\begin{figure*}[ht]
\centering
\addtocounter{figure}{-1}
\includegraphics[width=0.5\textwidth]{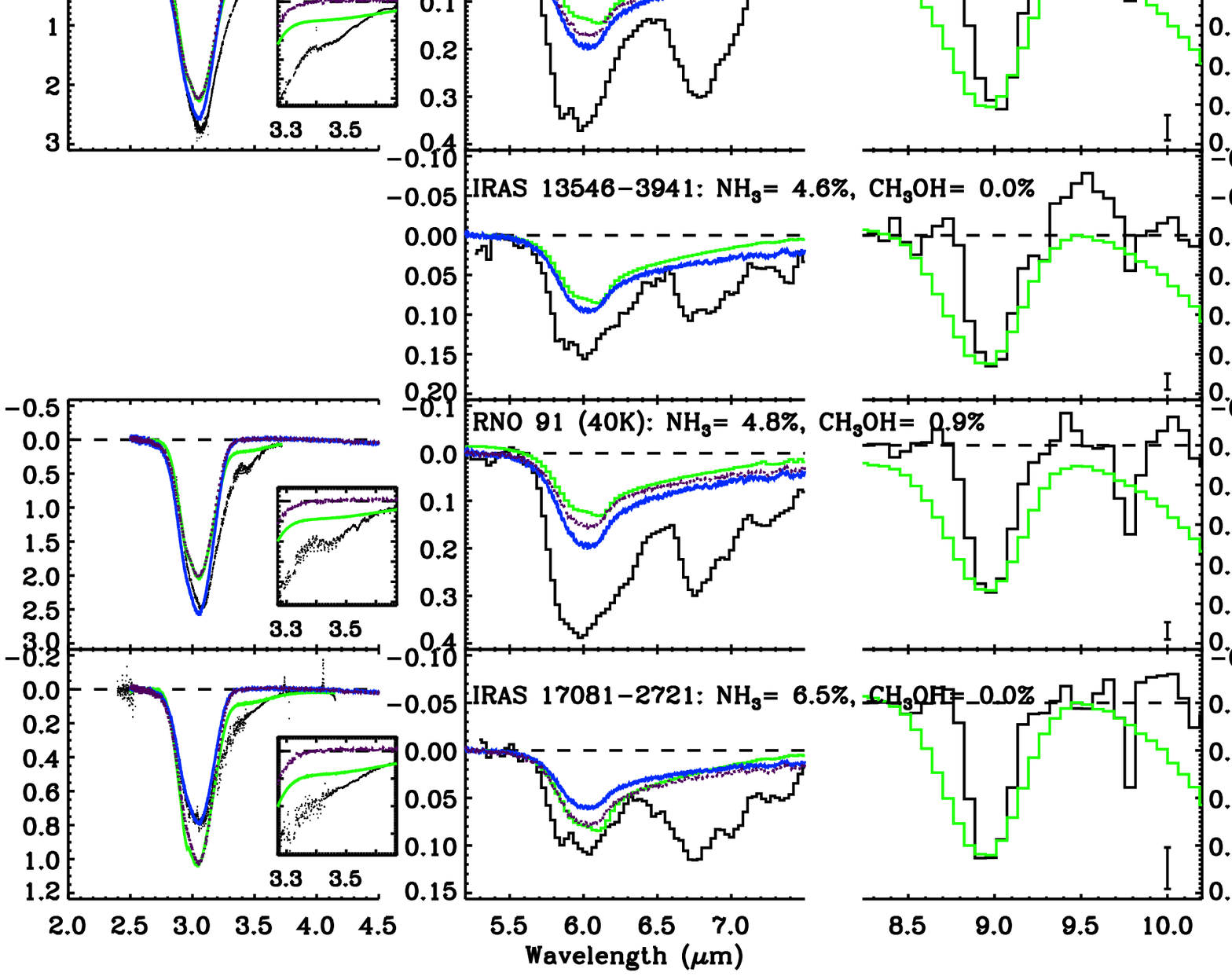}%
\includegraphics[width=0.5\textwidth]{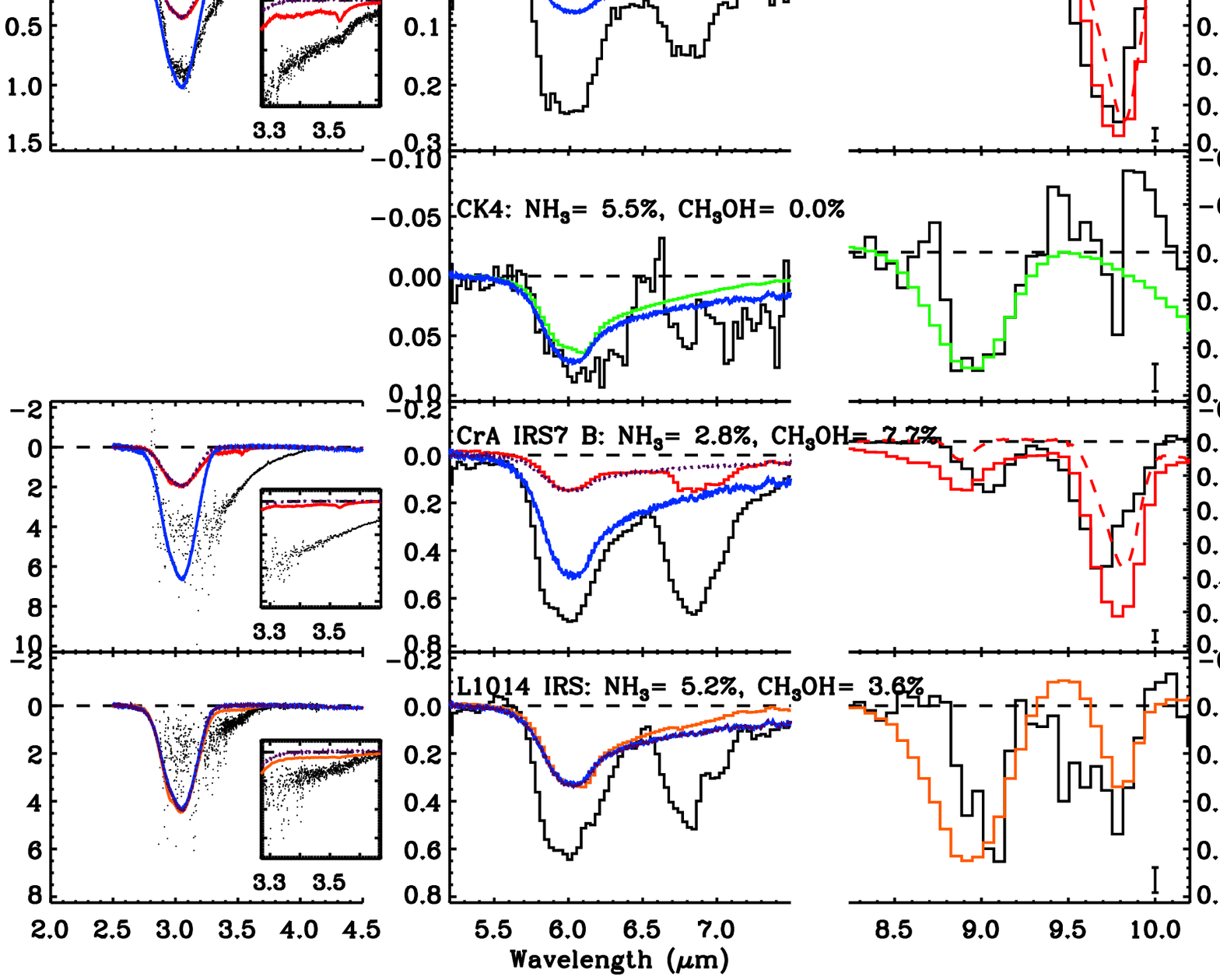}
\caption{\footnotesize(b) As (a) but for sources with no associated template, i.e. with the 10-\micron\
silicate feature subtracted via the local continuum method. Additionally, yellow
represents \water:\ammonia=4:1 (H. Frazer, priv. comm.).}
\end{figure*}

\clearpage

\bibliographystyle{/Users/bottinelli/Documents/biblio/apj} 
\bibliography{/Users/bottinelli/Documents/biblio/bib_sb}

\end{document}